\documentclass[prc,aps,float,twocolumn,showpacs,nofootinbib,superscriptaddress]{revtex4}

\usepackage{amsmath}
\usepackage{amssymb}
\usepackage{amsfonts}
\usepackage{graphicx}
\usepackage[usenames,dvipsnames]{color}
\usepackage{afterpage}
\usepackage{mathrsfs}

\allowdisplaybreaks

\renewcommand{\vec}[1]{{\bf #1}}
\newcommand{\nuc}[2]{$^{#1}${#2}}

\newcommand{\nn}{\nonumber}

\newcommand{\sigmavec}{\boldsymbol{\mathbf\sigma}}
\newcommand{\tauvec}{\boldsymbol{\mathbf\tau}}
\newcommand{\vnabla}{\boldsymbol{\mathbf\nabla}}
\newcommand{\vsigma}{\boldsymbol{\mathbf\sigma}}

\newcommand{\fourmat}[4]{{\begin{pmatrix}
                         {#1} & {#2} \\ {#3} & {#4}
                         \end{pmatrix}}}
\newcommand{\twospinor}[2]{{\begin{pmatrix}
                           {#1} \\ {#2}
                           \end{pmatrix}}}
\newcommand{\fourrow}[4]{{\begin{pmatrix}
                           {#1}\\{#2}\\{#3}\\{#4}
			   \end{pmatrix}}}

\newcommand{\emb}{{\rule{0cm}{0cm}}}

\newcommand{\dyn}{$\mathcal{J}^{(2)}$}	
\newcommand{\ho}{$\hbar\omega$}

\begin{document}
\title{The tensor part of the Skyrme energy density functional.\\
III. Time-odd terms at high spin\emb\ }

\author{V. Hellemans}
\affiliation{Universit\'e Libre de Bruxelles,Physique Nucl\'eaire
  Th\'eorique, CP229, B-1050 Bruxelles, Belgium}
 \affiliation{University of Notre Dame, Department of Physics,\\
    225 Nieuwland Science Hall, Notre Dame IN 46556-5670, USA}

\author{P.-H. Heenen}
\affiliation{Universit\'e Libre de Bruxelles,Physique Nucl\'eaire
  Th\'eorique, CP229, B-1050 Bruxelles, Belgium}

\author{M. Bender}
\affiliation{Univ. Bordeaux,
             Centre d'Etudes Nucl{\'e}aires de Bordeaux Gradignan,\\
             UMR5797, F-33170 Gradignan, France}
\affiliation{CNRS, IN2P3,
             Centre d'Etudes Nucl{\'e}aires de Bordeaux Gradignan,\\
             UMR5797, F-33170 Gradignan, France}

\begin{abstract}

This article extends previous studies on the effect of tensor terms in the Skyrme energy density functional by breaking of time-reversal invariance. We have systematically probed the impact of tensor terms on properties of superdeformed rotational bands calculated within the cranked Hartree-Fock-Bogoliubov approach for different parameterizations covering a wide range of values for the isoscalar and isovector tensor coupling constants. We analyze in detail the contribution of the tensor terms to the energies and dynamical moments of inertia and study their impact on quasi-particle spectra. Special attention is devoted to the time-odd tensor terms, the effect of variations of their coupling constants and finite-size instabilities.
\end{abstract}


\maketitle


\section{Introduction}
\label{sect:intro}

Recent years have seen a renewed interest in the role of the
effective nucleon-nucleon tensor force for nuclear structure,
sparked by the finding that it provides one of the possible sources
for the evolution of nuclear shell structure with neutron and
proton numbers. Indeed, the contribution of tensor interactions to
single-particle energies depends on the filling of shells. It (nearly)
vanishes in spin-saturated nuclei, whereas it might be significant
when only one out of two spin-orbit partner levels is filled
for one or even both nucleon species~\cite{Ots05a}.

Up to now, none of the standard parameterizations of any mean-field
approach considered an explicit tensor part, cf.\ Ref.~\cite{lesinski07}
for a historical overview.
The first studies of the effective tensor interaction within
self-consistent mean-field approaches concentrated on
single-particle spectra in chains of semi-magic spherical nuclei
covering all successful models, i.e.\ the non-relativistic Gogny
force~\cite{Ots06a,Ang11a} and Skyrme
interactions~\cite{Bro06a,colo07,lesinski07,zalewski08}, as well as
relativistic mean-field approaches~\cite{Lon08a,Lal09a}.
More recently, the impact of the tensor terms on
more complex structure properties has been studied as well,
such as the topography of deformation energy surfaces~\cite{bender09}
and various spin- and spin-isospin excitation modes in Quasiparticle
Random Phase Approximation (QRPA) using Skyrme functionals
\cite{Bai09a,Cao09a,Bai09b,cao10,Bai10a,Bai10b,Cao11a}
and Gogny interactions~\cite{Ang11a}.

These QRPA calculations deal with a very different aspect of
an effective tensor interaction than the analysis of single-particle
energies. This becomes most obvious when using the Skyrme
energy density functional (EDF). The Skyrme EDF can be separated into
two parts: the first one composed of densities and currents that are
even under time reversal such as the normal and kinetic densities,
and a second one grouping combinations of time-odd densities such as spin
density or current. The latter part of the EDF is usually called the
``time-odd'' part, although strictly speaking the EDF itself
is time-even by construction. While these time-odd densities are
zero for the HFB ground states of even-even nuclei, they become
non-zero for
\begin{enumerate}
\item[(i)]
blocked quasiparticle states, i.e.\ the self-consistent calculation
of non-collective low-lying states in odd-$A$ and odd-odd nuclei, or $K$
isomers in even-even ones. The time-odd terms contribute to the total
energy~\cite{Rut98a,Rut99a,Dug01a,schunk10,zalewski08,pototzky10,Afa10a}
and their presence can strongly modify the expectation values
of time-odd observables such as magnetic moments~\cite{Lip77a,Hof88a},
\item[(ii)]
rotational states calculated by the cranked HFB method
\cite{post85,bonche87,dobaczewski95,Zdu05a,Afa00a,afanasjev08,afanasjev10},
where they affect the alignment of single-particle levels with
the rotational axis and thereby the moments of inertia,
\item[(iii)]
time-dependent Hartree-Fock (-Bogoliubov) (TDHF(B))
\cite{engel75,Mar06a,Nes08a} and its linear response limit,
the (Quasiparticle) Random Phase Approximation ((Q)RPA),
\item[(iv)]
configuration mixing such as symmetry restoration or
Generator Coordinate Method calculations, or adiabatic
Time-dependent Hartree-Fock-Bogoliubov calculations (ATDHFB)
\cite{engel75,Bon90a,Hin06a}.
\end{enumerate}
The study of excitation modes of unnatural parity,
such as for example $M1$, spin-dipole or Gamov-Teller excitations in
(Q)RPA provides a sensitive benchmark for the time-odd
terms in the EDF. Indeed, for those the residual interaction is entirely
determined by the time-odd terms, cf., for example,
Refs.~\cite{Ber81a,Gia81a,bender02,Fra07a,Ves09a,Nes10a}
and references therein. For the other phenomena listed above, the
time-odd terms provide a correction to the dominant time-even terms
that might become substantial in some cases.
One such observable are the moments of inertia at high spin in
superdeformed (SD) rotational bands of heavy nuclei
\cite{janssens91a,wadsworth02a,satula05a}. These will be the
object of our study.

Skyrme's two-body tensor force contributes to the time-even and
time-odd parts of the EDF. Studies of the eigenvalue spectrum
of the single-particle Hamiltonian of even-even nuclei only probe
the contribution to the time-even part. The corresponding time-odd
terms affect how the nucleus responds to its collective rotation,
thereby modifying its moment of inertia and how it evolves with spin. Of course, the time-even tensor terms also influence the
moments of inertia.

The aim of the present study is to investigate the generic
influence of tensor terms on high-spin properties. The
following questions will be addressed:
\begin{itemize}
\item
How do the time-odd tensor terms behave when increasing the total
spin of the nucleus?
\item
How does the presence of time-even and time-odd tensor terms
influence the dynamical moments of inertia in
superdeformed rotational bands at high spin?
\item
How much of these changes is caused by the time-even part of
the EDF, i.e.\ the modification of the single-particle spectrum
at spin zero, and how much by the time-odd part of the EDF?
\item
How much of these changes is caused by the tensor terms
themselves, and how much is caused by the rearrangement of
all other terms during the fit of the parameterizations?
\end{itemize}
Studies of the impact of time-odd terms on the moments of inertia
in superdeformed bands have been performed before in the context
of Skyrme interactions~\cite{dobaczewski95,bender02}
and relativistic mean-field Lagrangians~\cite{Afa00a,afanasjev10},
but none of these studies considered time-odd terms associated
with genuine tensor interactions.

The present article complements the studies of spherical single-particle
energies of Ref.~\cite{lesinski07} and of the deformation energy curves
of Ref.~\cite{bender09}. We will refer to these references as
Articles~I and~II. The present Article is structured as follows:
In Section~\ref{sect:SkyrmeEDF} we briefly review the properties
relevant for our discussion of the Skyrme EDF including tensor terms.
In Sect.~\ref{sect:hg194}, we analyze in detail how sensitive is the
superdeformed rotational band in \nuc{194}{Hg} when
the coupling constants of the Skyrme EDF are varied.
In Sect.~\ref{sect:dy152}, we check the generality of our conclusions
for \nuc{194}{Hg} by similar calculations for the SD ground-state band
of \nuc{152}{Dy}, and Sect.~\ref{sect:conclusions} summarizes our results.
Appendices provide further technical information about the interrelations
between the coupling constants of the Skyrme EDF (Appendix~\ref{app:cpling}),
the representation of local densities and currents in our code
(Appendix~\ref{app:cr8}) and about the Landau-Migdal interaction
corresponding to the standard Skyrme EDF with tensor terms
(Appendix~\ref{app:LM interaction}).
%
%
\section{The self-consistent mean-field method}
\label{sect:SkyrmeEDF}

The energy of the atomic nucleus can be expressed by means of an
energy density functional \cite{chabanat97,chabanat98,bender03,perlinska04},
containing five parts: the kinetic energy, a Skyrme potential
energy functional modeling the strong force in the particle-hole
channel, a pairing energy functional, a Coulomb energy functional,
and terms to approximately correct for the spurious-motion
caused by broken symmetries
\begin{equation}
\label{eq:EDfunctional}
\mathcal{E}
=   \mathcal{E}_{\text{ kin}}
  + \mathcal{E}_{\text{Sk}}
  + \mathcal{E}_{\text{pairing}}
  + \mathcal{E}_{\text{Coulomb}} 		
  + \mathcal{E}_{\text{corr}}
\, .
\end{equation}
For the kinetic energy and the Coulomb energy functional comprising
a direct term and the exchange term in Slater approximation we
use the same expressions as presented in Ref.~\cite{chabanat98}.
For all parameterizations used throughout this Article, the
center-of-mass recoil effect is approximately taken into account
by subtracting
$\mathcal{E}_{\text{corr}} = \big\langle\sum_{k} p_{k}^{2}\big\rangle / 2m A$
from the total energy, which amounts to an $A$-dependent renormalization
of the nucleon mass.
In the following Sections we will introduce the
ingredients of the Skyrme EDF, its explicit form and the equations of motion.
%
%
\subsection{Densities and currents}
\label{subsection:local densities}
Under the assumption that the single-particle states are either neutron
or proton states, the Skyrme part of the energy density functional
$\mathcal{E}_{\text{Sk}}$ depends on the following local densities
and currents
\begin{subequations}
\begin{align}
\label{eq:locdensities:rho}
\rho_q (\vec{r})
& =  \rho_q (\vec{r},\vec{r}') \big|_{\vec{r} = \vec{r}'}
      \, ,
      \\
\label{eq:locdensities:tau}
\tau_q (\vec{r})
& =  \vnabla \cdot \vnabla' \; \rho_q (\vec{r},\vec{r}')
      \big|_{\vec{r} = \vec{r}'}
      \, ,
      \\
\label{eq:locdensities:J}
J_{q,\mu \nu}(\vec{r})
& =  - \tfrac{i}{2} (\nabla_\mu - \nabla_\mu^\prime) \;
      s_{q, \nu} (\vec{r},\vec{r}') \big|_{\vec{r} = \vec{r}'}
	\, , 	\\ \label{eq:locdensities:j} \vec{j}_q (\vec{r})
& = \tfrac{1}{2i}(\vnabla - \vnabla^\prime) \; \rho_q (\vec{r},\vec{r}')
    \big|_{\vec{r} = \vec{r}'} 	\, , 	\\
\label{eq:locdensities:s}
\vec{s}_q (\vec{r})
& = \vec{s}_q (\vec{r},\vec{r}') \big|_{\vec{r} = \vec{r}'}
      \, ,
      \\
\label{eq:locdensities:T}
\vec{T}_q (\vec{r})
& =  \vnabla \cdot \vnabla' \; \vec{s}_q (\vec{r},\vec{r}')
      \big|_{\vec{r} = \vec{r}'}
      \, ,
      \\
\label{eq:locdensities:F}
\vec{F}_{\mu,q} (\vec{r})
& =  \tfrac{1}{2} \sum_{\nu=x,y,z}
      (   \nabla_{\mu}  \nabla^\prime_{\nu}
            + \nabla^\prime_{\mu}  \nabla_{\nu}  )
       \,  s_{q,\nu} (\vec{r}, \vec{r}') \big|_{\vec{r} = \vec{r}'}
      \, .
\end{align}
\end{subequations}
which are the density $\rho_q (\vec{r})$, the kinetic
density $\tau_q (\vec{r})$, the spin-current (pseudo-tensor) density
$J_{q,\mu\nu}(\vec{r})$, the current (vector) density $\vec{j}_q (\vec{r})$,
the spin (pseudo-vector) density $\vec{s}_q (\vec{r})$, the spin-kinetic
(pseudo-vector) density $\vec{T}_q (\vec{r})$, and the
tensor-kinetic (pseudo-vector) density $\vec{F}_{q} (\vec{r})$ for protons
and neutrons ($q=n,p$). All  densities and currents can be recoupled to
isoscalar $t=0$ and isovector ($t=1$, $t_3 = 0$) densities (e.g.\
$\rho_0(\vec{r}) = \rho_p(\vec{r}) + \rho_n(\vec{r})$ and
$\rho_1(\vec{r}) = \rho_n(\vec{r})-\rho_p(\vec{r})$). They are
constructed from the density matrix in coordinate space and its
derivatives up to second order \cite{perlinska04,dobaczewski00}
\begin{align}
\label{eq:density}
\rho_q(\vec{r},\sigma,\vec{r}' \sigma')
& = \langle a^\dag(\vec{r}',\sigma',q) a(\vec{r},\sigma,q) \rangle \, ,
    \nn \\ 				
& =   \tfrac{1}{2} \rho_q (\vec{r},\vec{r}') \delta_{\sigma\sigma'}
    + \tfrac{1}{2} \langle \sigma|\hat{\vsigma}| \sigma' \rangle \cdot \vec{s}_q (\vec{r},\vec{r}') \, ,
\end{align}
with
\begin{align} \rho_q (\vec{r}, \vec{r}')
& =  \sum_{\sigma=\pm 1} \rho_q
(\vec{r} \sigma ,\vec{r}' \sigma) 	\, ,\\
\vec{s}_q (\vec{r},\vec{r}')
& =  \sum_{\sigma \sigma'=\pm 1} \rho_q (\vec{r} \sigma ,\vec{r}' \sigma') \;
      \langle \sigma' | \hat{\vsigma} | \sigma \rangle
      \, ,
\end{align}
and $\hat{\vsigma}$ are the Pauli spin matrices. From the properties
of the density and the spin density matrices under time-reversal
\cite{engel75}
\begin{align}
\rho_q^T (\vec{r},\vec{r}')
& = \rho_q (\vec{r}',\vec{r}) \, ,
    \\
\vec{s}_q^T (\vec{r},\vec{r}')
& = -\vec{s}_q (\vec{r}',\vec{r}) \, ,
\end{align}
it follows that
\begin{align}
 \rho_q^T(\vec{r})
& = \rho_q(\vec{r})
	\,  ,
\tau_q^T(\vec{r})
  = \tau_q(\vec{r}) 	\,  ,
J_{q,\mu\nu}^T(\vec{r})=J_{q,\mu\nu}(\vec{r}) 	\,  , 	 \nn\\
 \vec{s}_q^T(\vec{r})&=-\vec{s}_q(\vec{r})
	\,  ,\vec{j}_q^T(\vec{r})=-\vec{j}_q(\vec{r}) 	\,  ,
\vec{T}_q^T(\vec{r})=-\vec{T}_q(\vec{r}) 	\,  , 	\nn\\
\vec{F}_q^T(\vec{r})&=-\vec{F}_q(\vec{r})
	\,  . 
\end{align}
One notes that $\rho_q(\vec{r})$, $\tau_q(\vec{r})$, and
$J_{q,\mu\nu}(\vec{r})$ are time-even, whereas $\vec{s}_q(\vec{r})$,
$\vec{j}_q(\vec{r})$, $\vec{T}_q(\vec{r})$, and $\vec{F}_q(\vec{r})$
are time-odd. When time reversal is a self-consistent
symmetry, the time-odd densities vanish. When intrinsic time-reversal
invariance is broken, both time-even and time-odd densities contribute
to the energy and the single-particle Hamiltonian.

For the part of the EDF that describes the pairing correlations,
$\mathcal{E}_{\text{pairing}}$, we also need to introduce
the skew-symmetric pair tensor \cite{ring80}
\begin{equation}
\kappa_{q}(\vec{r},\sigma,\vec{r}',\sigma')
= \langle a(\vec{r}',\sigma',q) a(\vec{r},\sigma,q) \rangle \, ,
\end{equation}
or, equivalently, the pair density matrix \cite{dobaczewski84}
\begin{align}
\tilde{\rho}_{q}(\vec{r},\sigma,\vec{r}',\sigma')
& = -\sigma'\langle a(\vec{r}', -\sigma', q)a(\vec{r}, \sigma, q)\rangle
     \nn \\
& = -\sigma' \kappa_{q}(\vec{r},\sigma,\vec{r}',-\sigma')  \, .
\end{align}
The pair density matrix presents the interest that it enables to construct a local pair density
\begin{equation}
\tilde{\rho}_{q}(\vec{r})
= \sum_{\sigma=\pm 1} \tilde{\rho}_{q}(\vec{r},\sigma,\vec{r},\sigma)
\, ,
\end{equation}
that facilitates the construction of a local pairing EDF as used here. The pair density $\tilde{\rho}_{q}(\vec{r})$
is neither time-even nor time-odd and becomes complex when intrinsic time-reversal symmetry is broken.

%
%
\subsection{The energy density functional}
%
%
\subsubsection{The Skyrme energy functional}
\label{subsect:skyrmeEDF}

Skyrme functionals have been discussed extensively in the literature
\cite{vautherin72,engel75,bonche87,dobaczewski00,bender03,perlinska04,lesinski07}.
We restrict ourselves here to aspects relevant for this work.
As outlined in the Introduction, the Skyrme EDF can be separated into two parts
\begin{equation}
\mathcal{E}_{\text{Sk}}
= \int d^3r \sum_{t=0,1}
		\left[ \mathcal{H}^{\text{t.e.}}_t(\vec{r}) 		
  + \mathcal{H}^{\text{t.o.}}_t(\vec{r}) \right]
\, ,
\end{equation}
a part $\mathcal{H}_{t}^{\text{t.e.}}$ that contains only
time-even densities
\begin{subequations}
\begin{align}
\label{eq:EF:time-even}
\mathcal{H}_t^{\text{t.e.}}
 = \quad&    C^\rho_t[\rho_{0}] \rho_t^2
             + C^{\Delta \rho}_t  \rho_t \Delta \rho_t
             + C^\tau_t            \rho_t \tau_t
             + C^{\nabla \cdot J}_t  \rho_t \nabla \cdot \vec{J}_t
      \nn \\
&
             - C^{T}_t  \sum_{\mu, \nu = x}^{z} J_{t, \mu \nu} J_{t, \mu \nu}
      \nn \\
&
             - \tfrac{1}{2} C^{F}_t \Big[
                    \Big( \sum_{\mu = x}^{z} J_{t,\mu \mu}
                                 \Big)^2
                 +
                   \sum_{\mu, \nu = x}^{z} J_{t, \mu \nu} J_{t, \nu \mu}
                \Big]
\end{align}
and another one $\mathcal{H}_{t}^{\text{t.o.}}$
that contains bilinear combinations of time-odd densities
\begin{align}
\label{eq:EF:time-odd}
\mathcal{H}_t^{\text{t.o.}}
 = \quad&   C^s_t [\rho_{0}]   \vec{s}_t^2
             + C^{\nabla s}_t     (\nabla \cdot \vec{s}_t)^2
             + C^{\Delta s}_t     \vec{s}_t \cdot \Delta \vec{s}_t
             - C^\tau_t           \vec{j}_t^2
      \nn \\
   &
             + C^{T}_t   \vec{s}_t \cdot \vec{T}_t
             + C^{F}_t   \vec{s}_t \cdot \vec{F}_t
	     + C^{\nabla \cdot J}_t \vec{s}_t \cdot \nabla \times \vec{j}_t
\, .
\end{align}
\end{subequations}
The common practice to call these energy densities ``time-even'' and ``time-odd''
refers to the properties under time reversal of the densities and currents they are built
from and not to the properties of the energy density itself, as the EDF is  time-reversal invariant by construction~\cite{dobaczewski96}.

All coupling constants of \eqref{eq:EF:time-even} and \eqref{eq:EF:time-odd}
could be chosen to be density dependent.
\footnote{
If all coupling constants are density-dependent, there appear additional terms containing the nabla or Laplacian operator acting on a local density, as some
bilinear forms are not equivalent under partial integration anymore~\cite{dobaczewski96}.
}
In practice, however, the density-dependence is usually restricted to
the $C^{\rho}_t$ and $C^{s}_t$ coupling constants, and chosen to be
a non-integer power of the isoscalar density \cite{vautherin72,beiner75}
\begin{align}
C^{\rho}_{t}[\rho_{0}]
& = C^{\rho}_{t}[0] + \Big( C^{\rho}_{t}[\rho_{\text{nm}}] - C^{\rho}_{t}[0] \Big)
    \left( \frac{\rho_{0}}{\rho_{\text{nm}}}\right)^{\alpha} \ , \\
C^{s}_{t}[\rho_{0}]
& = C^{s}_{t}[0] + \Big( C^{s}_{t}[\rho_{\text{nm}}] - C^{s}_{t}[0] \Big)
    \left( \frac{\rho_{0}}{\rho_{\text{nm}}}\right)^{\alpha}
\ ,
\end{align}
where $\rho_{\text{nm}}$ is the value of the isoscalar density $\rho_{0}$ in
saturated infinite nuclear matter.

%
%
\subsubsection{Choice of independent coupling constants in the EDF}
\label{subsect:choice:cpling}

The Skyrme energy density functional (EDF) can be introduced in two
non-equivalent ways~\cite{bender03}. One can start from a density-dependent
zero-range two-body interaction as proposed by Skyrme~\cite{skyrme56,skyrme58}
and define the Skyrme EDF as its Hartree-Fock expectation
value~\cite{vautherin72,engel75,perlinska04}. We refer to this approach
as force-generated EDF. It is customary that the pairing functional is
calculated from a different pairing interaction, such that only the direct and exchange terms
in the particle-hole channel refer to the same interaction, but not those in the pairing
channel. The same expression for the energy densities $\mathcal{H}_{t}^{\text{t.e.}}$ and $\mathcal{H}_{t}^{\text{t.o.}}$ is obtained
when writing down all possible bilinear combinations of the local densities
\eqref{eq:locdensities:rho}-\eqref{eq:locdensities:F} up to second order in the derivatives
that are invariant under parity, time reversal, rotations and Galilean transformations
\cite{dobaczewski95,dobaczewski96}.
Then, the coupling constants of most time-odd terms can be chosen independently
of those of the time-even terms. By contrast, in a force-generated EDF, the coupling
constants of time-odd terms depend on those of the time-even ones; the 9 free
parameters of the most general density-independent
two-body Skyrme force determines all 18 coupling constants $C$ of the
corresponding EDF (\ref{eq:EF:time-even}-\ref{eq:EF:time-odd}). The various possible
choices for the coupling constants of time-odd terms that correspond to the same
time-even part of the EDF have been reviewed in Article~II.

Galilean invariance is a necessary constraint to obtain a functional that depends
only on the relative momenta of the nucleons, but not on the total momentum.
This invariance is particularly important for dynamical calculations such as TDHF
and TDHFB \cite{Mar06a} or cranked HFB such as performed here~\cite{Tho62a,Fle79a,Nak96a}
since it ensures that the results of the calculation will not depend on the frame
of reference. In relativistic approaches, the EDF has to
be constructed to be Lorentz invariant instead \cite{Afa00a,afanasjev10}.
Galilean invariance is automatically
fulfilled for an EDF generated from the Skyrme force. In a functional approach, it is automatically fulfilled for some
terms and has to be imposed by taking  specific combinations of other terms~\cite{engel75, dobaczewski95}.
This results in the presence  of $C^{\tau}_{t}$,
$C^{\nabla\cdot J}_{t}$, $C^{T}_{t}$, and $C^{F}_{t}$ in both the time-even
and the time-odd energy density (\ref{eq:EF:time-even}-\ref{eq:EF:time-odd}).
Since several time-odd terms of an EDF are not constrained by symmetry requirements, the number of
independent coupling constants is always smaller for a force-generated EDF. This renders the
mapping of an arbitrary local EDF onto a density-dependent Skyrme force \emph{a priori}
impossible when time-reversal invariance is broken.

In the literature, there also exist hybrid approaches
which adopt the larger freedom of functionals for 
some terms in the functional only and follow the ``force-generated''
philosophy for all others.
We will come back to this
in Sect.~\ref{sect:parameterizations}.

%
%
\subsubsection{Skyrme's tensor interaction}
\label{subsect:skyrme:tensor}

A zero-range tensor force with two terms  was originally proposed by Skyrme
\cite{skyrme56,skyrme58}
\footnote{Skyrme's tensor force is sometimes given in a different, slightly simpler form \cite{stancu77,lesinski07,bender09,wang11,dong11}. Both forms give rise to the same EDF, but not to the
same residual interaction, for example, in QRPA.}

\begin{align}
\label{eq:Skyrme:tensor}
v^{\text{t}} (\vec{r})
 =  \frac{t_{e}}{2}
    &\Big\{
    \big[ 3 ( \vsigma_1 \cdot \vec{k}' )  ( \vsigma_2 \cdot \vec{k}' )
          - ( \vsigma_1 \cdot \vsigma_2 ) \vec{k}^{\prime 2}
    \big] \delta (\vec{r})
    \nn \\
& +
     \delta (\vec{r})
      \big[ 3 ( \vsigma_1 \cdot \vec{k} ) ( \vsigma_2 \cdot \vec{k} )
            -  ( \vsigma_1 \cdot \vsigma_2)  \vec{k}^{2}
      \Big]
    \Big\}
   \nn \\
 +  \frac{t_{o}}{2}
     &\Big\{
        3( \vsigma_1 \cdot \vec{k}' ) \delta (\vec{r})
          ( \vsigma_2 \cdot \vec{k} )
          -  ( \vsigma_1 \cdot \vsigma_2 ) \vec{k}' \cdot
        \delta (\vec{r}) \vec{k} \nonumber\\
       &   +  3 ( \vsigma_2 \cdot \vec{k}' ) \delta (\vec{r})
          ( \vsigma_1 \cdot \vec{k} )
           -  ( \vsigma_1 \cdot \vsigma_2 ) \vec{k} \cdot
        \delta (\vec{r}) \vec{k}'
                \Big\} \, .
\end{align}

The inclusion of this tensor force gives rise to new terms in
the force-generated energy density~\cite{lesinski07}
\begin{align}
\label{eq:EF:tensor}
\mathcal{H}^{t}
 = &   B^{T}_t \Big(  \vec{s}_t \cdot \vec{T}_t
                    - \sum_{\mu, \nu = x}^{z} J_{t, \mu \nu} J_{t, \mu \nu} \Big)
      \nn \\
   & + B^{\Delta s}_t \vec{s}_t \cdot \Delta \vec{s}_t
     + C^{\nabla s}_t (\nabla \cdot \vec{s}_t)^2
      \nn \\
   &    + C^{F}_t \Big[  \vec{s}_t \cdot \vec{F}_t
                 - \tfrac{1}{2} \Big( \sum_{\mu = x}^{z} J_{t,\mu \mu}
                                \Big)^2
                 - \tfrac{1}{2}
                   \sum_{\mu, \nu = x}^{z} J_{t, \mu \nu} J_{t, \nu \mu}
                \Big]
\, .
\end{align}
The first two terms also appear in the EDF constructed from a central
Skyrme force, whereas the latter two occur for genuine two-body tensor
forces only. They differ in the way  derivatives and Pauli matrices are
coupled. In the first two terms, the scalar products are between one
derivative and the other, and between one Pauli matrix and the other.
In the last two terms, the scalar products are between derivatives and
Pauli matrices.

In a force-generated EDF, each coupling constant $C^{T}_t$ and
$C^{\Delta s}_t$ results from two contributions. Following Article I, we label the
contributions coming from the central Skyrme interaction by the
letter $A$ and those generated by the tensor part of the interaction
by $B$. The inclusion of a tensor force thus increases the
flexibility for the choice of the coupling constants $C^{T}_t$ and
$C^{\Delta s}_t$ in a force-based approach. At the same time, the
tensor force introduces additional terms in the EDF that couple
derivatives and Pauli-spin matrices in a unique manner.

The most appropriate way would be to label tensor terms as
those generated by the tensor force, Eq.~\eqref{eq:Skyrme:tensor}.
However, the above discussion shows that these terms cannot be
easily singled out with respect to similar terms generated by the
central part of the EDF. Therefore, throughout this article, we
will call ``tensor terms'' those terms in
the EDF that couple two Pauli matrices and two derivatives. Although not
all of them are related to a two-body tensor force, they are of the same
order as the terms bilinear in the spin-current tensor density that have
been called tensor terms in Articles~I and~II.

%
%
\subsubsection{The pairing energy}

As in our previous studies, we have chosen a density-dependent
zero-range interaction to describe the pairing correlations
\cite{dobaczewski84,terasaki95,bender03}, which leads to a functional
of the form
\begin{equation}
\label{eq:Epair}
\mathcal{E}_{\text{pairing}}
= \sum_{q=p,n} \frac{V_{q}}{4}\int d^{3}\vec{r} 			
  \left[1-\frac{\rho_{0}(\vec{r})}{\rho_{c}}\right] 			 
  \tilde{\rho}_{q}(\vec{r}) \, \tilde{\rho}_{q}^{\ast}(\vec{r})
\ ,
\end{equation}
where the switching density $\rho_{c}$ determines whether the pairing
is more active in the volume of the nucleus or on its surface. The functional
depends on the local pair density $\tilde{\rho}_{q}$ and the local density
$\rho_{q}$.

%
%
\subsection{The cranked HFB method}

The superdeformed rotational bands will be calculated by the
self-consistent cranked HFB approach. This method can be seen as a
semi-classical description of the collective rotation of a finite
system with a constant angular velocity $\omega$. In particular, it
takes into account the distortion of the nucleus' intrinsic state
due to the centrifugal and Coriolis forces that are
induced by the collective rotation
\cite{Tho62a,rowe70,ring80,szymanski83}.

The variation of the EDF including constraints on
particle number, orthonormality of the quasiparticle states
and rotational frequency leads to the cranked HFB equation
\begin{equation}
\label{eq:HFB}
\fourmat{h - \lambda - \omega_z J_z}{\Delta}
        {-\Delta^{\ast}}{-h^{\ast} + \lambda + \omega_z J_z^{\ast}}
\twospinor{U_\mu}{V_\mu}
= E_\mu \twospinor{U_\mu}{V_\mu}
\, ,
\end{equation}
where $U_\mu$ and $V_\mu$ are the two components of the quasiparticle
wave functions and $E_\mu$ the quasiparticle energies, often called Routhian
in this context. The effective interaction enters the HFB Hamiltonian
through the mean-field Hamiltonian $h$ and the pairing field $\Delta$
\begin{equation}
\label{eq:h}
h_{ij}
= \frac{\delta\mathcal{E}}{\delta\rho_{ji}} \, ,
\quad
\Delta_{ij}
= \frac{\delta\mathcal{E}}{\delta\kappa^{\ast}_{ij}}
\, .
\end{equation}
The Fermi energies $\lambda$ for protons and neutrons and the rotational frequency $\omega_z$
in Eq.~(\ref{eq:HFB}) are the Lagrange multipliers of the constraints,
which are self-consistently adjusted to fulfill auxiliary conditions
for the mean values of the particle number and of the projection along $z$ of the angular
momentum. The component $J_{z}$ of the angular momentum $J$ is chosen along the
axis perpendicular to the axes of longest elongation.
 At high spins and large deformation, the solution of Eq.~(\ref{eq:HFB}) can be shown to be an
approximation of a variation after projection on angular
momentum \cite{kamlah68a}. As such, the model is particularly well
adapted for the description of superdeformed bands.

All calculations have been carried out using the triaxial self-consistent
cranking code \texttt{CR8} documented in Refs.~\cite{bonche87,gall94,terasaki95}.
The HFB equation is complemented by the Lipkin-Nogami (LN) prescription
to avoid a sudden breakdown of pairing correlations as a function of
rotational frequency.

We recall that for constrained calculations, as discussed below, the
constraints do not contribute to the observable total energy, whereas
the eigenvalues $E_\mu$ of the HFB Hamiltonian used
to construct the quasiparticle Routhians contain a contribution
from the constraint.

%
%
%
\subsection{The single-particle Hamiltonian}
The isospin representation of the
EDF is the most appropriate one to discuss its physics content.
However, a representation where proton and neutron densities are explicitly
used is more convenient for numerical implementations.
In this case, one possibility to write the Skyrme EDF is \cite{bonche87}
\begin{equation}
\label{eq:func:Skyrme:pn}
\mathcal{E}_{\text{Sk}}
= \int \! d^3r \; \Big[  \mathcal{H}(\vec{r})
                       + \sum_{q=p,n} \mathcal{H}_{q}(\vec{r}) \Big]
\end{equation}
with
\begin{align}
\mathcal{H}(\vec{r})
= &
      	   \;  b_1 \; \rho^2
             + b_3 \; \big( \rho \tau - \vec{j}^2 \big)
             + b_5 \; \rho \Delta \rho
              + b_{7} \; \rho^{2 + \alpha}
       \nn\\
  &
             + b_{9}  \; \big(   \rho \vnabla \cdot \vec{J}
                           + \vec{j} \cdot \vnabla \times \vec{s} \big)
             + b_{10} \; \vec{s}^2
             + b_{12} \; \rho^{\alpha} \vec{s}^2
       \nn\\
   &        + b_{14}\; \Big( \sum_{\mu, \nu = x}^{z} J_{\mu \nu} J_{\mu \nu}
                            - \vec{s} \cdot \vec{T} \Big)
      \nn \\
   &
             + b_{16} \; \Big[  \Big( \sum_{\mu = x}^{z} J_{\mu \mu} \Big)^2
                           + \sum_{\mu, \nu = x}^{z} J_{\mu \nu} J_{\nu \mu}
                          -  2 \; \vec{s} \cdot \vec{F}   \Big]
      \nn\\
      &    + b_{18} \; \vec{s} \cdot \Delta \vec{s}
             + b_{20} \; \big( \vnabla \cdot \vec{s} \big)^2
             \ ,
\end{align}
containing the total density $\rho=\rho_{p}+\rho_{n}$ and similar for
all the other local densities and currents,
\footnote{Even though the
``total'' local densities and currents are identical to the ``isoscalar''
local densities and currents, we use a different notation to clearly
distinguish between the isospin representation and the proton-neutron
representation used in our codes.
}
and
\begin{align}
\mathcal{H}_{q}(\vec{r})
= &
       \; b_2 \;\rho^2_q
        + b_4 \; \big( \rho_q \tau_q - \vec{j}_q^2 \big)
        + b_6 \;\rho_q \Delta \rho_q
        + b_{8} \; \rho_0^{\alpha} \; \rho^2_q
      \nn\\
      &
        + b_{9q}  \; \big( \rho_q \vnabla \cdot \vec{J}_q
                  + \vec{j}_q \cdot \vnabla \times \vec{s}_q \big)
        + b_{11}  \; \vec{s}^2_q
        + b_{13} \; \rho^{\alpha} \vec{s}^2_q
       \nn\\
       &
        + b_{15}\; \Big( \sum_{\mu, \nu = x}^{z} J_{q,\mu \nu} J_{q,\mu \nu}
                       - \vec{s}_q \cdot \vec{T}_q \Big)
       \nn\\
       & + b_{17} \Big[  \Big( \sum_{\mu = x}^{z} J_{q,\mu \mu} \Big)^2
                        + \sum_{\mu, \nu = x}^{z}
                          J_{q, \mu \nu} J_{q, \nu \mu}
                     -  2 \; \vec{s}_q \cdot \vec{F}_q
                  \Big]
        \nn\\
        &
         + b_{19} \; \vec{s}_q \cdot \Delta \vec{s}_q
         + b_{21} \; \big( \vnabla \cdot \vec{s}_q \big)^2
\, ,
\end{align}
containing the proton and neutron local densities and currents.
The relation between the coupling constants in the isospin representation,
Eq.~\eqref{eq:EF:time-even} and~\eqref{eq:EF:time-odd}, and the parameters
in the proton-neutron representation, Eq.~\eqref{eq:func:Skyrme:pn}, is
given in Appendix~\ref{app:cpling}. The single-particle Hamiltonian for
protons and neutrons is given by
\begin{align}
\label{eq:sphamiltonian}
\hat{h}_q
 = &   - \vnabla \cdot B_q (\vec{r}) \vnabla
       +  U_q (\vec{r})
       + \vec{S}_q (\vec{r}) \cdot  \hat{\sigmavec}
   \nn \\
   &
    - \frac{i}{2}  \sum_{\mu,\nu=x}^{z}[W_{q,\mu\nu} (\vec{r}) \nabla_{\mu}\sigma_{\nu}
                     + \nabla_{\mu}\sigma_{\nu} W_{q,\mu\nu} (\vec{r}) ]
  \nn \\
   &
     - \frac{i}{2} [   \vec{A}_q (\vec{r}) \cdot \vnabla
                       + \vnabla \cdot \vec{A}_q (\vec{r}) ]
   \nn\\
   &
     - \vnabla  \cdot [ \hat{\sigmavec} \cdot \vec{C}_q (\vec{r}) ] \vnabla
     - \vnabla \cdot \vec{D}_q (\vec{r}) \, \hat{\sigmavec} \cdot \vnabla
\, .
\end{align}
The expressions obtained in \cite{bonche87} for the inverse effective mass
$B_q(\vec{r})$, the single-particle potential $U_q(\vec{r})$, and
the time-odd field $\vec{A}_q(\vec{r})$ are not affected by the
introduction of tensor terms. The local potentials that contain
contributions from the tensor terms are given by
\begin{subequations}
\begin{align} \label{eq:locpot:W}
W_{q,\mu \nu} (\vec{r})
 = &  - \sum_{\kappa=x}^{z} \epsilon_{ \kappa \mu \nu}
        \big(    b_{9}  \, \nabla_\kappa \rho
               + b_{9q} \, \nabla_\kappa \rho_q \big)
            \nn \\
   &  + 2 \, b_{14} \, J_{\mu \nu}
      + 2 \, b_{15} \, J_{q,\mu \nu}
      \nn \\
   &  + 2 \, b_{16} \,\Big( J_{\nu \mu}
      + \sum_{\kappa=x}^{z} J_{\kappa \kappa}  \delta_{\mu \nu}\Big)
      \nn \\
   &  + 2 \, b_{17} \, \Big( J_{q,\nu \mu}
      + \sum_{\kappa=x}^{z} J_{q, \kappa \kappa} \delta_{\mu \nu} \Big)
	\, ,
     \\
\label{eq:locpot:S}
S_{q,\mu} (\vec{r})
 = &  -  \big(    b_{9}  \, \vnabla \times \vec{j}
               + b_{9q} \, \vnabla \times \vec{j}_{q} \big)_\mu
      \nn \\
   &  + 2 \, b_{10} \, s_{\mu}
      + 2 \, b_{11} \, s_{q,\mu}   \nn\\
   &   + 2 \, b_{12} \, \rho^{\alpha}  \, s_{\mu}
       + 2 \, b_{13} \, \rho^{\alpha}  \, s_{q,\mu} \nn\\
   &  -      b_{14}  \, T_{\mu}
      -      b_{15}  \, T_{q,\mu}
      \nn \\	
   &  - 2 \, b_{16}  \, F_{\mu}
      - 2 \, b_{17}  \, F_{q,\mu} \nn\\
   &  + 2 \, b_{18} \, \Delta s_{\mu}
      + 2 \, b_{19} \, \Delta s_{q,\mu} \nn \\
   & - 2 \, b_{20}  \, \nabla_{\mu} (\vnabla \cdot \vec{s})
     - 2 \, b_{21}  \, \nabla_{\mu} (\vnabla \cdot \vec{s}_q )
	\, ,
      \\
\label{eq:locpot:C} C_{q,\mu} (\vec{r})
  =&  - b_{14} \, s_{\mu}
      - b_{15} \, s_{q,\mu}
	\, ,
      \\
\label{eq:locpot:D} D_{q,\mu} (\vec{r})
  = & - 2 \, b_{16} \, s_{\mu}
      - 2 \, b_{17} \, s_{q,\mu}
\, ,
\end{align}
\end{subequations}

The scalar central potential
$U_q(\vec{r}) \equiv \delta \mathcal{E} / \delta \rho_q(\vec{r})$,
the position-dependent inverse effective mass
$B_q(\vec{r}) = \hbar^{2} / 2 m^{\ast}_{q} (\vec{r}) \equiv \delta \mathcal{E} / \delta \tau_q(\vec{r})$,
and the spin-current tensor potential
$W_{q,\mu\nu}(\vec{r}) \equiv \delta \mathcal{E} / \delta J_{q,\mu \nu} (\vec{r})$
are all time-even fields, whereas
$\vec{A}_q (\vec{r}) \equiv \delta \mathcal{E} / \delta \vec{j}_q (\vec{r})$ and
$\vec{S}_q (\vec{r}) \equiv \delta \mathcal{E} / \delta \vec{s}_q (\vec{r})$,
$\vec{C}_q (\vec{r}) \equiv \delta \mathcal{E} / \delta \vec{T}_q (\vec{r})$, and
$\vec{D}_q (\vec{r}) \equiv \delta \mathcal{E} / \delta \vec{F}_q (\vec{r})$ are
time-odd fields. The vector potentials $\vec{A}_q (\vec{r})$ and $\vec{S}_q (\vec{r})$
are nuclear counterparts of electromagnetic potentials that couple orbital
movement and spin to magnetic fields. The field $\vec{C}_q (\vec{r})$
introduces a spin dependence of the position-dependent effective masses
of protons and neutrons. Finally, the field $\vec{D}_q (\vec{r})$
contributes to a non-diagonal tensor effective mass that is position- and spin-dependent.
As long as time-reversal invariance is not broken, the time-odd
fields remain zero.

For density-independent $C_{t}^{\tau}$, $C_{t}^{T}$, and $C_{t}^{F}$,
in a static calculation, and for our choice of symmetries (see
Appendix~\ref{app:cr8} on conserved symmetries)
the single-particle Hamiltonian \eqref{eq:sphamiltonian} can be reduced to
\begin{align}
\label{eq:redsphamiltonian}
\hat{h}_q
 = & - \vnabla \cdot B_q (\vec{r}) \vnabla
     + U_q (\vec{r})
     + \vec{S}_q (\vec{r}) \cdot  \hat{\sigmavec}
   \nn \\
   &
    - i \sum_{\mu,\nu=x}^{z} W_{q,\mu\nu} (\vec{r}) \nabla_{\mu}\sigma_{\nu}
    - i \vec{A}_q (\vec{r}) \cdot \vnabla
   \nn\\
   &
     - \vnabla \cdot \big[ \hat{\sigmavec} \cdot \vec{C}_q (\vec{r}) \big] \vnabla
     - \vnabla \cdot \vec{D}_q (\vec{r}) \, \hat{\sigmavec} \cdot \vnabla
\, .
\end{align}

As already mentioned in the Introduction, the contribution of the tensor interaction to the eigenvalues of the single-particle Hamiltonian depends on the filling of shells. This is due to the near cancelation of the contributions of two spherical spin-orbit partners to the spin-current
tensor $J_{\mu \nu}$ when both levels are filled. As a consequence, the
time-even terms bilinear in $J_{\mu \nu}$ (nearly) vanish in
spin-saturated nuclei, whereas they might be quite large when only
one level out of two spin-orbit partners is filled, cf.\ the
discussion in Articles~I and~II and references given therein.
It is noteworthy that none of the various time-odd terms associated
with the tensor force has such property. Apart from the usual
cancelation of Kramers-degenerate levels that are connected by
time reversal, there is no additional dependence of these time-odd
terms on shell structure as such.

Further technical information about the detailed form and symmetries
of the local densities and fields as implemented in the \texttt{CR8}
code is presented in Appendix~\ref{app:cr8}.

%
%
\subsection{Landau parameters}
\label{subsect:Landau}

The gross properties of the spin-spin interaction in nuclei are often
characterized by the so-called Landau-Migdal parameters
\cite{backman75,Gia81a,liu91,bender02,cao10}.
In Landau theory for normal Fermi liquids \cite{olsson02}, the
Landau-Migdal parameters represent the strength of the residual interaction
between particles on the Fermi surface. Being a simple number in each
partial wave and spin-isospin channel of the central and tensor interaction,
they cannot represent all details of the effective interaction in a finite
system. Still, they provide an often useful first indication about its
relative strength. Relevant for our present study are the Landau parameters
in the spin- and spin-isospin channels of the central residual interaction
and in the tensor channel
\begin{align}
\label{eq:landau:param}
g_0
& = 2 N_0 \big[ C^s_0 + \big(C^T_0 + \tfrac{1}{3} \, C^F_0 \big) \, k_F^2
         \big] \, ,
       \nonumber  \\
g^\prime_0
& = 2 N_0 \big[ C^s_1 + \big(C^T_1 + \tfrac{1}{3} \, C^F_1 \big) \, k_F^2
         \big] \, ,
       \nonumber  \\
g_1
& =  - 2 N_0 \; \big( C^T_0 + \tfrac{1}{3} \, C^F_0 \big) \, k_F^2 \, ,
       \nonumber  \\
g^\prime_1
& =  - 2 N_0 \; \big( C^T_1 + \tfrac{1}{3} \, C^F_1 \big) \, k_F^2 \, ,
       \nonumber  \\
h_0
& =   \tfrac{1}{3} \, N_{0}  \, C^F_0 \, k_F^2 \, ,
        \nonumber \\
h^\prime_0
& =   \tfrac{1}{3} \, N_{0}  \, C^F_1 \, k_F^2 \, .
\end{align}
We use the convention of Refs.~\cite{Gia81a,bender02} where the
normalization factor is defined as the average level density
$N_{0} \equiv 2 k_{F} m^{*}_0 / \hbar^{2}\pi^{2}$ at the Fermi momentum
$k_{F} = (\tfrac{3}{2}\pi^{2}\rho_{0})^{1/3}$,
in which $m^{*}_0$ is the isoscalar effective mass associated with a given
parameterization, but other choices are sometimes found in the literature.
All higher Landau parameters are zero by construction for a Skyrme EDF
that contains only terms up to second order in derivatives.

In a force-based framework, the central and tensor parts remain
separated in the residual Landau interaction, such that $h_0$ and $h_0'$
are entirely determined by $t_e$ and $t_o$ of Eq.~(\ref{eq:Skyrme:tensor}),
which at the same time do not contribute to $f_\ell$, $f^\prime_\ell$,
$g_\ell$ or $g^\prime_\ell$, cf.\ the expressions given in
Ref.~\cite{cao10}. In a functional-based framework as assumed in
Eq.~(\ref{eq:landau:param}), however, this clear separation is lost,
as can be seen from the appearance of $C^F_t$ in all six Landau
parameters. The reason is that one has to combine contributions from the
$J_{\mu\nu}J_{\mu\nu}$, $J_{\mu\nu}J_{\nu\mu}$, $J_{\mu\mu}$,
$\vec{s} \cdot \vec{F}$, and $\vec{s} \cdot \vec{T}$ terms to recover
the structure of the tensor operator that multiplies $h_0$ and $h_0'$, cf.\ Appendix~\ref{app:LM interaction} for details of the derivation.

%
%
\subsection{Parameterizations}
\label{sect:parameterizations}

\begin{figure}[!tb]
\includegraphics {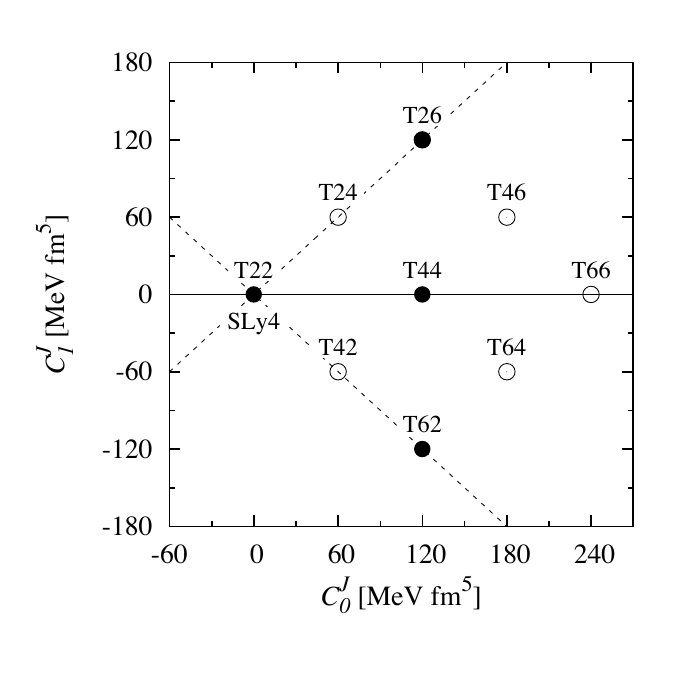}
\caption{\label{fig:general:tij}
Coupling constants $C^{J}_{0}$ and $C^{J}_{1}$, where
$C^{J}_{t}=-C^{T}_{t}+\tfrac{1}{2}C^{F}_{t}$, for the
parameterizations discussed in the article.
}
\end{figure}

\begin{figure}[b]
\centerline{\includegraphics{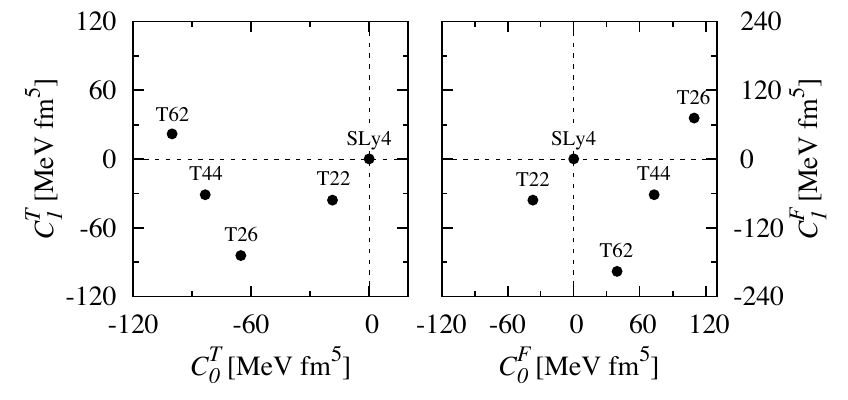}}
\caption{
\label{fig:general:ctf}
Coupling constants $C^{T}_{t}$ and $C^{F}_{t}$ for the
parameterizations discussed in the article.
}
\centerline{\includegraphics{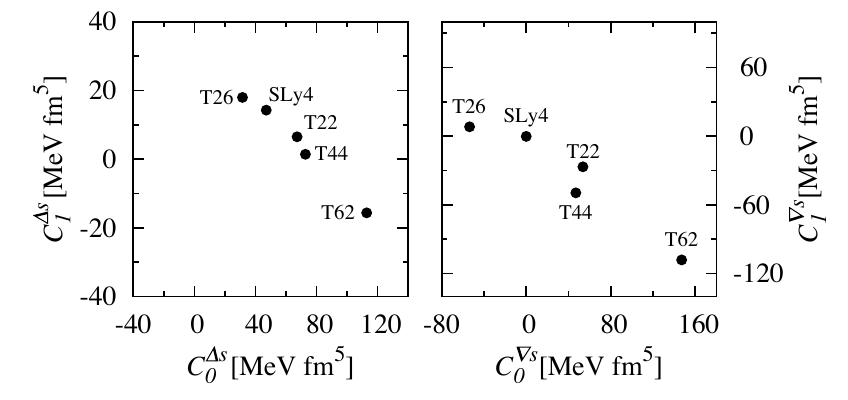}}
\caption{
\label{fig:general:cnsLs}
Coupling constants $C^{\Delta s}_{t}$ and $C^{\nabla s}_{t}$ as
obtained in a force-generated EDF for the parameterizations discussed
in the article.
}
\end{figure}

In Article~I, a set of 36 parameterizations
for the Skyrme interaction including a zero-range tensor force has been determined
using a fitting protocol almost identical to
the one used for the SLy\textit{x} parameterizations
\cite{chabanat97,chabanat98}. These parameterizations, labeled T$IJ$,
systematically cover a wide range of
the $C^{J}_{t} \equiv -C^{T}_{t}+\tfrac{1}{2}C^{F}_{t}$
coupling constants of the tensor terms in spherical symmetry,
see  Articles~I and~II for further details.
In the present study, we restrict ourselves to a subset of four of these
parameterizations, T22, T26, T44, and T62 (see Figs.~\ref{fig:general:tij}
and~\ref{fig:general:ctf}). T22 has been constructed to
give vanishing contributions of the tensor terms in spherical symmetry and time reversal invariance. It is aimed to have properties close to those of SLy4, which does not include a tensor interaction and for which the contributions of
the central part of the interaction to tensor terms have been neglected. However, the tensor terms of T22 can be different from zero when spherical symmetry or time-reversal invariance are broken, cf. Article II for the breaking of spherical symmetry. In the same way, time-odd terms $C^{T}_{t} \vec{s}_t \cdot \vec{T}_t$
and $C^{F}_{t} \vec{s}_t \cdot \vec{F}_t$ do \emph{a priori}
not cancel each other.

The T22 and T44 parameterizations have an isovector $C^{J}_{1}$  coupling
constant equal to $0$. The isoscalar coupling constant $C^{J}_{0}$ has the
same value 120 MeV fm$^{5}$ for T26, T44, and T62,
while the isovector $C^{J}_{1}$ coupling constants respectively take
the values 120 MeV fm$^{5}$, 0 MeV fm$^{5}$, and $-120$ MeV fm$^{5}$.
As a consequence, the tensor terms in spherical symmetry
are purely between particles of same isospin for T26,
purely proton-neutron for T62 and a mixture of both for T44.

As will be discussed in Sect.~\ref{subsect:results:finite size},
the force-generated values of at least one of the $C^{\Delta s}_{t}$
and $C^{\nabla s}_{t}$ coupling constants of all T$IJ$ parameterizations
leads to an unphysical solution of the cranked HFB equation at high spin.
Unless noted otherwise, we have set $C^{\Delta s}_{t}$ and $C^{\nabla s}_{t}$ to zero for those
parameterizations.

As a reference without tensor terms,
we have also performed calculations with the SLy4 parameterization
\cite{chabanat98} for which $C^T_t$ is put to zero during the fit.
To fix the time-odd terms, we have adopted a hybrid point of view that
has already often been used in the literature before \cite{bonche87,rigollet99,chatillon06,chatillon07,ketelhut09} :
a force-generated point of view is taken for the coupling constant $C^s_{t}$,
whereas we use the functional point of view to set the coupling constant
of the $C^{\Delta s}_{t} \vec{s}_{t} \cdot \Delta\vec{s}$ to zero
such that all three terms of second order in the derivatives and Pauli
matrices vanish $C^{\Delta s}_t = C^T_t = 0$. This choice is not unique.
Other groups use SLy4 in their cranked HFB or QRPA calculations by setting
just $C^T_{t} = 0$, keeping $C^s_{t}$ and $C^{\Delta s}_{t}$ at their Skyrme-force
values \cite{dobaczewski95,bender02,schunk10}.

For the the pairing EDF \eqref{eq:Epair}, we choose a
surface-type interaction with $\rho_{c} = 0.16$~fm$^{3}$ and
a strength of $V_q = -1250$~MeV~fm$^{-3}$ for both protons
and neutrons, together with a 5~MeV cut-off above and below
the Fermi level as explained in \cite{rigollet99}.

%
%
\section{Results for superdeformed bands in \nuc{194}{Hg}}
\label{sect:hg194}
\subsection{General comments}

In contrast to superdeformed bands in nuclei around $A = 150$, the high-spin
properties of nuclei in the Hg region are sensitive to pairing
correlations. In cranked Woods-Saxon and Nilsson-model
calculations~\cite{riley90,drigert91},
the gradual increase of the dynamical moments of inertia $\mathcal{J}^{(2)}$
as a function of the rotational frequency results from both the alignment of
the intruder orbitals and from a gradual disappearance of pairing correlations.
The properties of the ground superdeformed (SD) bands have been studied
extensively within self-consistent cranked HFB models using an effective
EDF. In general, a very good agreement with experiment is obtained for the
Hg region \cite{flocard93,gall94,girod94,terasaki95,heenen98}.
In view of this success, we choose the ground SD band in \nuc{194}{Hg}
as a laboratory for the study of tensor terms on high-spin properties.

Our discussions are mainly based on the behavior of the dynamical
moment of inertia \dyn\ as a function of $\hbar \omega$
\begin{equation}
\label{eq:results:dynmom1}
\mathcal{J}^{(2)}
= \frac{\partial \langle J_{z}\rangle}{\partial\omega}
\, ,
\end{equation}
where $\langle J_{z} \rangle$ is the average value of the projection of the angular
momentum on the rotation axis. The relevance of this quantity for the
purpose of our study becomes clearer when realizing that \dyn\ is
proportional to the derivative of the EDF with respect to rotational
frequency \cite{szymanski83,dudek92,frauendorf01}
\begin{equation}
\label{eq:results:dynmom2}
\mathcal{J}^{(2)}
= \frac{1}{\omega} \, \frac{\partial \mathcal{E}}{\partial\omega}
\, ,
\end{equation}
which allows to calculate the contribution of each term in the EDF
(\ref{eq:EF:time-even}-\ref{eq:EF:time-odd}) to \dyn\ separately. The
moment of inertia can also be decomposed into the neutron and proton contributions to $J_z$ using Eq.~\eqref{eq:results:dynmom1}.
These  contributions, however, do not
correspond to the decomposition of the EDF into neutron-neutron,
proton-proton and proton-neutron terms.

The numerical determination of \dyn\ is
far from being trivial. To obtain a smooth dependence of \dyn\ as a function
of $\hbar \omega$, it requires a very high degree of convergence of the
calculations. The derivatives with respect to $\omega$ are determined
by finite differences formulas.
It should be noted that Eqns. \eqref{eq:results:dynmom1} and \eqref{eq:results:dynmom2}
might lead to slightly different \dyn\,, especially when convergence to a very high degree is difficult to attain. In general, the $\langle J_{z}\rangle$
are converged to a higher degree than the energy $\mathcal{E}$, hence Eq. \eqref{eq:results:dynmom1} is
expected to be more stable.

The experimental value of $\hbar \omega$ is given by $E_{\gamma}/2$,
and the one of the dynamical moment of inertia by
$\mathcal{J}^{(2)} = 4 \hbar^2 / \Delta E_{\gamma}$, where
$\Delta E_{\gamma}$ is the difference between two successive $\gamma$-ray
energies populating and depopulating a level. Note that both can be
determined without an angular momentum assignment of the level, as
is the case for the ground-state SD band in \nuc{194}{Hg}.

%
%
\subsection{Finite-size instabilities}
\label{subsect:results:finite size}

\begin{figure}[!tbp]
\includegraphics {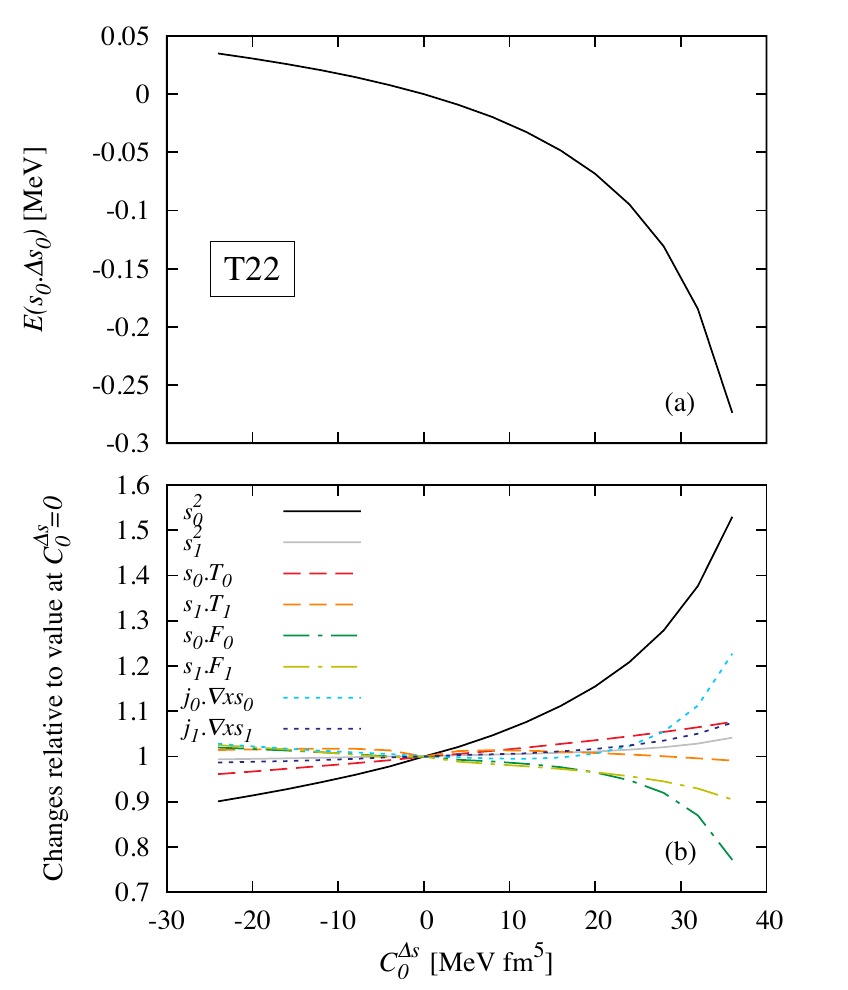}
\caption{\label{fig:instabilities:laplacian isoscalar}
(color online)
(a) Dependence of the
$C^{\Delta s}_{0} \vec{s}_{0}\cdot\Delta\vec{s}_{0}$ term of a
variant of the T22 parameterization on the value of $C^{\Delta s}_{0}$
for the $J_{z}=54\hbar$ state in the ground-state superdeformed
band of $^{194}$Hg (see text).
(b) Dependence of all other time-odd terms containing the
spin density $\vec{s}_{t}$ relative to their value at $C^{\Delta s}_{0} = 0$
in the same calculations.
}
\end{figure}

In our calculations of the SD rotational bands with T$IJ$
parameterizations mentioned before, we systematically encountered non-convergence
of the code
when using force-based coupling constants
for all time-odd terms. After careful analysis, this behavior turned
out not to be a numerical problem, but a property of these parameter
sets. Switching to a functional framework, systematic variation of
coupling constants reveals that large positive or negative values
of either $C_{t}^{\Delta s}$ or $C^{\nabla s}_{t}$ lead to
an unphysical finite-size instability of a given parameterization
of the interaction.

As a typical example, Fig.~\ref{fig:instabilities:laplacian isoscalar}
presents the energies of time-odd terms containing the spin density
for the SD $J_{z}=54\hbar$ state in \nuc{194}{Hg} as a function of the value
of $C^{\Delta s}_{0}$ for a variant of the T22 parameterization.
Panel (a) displays the absolute energy of the
$C^{\Delta s}_{0} \, \vec{s}_{0} \cdot \Delta\vec{s}_{0}$ term, whereas
panel (b) presents the evolution of other time-odd terms that
contain the spin density relative to their value at $C^{\Delta s}_{0} = 0$.
In this calculation, the coupling constants of all time-even and time-odd
terms are set to their force-based values, except for
$C^{\Delta s}_{1} = C^{\nabla s}_{0} = C^{\nabla s}_{1} = 0$
which are set to zero,
and $C^{\Delta s}_{0}$ that is systematically varied.
For larger or smaller values of $C^{\Delta s}_{0}$ than those shown in
Fig.~\ref{fig:instabilities:laplacian isoscalar}
our calculations do not converge. When approaching
$C^{\Delta s}_{0} \approx 36$ MeV~fm$^5$, the energy of the
$C^{\Delta s}_{0} \, \vec{s}_{0} \cdot \Delta \vec{s}_{0}$ term displays a
steep downwards slope. Simultaneously, all other terms containing
the spin density are strongly amplified, in particular the
$C_0^s \, \vec{s}^{2}_{0}$ term, indicating a strong change
in spin polarization. Still, in spite of their strong variation,
the absolute contribution of all these terms to the total energy of
\nuc{194}{Hg} remains less than 0.3~$\%$
even at the threshold of the finite-size instability. Also,
a strong dependence of the spin terms on the coupling constant
is not a necessary condition for the onset of an instability of the
$C^{\Delta s}_{0} \, \vec{s}_{0} \cdot \Delta \vec{s}_{0}$ term.
This can be seen when approaching $C^{\Delta s}_{0} \approx -24$ MeV~fm$^5$,
beyond which the interaction also becomes unstable. Similar results are
obtained for the variation of the $C^{\Delta s}_{1}$; an
instability sets in at the same values as for $C^{\Delta s}_{0}$.
In fact, the instability of the
$C^{\Delta s}_{t} \, \vec{s}_{t} \cdot \Delta \vec{s}_{t}$ terms
at large positive values of $C^{\Delta s}_{t} \gtrapprox 36$ MeV~fm$^5$
has already been pointed out earlier \cite{lesinski06,schunk10}.

In a similar manner, we find that values of $C_{t}^{\nabla s}$
outside the interval \mbox{$[-56,92]$} MeV~fm$^{5}$ lead to
instabilities as well. The limits of the stable regions, however,
should be taken with a grain of salt because their values are sensitive
to the details of the calculation, the mass number of the nucleus,
or the other parameters of the EDF. In fact, even when using
an \emph{a priori} unstable parameterization, a finite-size instability
might fortuitously remain undetected in the calculation of a finite
nucleus, depending on convergence criteria, cutoffs in the numerical
representation, the initial conditions of the calculations and other
numerical choices made. An unambiguous way to identify one class of finite-size
instabilities is through the calculation of the response function of
the model system of isotropic homogeneous infinite nuclear matter (INM)
to perturbations of the density in random phase  approximation (RPA)
\cite{lesinski06}. When an instability occurs at infinite wavelength,
the entire bulk of homogeneous nuclear matter undergoes a transition into a
different homogeneous state of nuclear matter. This kind of instability
can be identified from the values of the Landau parameters discussed in
Sect.~\ref{subsect:Landau} \cite{backman75,lesinski06,cao10}. If instead
the instability occurs at a finite wavelength, the homogeneous nuclear
matter can undergo a phase transition into an inhomogeneous phase, i.e.\
it exhibits a finite-size instability. The former instabilities are driven
by the bulk terms in the EDF, whereas the latter are driven by the terms
that contain a nabla or Laplacian acting on a density and that
are zero in homogeneous INM. Recently, the calculation of the linear response of the
full Skyrme EDF~(\ref{eq:EF:time-even}-\ref{eq:EF:time-odd}) with tensor
terms in force-based~\cite{davesne09} and general functional~\cite{pastore11}
frameworks has become available. There might also be, however, a second class
of instabilities which are related to surface modes~\cite{esb84a}.
 A more detailed analysis of
the finite-size instabilities will be reported
elsewhere~\cite{PastHell}.

In a force-base framework and for the forces considered here,
there is at least one of the $C^{\Delta s}_{t} \, \vec{s}_{t} \cdot \Delta \vec{s}_{t}$
and \mbox{$C^{\nabla s}_{t} \, (\vnabla\cdot\vec{s}_{t})^{2} $} terms causing
non-convergence (see also Fig.~\ref{fig:general:cnsLs}).
To suppress this unphysical behavior,
these four coupling constants are set to zero in all calculations
reported below unless stated otherwise.

%
%
\subsection{General features}

%
\begin{figure}[!tb]
\includegraphics {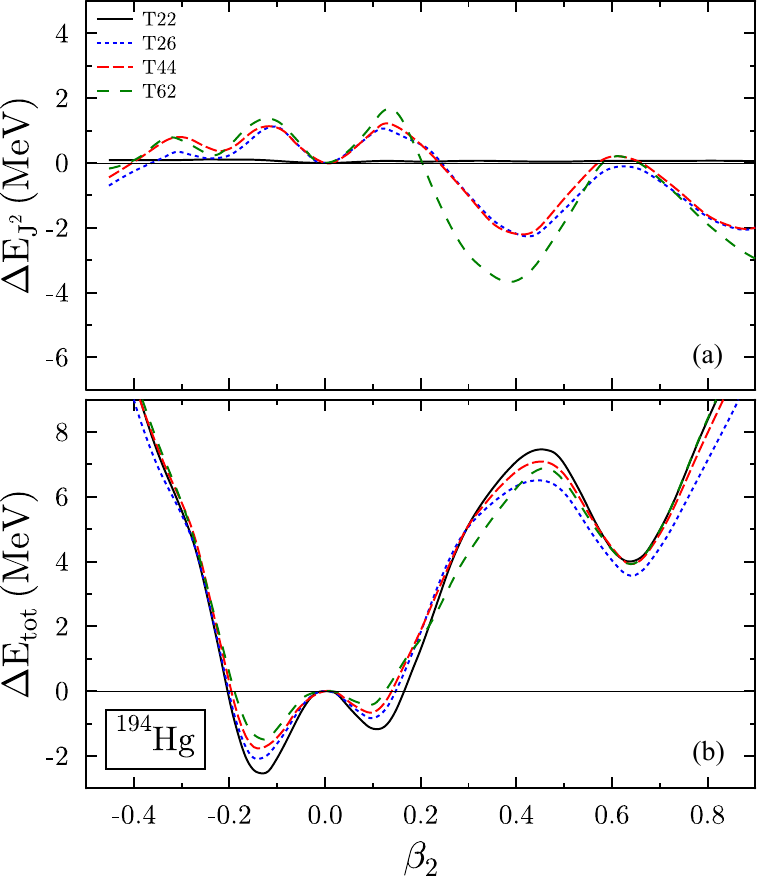}
\caption{(color online) Change of the (a) contribution from the tensor terms relative to the values at the spherical shape and (b) deformation energy relative to the spherical shape for \nuc{194}{Hg} obtained with the parameterizations T22, T26, T44, and T62. The energy scale is the same for the two panels. }
\label{fig:194hg:energysurface}
\end{figure}
%

\begin{figure}[!tb]
\includegraphics {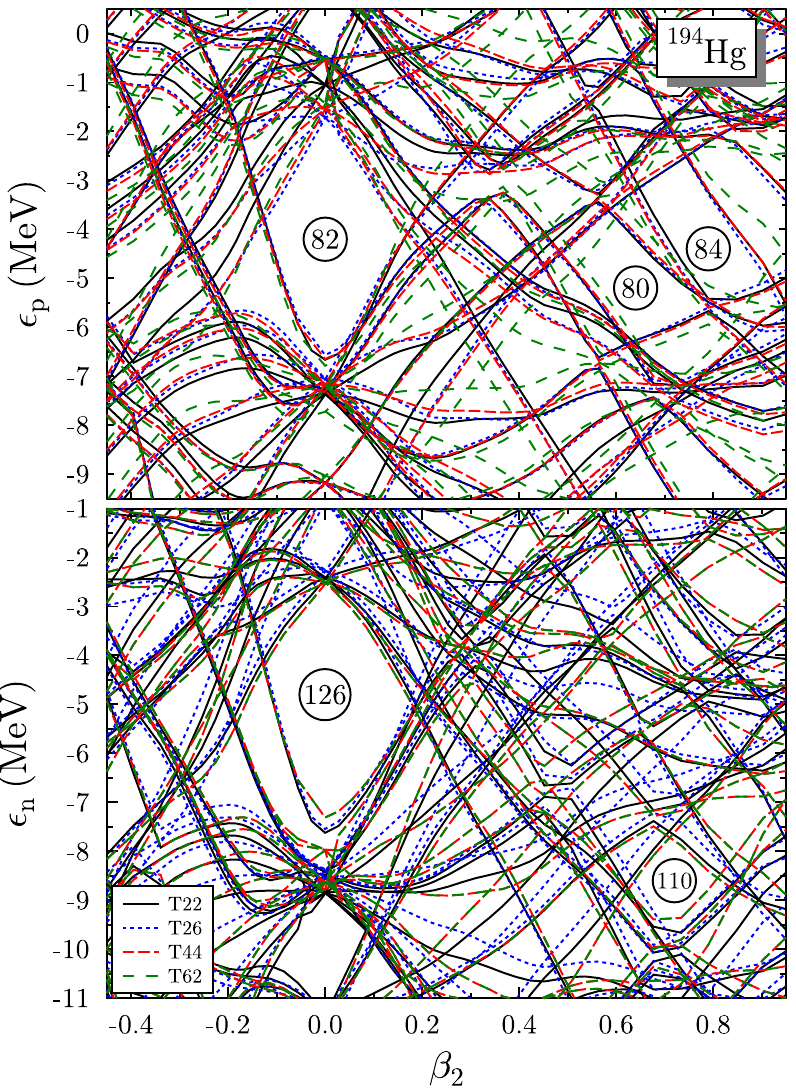}
\caption{(color online) Proton and neutron Nilsson diagrams in \nuc{194}{Hg} for the T22, T26, T44, and T62 parameterization. }
\label{fig:194hg:nilsson}
\end{figure}
Before discussing the rotational properties of \nuc{194}{Hg}, we analyze the evolution of the total energy and the single-particle spectra as a function of deformation.

The total contribution of the tensor terms relative to their value at spherical shape is plotted in the panel (a) of Fig. \ref{fig:194hg:energysurface}. For T22, the time-even tensor $E_{JJ}$ contribution is close to zero for all deformations.
For the other parameterizations, it increases relative at small deformations and then follows an almost 'oscillatory' pattern.
As can be seen from panel (b),
the differences between the parameterizations seen in panel (a) are strongly attenuated in the energy curves.
As discussed in Article II, this last result comes from an intricate compensation between all energy contributions to the EDF. The location and depth of the superdeformed minimum around $\beta_{2}=0.65$ that is the key point for the subsequent discussion is very similar for all parameterizations.

The proton and neutron Nilsson diagrams are presented in Fig. \ref{fig:194hg:nilsson}. We recall the conclusion of Article~II on deformed nuclei  that for parameterizations with different strength of the tensor terms  the differences between the single-particle spectra at sphericity are almost compensated at large deformation by the changes in slope for of the single-particle energies in the Nilsson diagram.
Hence, the single-particle spectra around the Fermi energy for strongly deformed nuclei are often found to be very close for all T$IJ$ tensor parameterizations in spite of significant changes at sphericity. The same property is observed in the case of \nuc{194}{Hg}. At the $Z=80$ superdeformed shell gap, the single-particle spectra for all parameterizations except T62 lie on top of each other. The neutron single-particle spectra follow each other closely for T22 and T44 with some small deviations for T26 and T62.

%
%
\subsubsection{SLy4, T22, and T44}
\label{subsect:194hg:sly4t22t44}

\begin{figure}[!tb]
\includegraphics {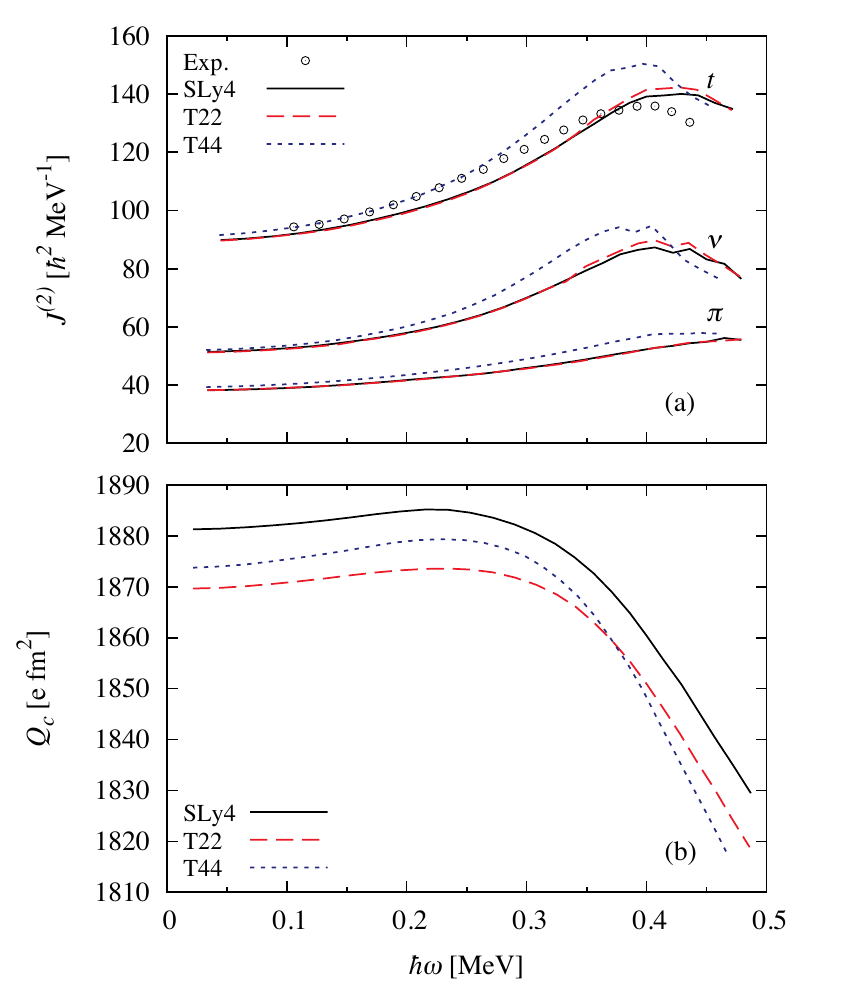}
\caption{
\label{fig:194hg:sly4t22t44:comparison J2 & Q}
(color online)
(a): Proton ($\pi$), neutron ($\nu$), and
total ($t$) dynamical moments of inertia 
as a function of the rotational frequency for the SD band
in $^{194}$Hg with the SLy4, T22, and T44
parameterization.
(b):
The charge quadrupole moment in as a function
of the rotational frequency for the SLy4, the T22, and
the T44 parameterization.
  }
\end{figure}

We first compare results obtained with the SLy4, T22 and T44
parameterizations.

The  dynamical moments of inertia \dyn are plotted as a function of $\hbar\omega$
in panel (a) of Figure~\ref{fig:194hg:sly4t22t44:comparison J2 & Q} and  the charge quadrupole moments $Q_c$ in
panel (b).
The differences between the \dyn\ calculated with SLy4 and T22 are marginal, even though they correspond to very different
coupling constants $C^{T}_{t}$ and $C^{F}_{t}$.
The \dyn\ obtained with T44 increases slightly faster with a plateau appearing for a smaller value of $\hbar\omega$ but, overall, the moments of inertia obtained with three interactions present the same behavior. As discussed in
Article~II, the position of deformed minima may depend on the parameterization but this is not the case here as can be checked from panel (b) of Fig.~\ref{fig:194hg:sly4t22t44:comparison J2 & Q} and also Fig. \ref{fig:194hg:energysurface}.
The charge quadrupole moments $Q_c$ values obtained
with the three parameterizations differ only by about $1\%$ at all spins.
This indicates that the differences between the moments of inertia
are mainly due to the differences in the relative weight of the contributions
in the EDF, and not  to a change in the  shape of the nucleus.

\begin{figure}[!tbp]
\includegraphics{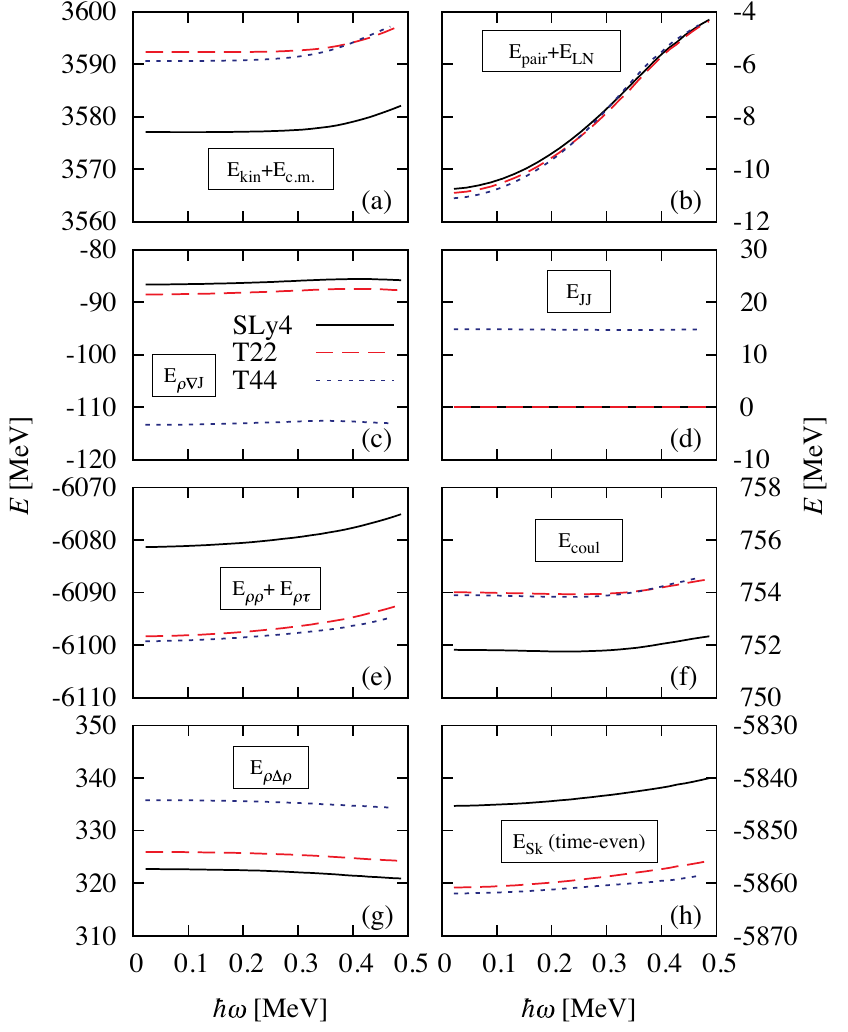}
\caption{
\label{fig:194hg:sly4t22t44:energyte}
(color online)
Evolution of the time-even terms in the EDF as a function of the
rotational frequency $\hbar \omega$ for SLy4, T22, and
T44 for the ground-state superdeformed band of
\nuc{194}{Hg}.
}
%
%
\includegraphics{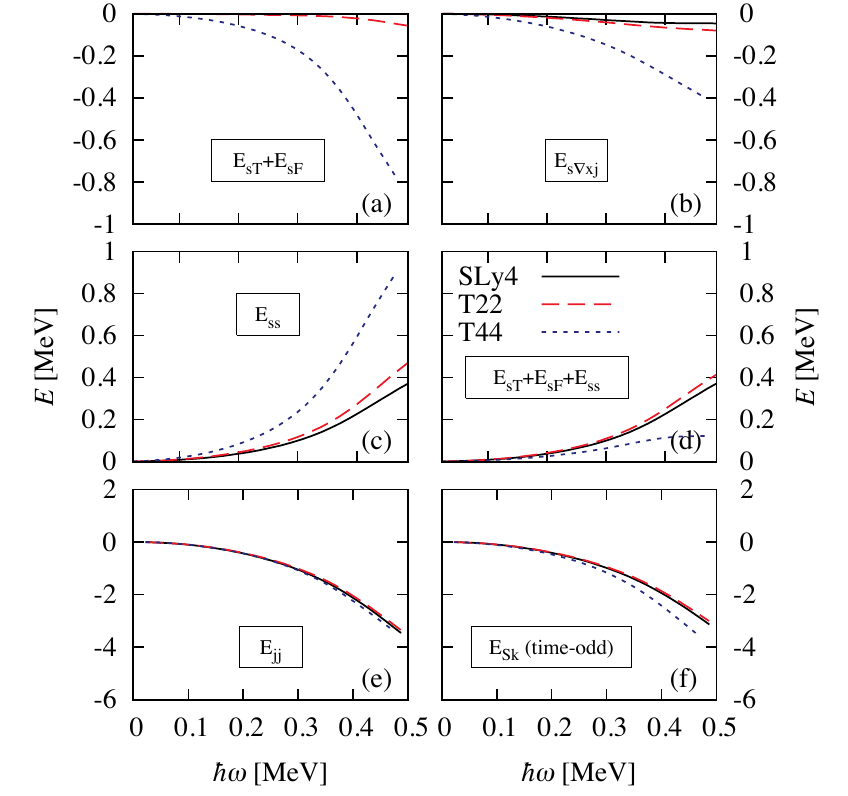}
\caption{
\label{fig:194hg:sly4t22t44:energyto}
(color online)
Same as Fig.~\ref{fig:194hg:sly4t22t44:energyte},
but for the time-odd terms in the Skyrme EDF.
}
\end{figure}

In Figs.~\ref{fig:194hg:sly4t22t44:energyte}
and~\ref{fig:194hg:sly4t22t44:energyto},
we present the contributions of various time-even and time-odd terms to the total energy as a function of $\hbar \omega$. Their labels refer to their density content in the EDF. The total energy is the sum of the kinetic ($E_{kin}+E_{c.m.}$), the pairing  ($E_{pair}+E_{LN}$), the time-even ($E_{Sk}$ (time even)) and
the time-odd ($E_{Sk}$ (time odd)) Skyrme parts of the EDF.

%

The excitation energy at $\langle J_z \rangle = 54 \, \hbar$, which is the value at which we stopped the calculations, is 14.1~MeV
for SLy4 and T22 and 13.6~MeV for T44. All parts
of the EDF contribute to this excitation energy in a very similar way for all parameterizations, around
7~MeV for the pairing energy, 5~MeV for kinetic energy, 2~MeV for  the  Skyrme EDF and 0.5~MeV for Coulomb. The Skyrme contribution results from a cancelation between the time-even (5~MeV) and time-odd (-3~MeV) parts.

As expected
from the Coriolis-anti pairing effect~\cite{ring80}, the pairing energy shows the largest variation with $\hbar \omega$, decreasing to less than half its value at spin zero.
The energies of the kinetic, Coulomb and
time-even terms of the Skyrme EDF change only by a small fraction of their absolute values when going from $\langle J_z \rangle = 0 \, \hbar$ to $\langle J_z \rangle = 54 \, \hbar$. The time-odd terms of the EDF start from zero at spin zero and increase in absolute value with
$\hbar \omega$. The sum of all time-odd contributions is negative and
cancels more than half
of the contribution brought by the time-even terms in the Skyrme EDF.


\begin{figure}[!tbp]
\includegraphics{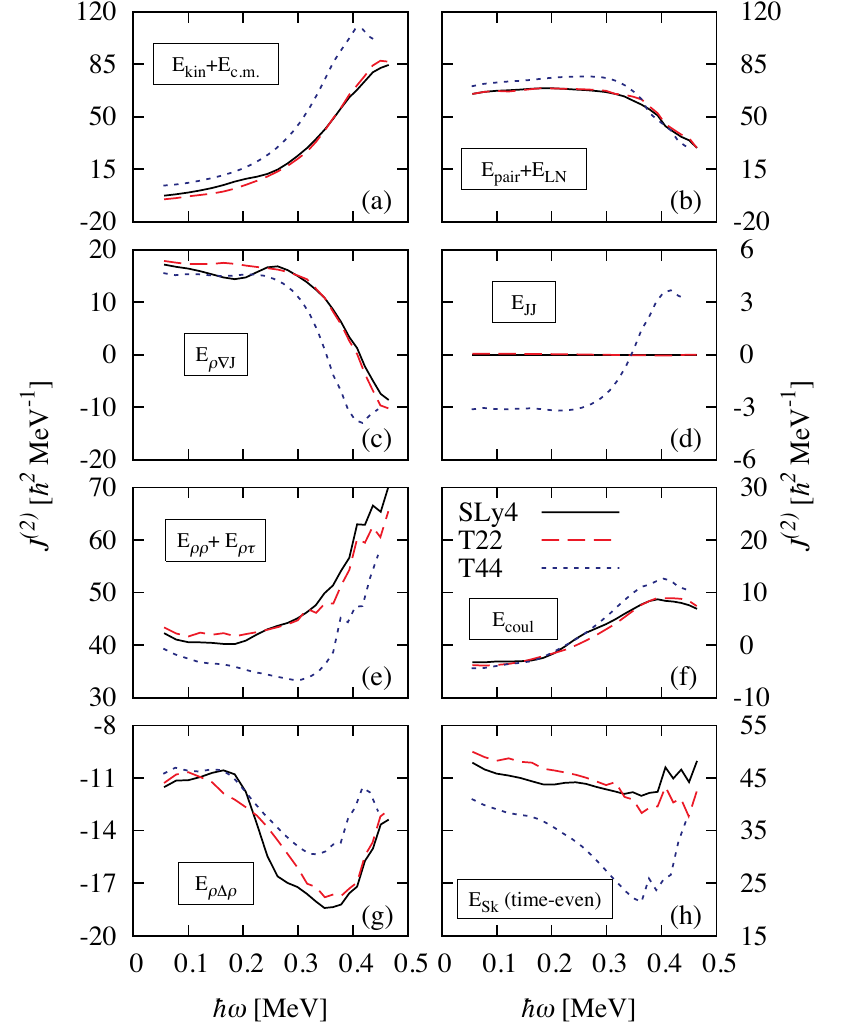}
\caption{
\label{fig:194hg:sly4t22t44:dynte}
(color online)
Dynamical moments of inertia of different terms in the EDF as a
function of the rotational frequency for SLy4, T22, and
T44 in the calculation of the ground-state
superdeformed band of $^{194}$Hg.
}
%
%
\includegraphics{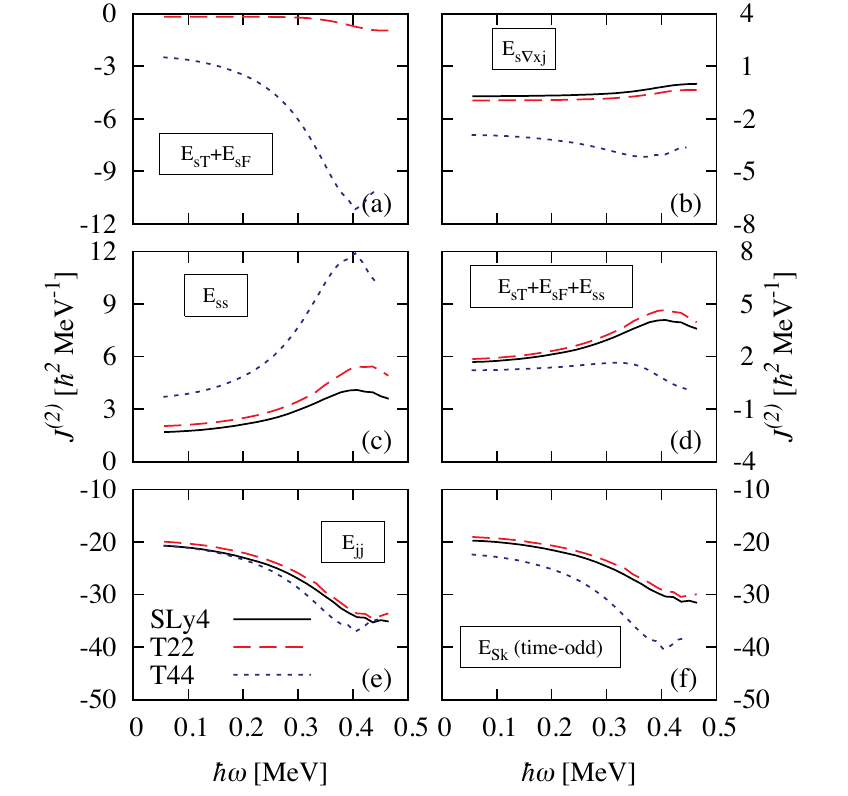}
\caption{
\label{fig:194hg:sly4t22t44:dynto}
(color online)
Same as Fig. \ref{fig:194hg:sly4t22t44:dynte},
but for the time-odd terms in the Skyrme EDF.
}
\end{figure}

 Let us first discuss the energy contributions of the time-even terms,
plotted in Fig.~\ref{fig:194hg:sly4t22t44:energyte}. As discussed in Article~II, the relative contributions of all terms in the EDF differ between each of the T$IJ$ parameterizations.
As a result, the magnitude of the various energy contributions is slightly different for the three  parameterizations. Their $\hbar \omega$ dependence, however, is very similar.
By construction, the tensor contribution $E_{JJ}$ is exactly zero for SLy4 and almost zero for T22 at all angular momenta.
One can note that the energy contribution $E_{\rho \nabla J}$ is more attractive for T44 than for  T22 and SLy4.
As discussed in Article~I, this results from the strong correlation between the spin-orbit and tensor coupling constants. In fact, the ratio of
$E_{\rho \nabla J}(T22)/E_{\rho \nabla J}(T44)$ is
very close to the ratio of the coupling constants
$C^{\nabla J}_{t}(T22)$/$C^{\nabla J}_{t}(T44)$.

The time-odd contributions to the EDF are plotted in Fig.~\ref{fig:194hg:sly4t22t44:energyto}. From top to bottom and left to right are shown
(a) the spin  terms $E_{ss}$, (b) the time-odd terms coupled by Galilean invariance to the time-even effective mass $E_{jj}$, (c) the spin-current tensor $E_{sT} + E_{sF}$  and
(d) the time-odd spin-orbit $E_{s \nabla \times j}$.
The sum of the spin terms that contribute to the
equation of state of infinite homogeneous spin-polarized nuclear matter is shown in panel (e), and the sum of all time-odd contributions in panel (f).  We recall that the $E_{s \Delta s}$ and $E_{\nabla s \, \nabla s}$ terms not shown in the figure are set to zero as they cause finite-size instabilities, see Sect. \ref{subsect:results:finite size}.
All time-odd terms start from zero and change rapidly with increasing $\hbar \omega$. The $E_{sT} + E_{sF}$ term is zero by construction for SLy4 and is negligible for T22. For T44, it decreases down to $-800$~keV at $\langle J_z \rangle = 54 \, \hbar$.

All other time-odd contributions are very
similar for SLy4 and T22. This is not surprising since
the coupling constants of these terms are
very similar.
It indicates also that the additional
$C^T_t \vec{s}_t \cdot \vec{T}_t$ and $C^F_t \vec{s} \cdot \vec{F}_t$
terms of T22 do not introduce a large polarization.
The situation is different for T44, for which all time-odd terms
but $-C^\tau_t \vec{j}^2_t$ take very different values.
These changes are due to the larger $C_{t}^{T}$ and $C_{t}^{F}$
coupling constants and the increased spin-polarization they induce.
All time-odd terms containing the spin density, however, tend to cancel
each other for all parameterizations, such that the sum of all time-odd
terms is very close to $E_{jj}$.

Using Eq.~\eqref{eq:results:dynmom2},
 the dynamical moment of inertia can be decomposed into various contributions to the EDF  that are plotted in Figs.~\ref{fig:194hg:sly4t22t44:dynte}
and~\ref{fig:194hg:sly4t22t44:dynto}.
The offset in total energy between the parameterizations has
no effect on the moments of inertia.

The evolution with \ho\, of the energy contributions being similar for all parameterizations, the general pattern of the contributions to \dyn\ is the same for most terms. At low \ho, the main contribution to \dyn\ is provided by the pairing energy $E_{pair}+E_{LN}$ and represents about 75\%. The Skyrme contribution brings the remaining 25\%, the contributions from $E_{kin}+E_{c.m.}$ and $E_{coul}$  being very small. These two contributions grow rapidly with \ho\, reaching about 65\% at high spin whereas the pairing contributions shrinks to approximately 25\%. The Coulomb contribution grows slowly with spin but never exceeds 10\% of the \dyn\,.

The time-even and time-odd contributions coming from the Skyrme EDF have opposite signs and to large extent cancel each other at high spin as the total contribution of the Skyrme EDF to \dyn\, does not exceed $\pm$10\%. This cancelation however, is not a generic feature, as will be seen with the examples of other parameterizations and other nuclei, although the $E_{Sk}$ (time-even) and $E_{Sk}$ (time-odd) counteract each other in all cases encountered. For all parameterizations $E_{jj}$ is the largest time-odd contribution. This was already found in an earlier study of the time-odd components of the Skyrme EDF \cite{dobaczewski96}, where it was concluded that the cranking term mainly induces a nonzero flow of nuclear matter as measured by the $E_{jj}$ term.

Concentrating on the tensor contributions, the time-even $E_{JJ}$ for T44 is small and varies between -4\% at low spin and 2\% of the \dyn\, at high spin. On a similar scale, the time-odd $E_{sT}+E_{sF}$ contribution lowers the \dyn by at most 7\%. Moreover, $E_{ss}$ and $E_{sT}+E_{sF}$ tend to counteract each other for T44, the total contribution $E_{ss}+E_{sT}+E_{sF}$ being less than 1\%. As already argued in Articles I and II, the presence of the tensor terms has an impact through rearrangements of other coupling constants in the fit and through self-consistency of the HFB. This combined effect modifies all time-even and time-odd Skyrme \dyn\, contributions, whereas the pairing and Coulomb contributions do not change.

\begin{figure}[!htbp]
\centerline{\includegraphics{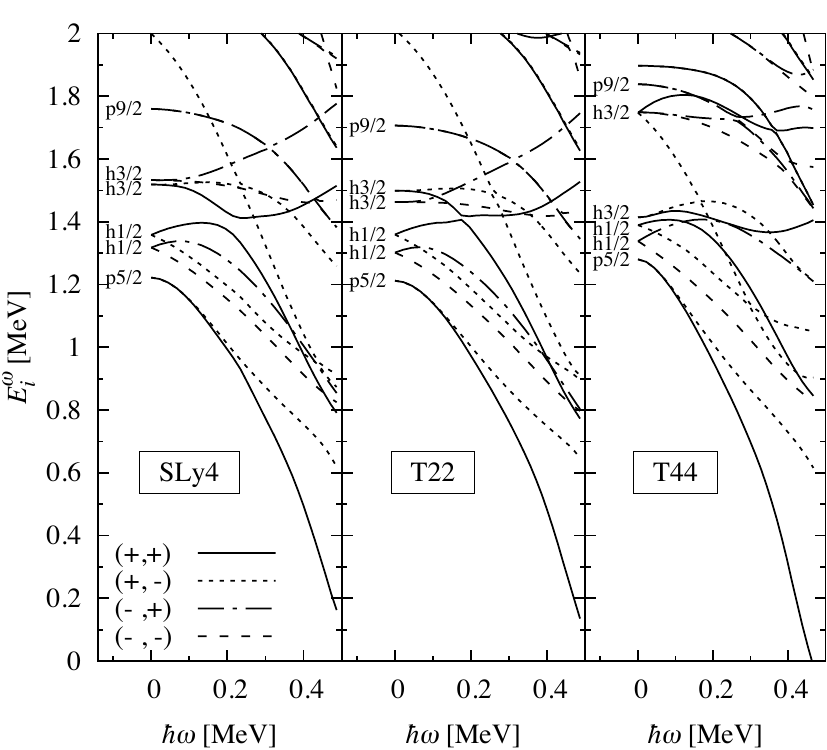}}
\caption{
\label{fig:194hg:sly4t22t44:protonrouthians}
Proton quasi-particle Routhians for the ground-state superdeformed
band of $^{194}$Hg for the SLy4, the T22, and the T44 parameterization. The
(parity,signature) combinations are indicated in the figure. All Routhians are additionally
characterized by the  $j$-component of the dominant single-particle state in the quasi-particle
wave function along the axis of largest elongation at \ho=0 and by their particle (p) or hole (h) character.
}
\centerline{\includegraphics{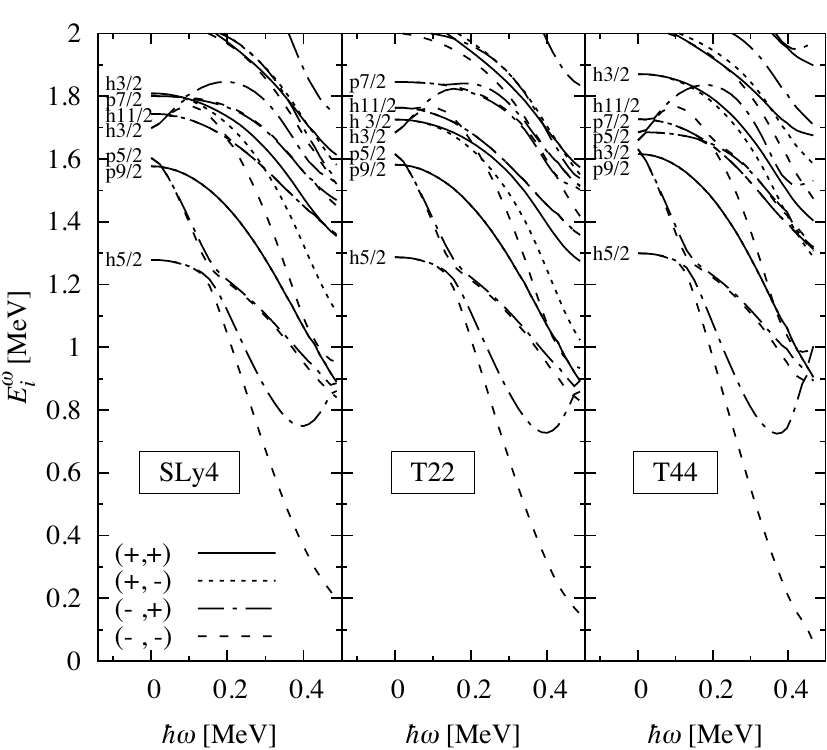}}
\caption{
\label{fig:194hg:sly4t22t44:neutronrouthians}
Same as Fig. \ref{fig:194hg:sly4t22t44:protonrouthians} but for neutrons.
}
\end{figure}

In Figs.~\ref{fig:194hg:sly4t22t44:protonrouthians} and \ref{fig:194hg:sly4t22t44:neutronrouthians}, we present the proton and
neutron quasiparticle Routhians, \textit{i.e.} the eigenvalues of Eq. \eqref{eq:HFB}, as a function of \ho\, for SLy4, T22 and T44. All Routhians are characterized by their parity and signature, by the  $j$-component of the dominant single-particle state in the quasi-particle
wave function along the axis of largest elongation at \ho=0, and by their particle (p) or hole (h) character. As discussed before, the proton and neutron Nilsson diagrams of T22 and T44 display only small differences (Fig. \ref{fig:194hg:nilsson}) . This is reflected in the quasi-particle spectra in
Fig.~\ref{fig:194hg:sly4t22t44:protonrouthians} and \ref{fig:194hg:sly4t22t44:neutronrouthians}. At \ho=0, the ordering of the lowest quasi-particle states is the same for SLy4, T22, and T44, although
the values of the energy are parameterization-dependent.
For the protons, the three lowest quasi-particle Routhians are
the \mbox{p 5/2$^{+}$} and the \mbox{h 1/2$^{-}$} and \mbox{h 1/2$^{+}$}.
The  first significant difference between T44 and the two other parameterizations is the position of the \mbox{h 3/2$^{+}$}
state, which is much closer to the \mbox{h 1/2$^{+}$} level for T44. This position affects the alignment of the \mbox{p 5/2$^{+}$} Routhian with increasing \ho, making it slightly faster for T44 because of  a stronger interaction with the higher-lying positive-parity states.
In the same way, for the neutrons, the exchange in position of the \mbox{h 3/2$^{-}$} and the \mbox{p 5/2$^{-}$} quasiparticle states at \ho=0 has an impact on the alignment of the
\mbox{h 5/2$^{-}$} state. Indeed, the \mbox{h 3/2$^{-}$} quasiparticle Routhian occurs lower in energy for T44. Since it aligns more rapidly
than the \mbox{p 5/2$^{-}$} state, the interaction with the
\mbox{h 5/2$^{-}$} state takes place at a lower frequency, which affects the alignment of this last quasiparticle Routhian.

%
%
\subsubsection{T26, T44, and T62}

\begin{figure}[!htbp]
\includegraphics {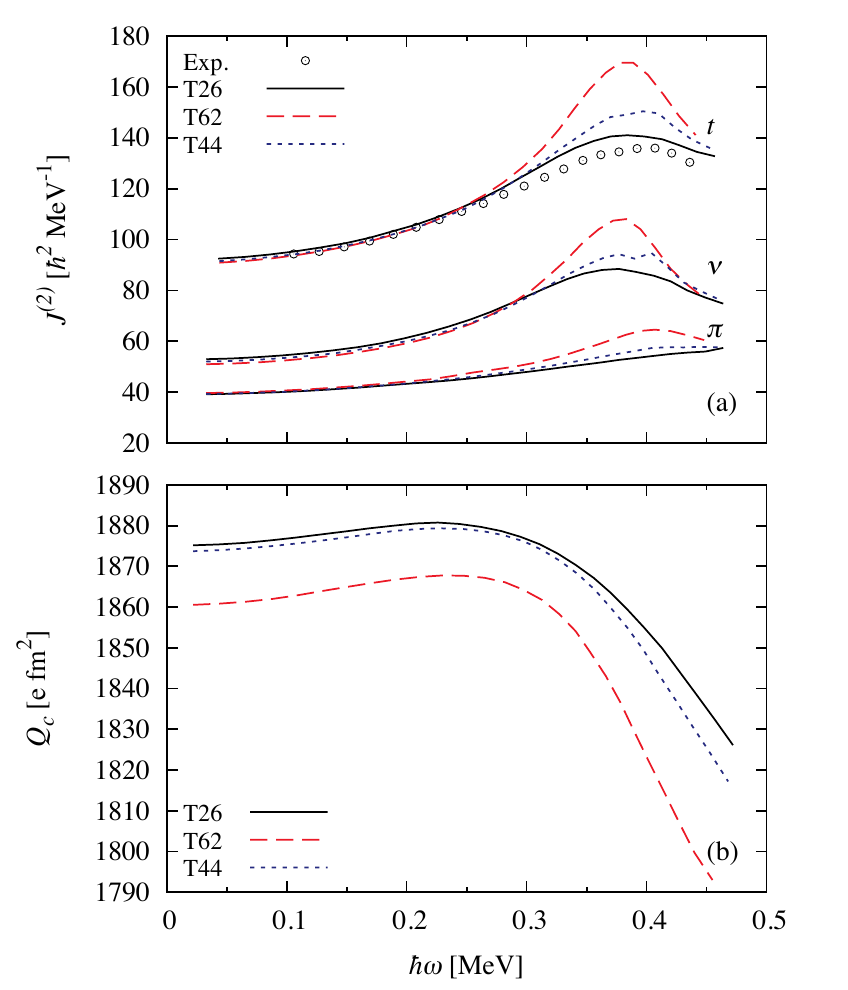}
\caption{(color online) (a) Proton ($\pi$), neutron ($\nu$), and  total ($t$) dynamical moments of inertia in $^{194}$Hg as a
function of the rotational frequency for the SD band in $^{194}$Hg with the T26, the T44, and the T62 parameterization. (b) The charge
quadrupole moment in $^{194}$Hg as a function of the rotational frequency for the T26, the T44, and the T62 parameterization.  }
\label{fig:194hg:t26t44t62:comparison J2 & Q}
\end{figure}

We now proceed to a comparison between results obtained using the T26, T44, and T62 parameterizations. They differ by their value of the isovector $C^{J}_{1}$ coupling constant while they share the isoscalar $C^{J}_{0}=120$ MeV fm$^{5}$ constant. This comparison probes the direct impact of a variation of $C^{J}_{1}$, but also its indirect impact due to the changes of the other coupling constants that are readjusted for each parameterization.

The dynamical moments of inertia \dyn\, and the charge quadrupole moments are presented in respectively panel (a) and (b) of Fig.
\ref{fig:194hg:t26t44t62:comparison J2 & Q}, respectively.

\begin{figure}[!htbp]
\centerline{\includegraphics{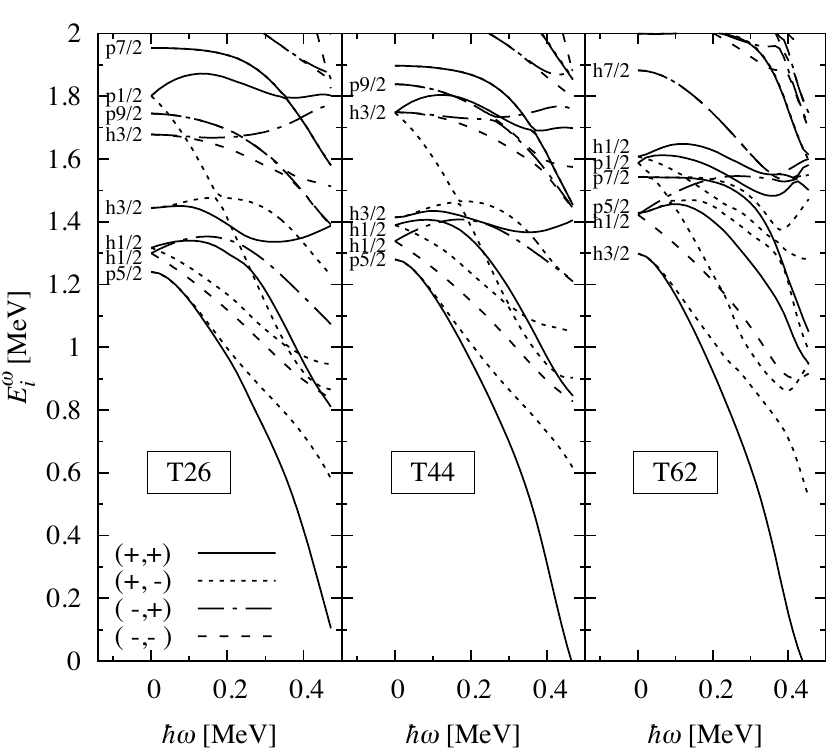}}
\caption{
\label{fig:194hg:t26t44t62:protonrouthians}
Same as Fig. \ref{fig:194hg:sly4t22t44:protonrouthians}, but for the
T26, T44, and T62 parameterization.
}
\centerline{\includegraphics{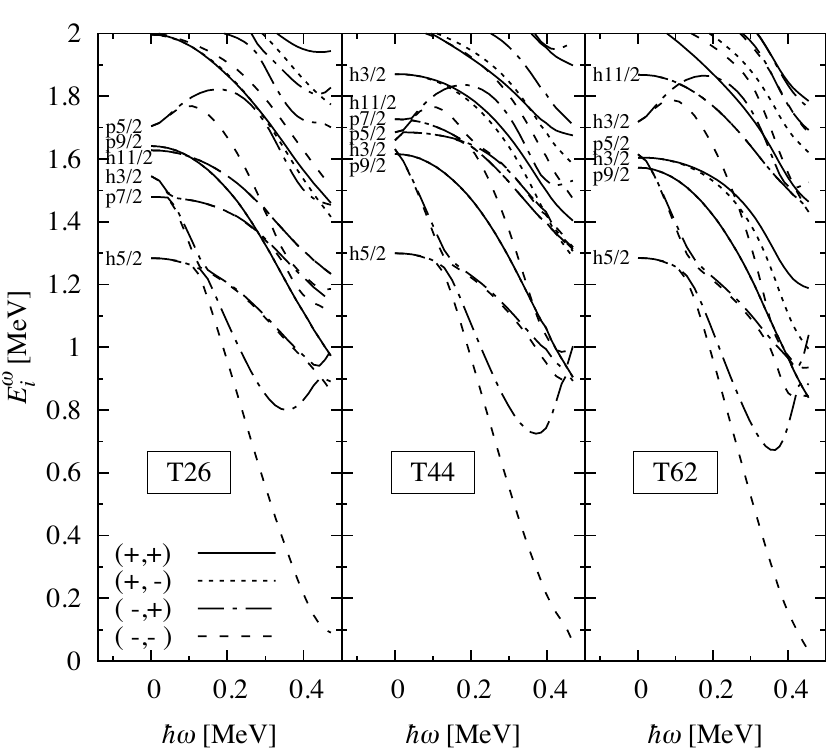}}
\caption{
\label{fig:194hg:t26t44t62:neutronrouthians}
Same as Fig. \ref{fig:194hg:t26t44t62:protonrouthians}, but for neutrons.
}
\end{figure}

Whereas the \dyn\, obtained with the T22 and T44 parameterizations (Fig. \ref{fig:194hg:sly4t22t44:comparison J2 & Q}) have
a different slope,  the behavior of the \dyn\, for T26, T44 and T62 is similar up to 0.3\ho\, and differs at large spins by the height of the plateau.
The proton moment of inertia displays a peak for T62, which is absent for the other parameterizations.
This difference can be related to the proton quasiparticle Routhians, presented in Fig. \ref{fig:194hg:t26t44t62:protonrouthians}.
In contrast to the other parameterizations, it is not the p $5/2^{+}$ but the h $3/2^{+}$ Routhian that is
the lowest quasiparticle state for T62. The \mbox{p $1/2^{+}$}, is at 1.8 MeV for T26 and higher for
all other parameterizations but T62 for which it is only at 1.6 MeV. With increasing \ho, the energy of its negative signature partner decreases quickly up to a bending around 0.4 \ho, which is
the frequency at which the moment of inertia displays a peak for T62.
The charge quadrupole moments differ by less than 1\%, see Fig. \ref{fig:194hg:t26t44t62:comparison J2 & Q}(b).

\begin{figure}[!htbp]
\includegraphics {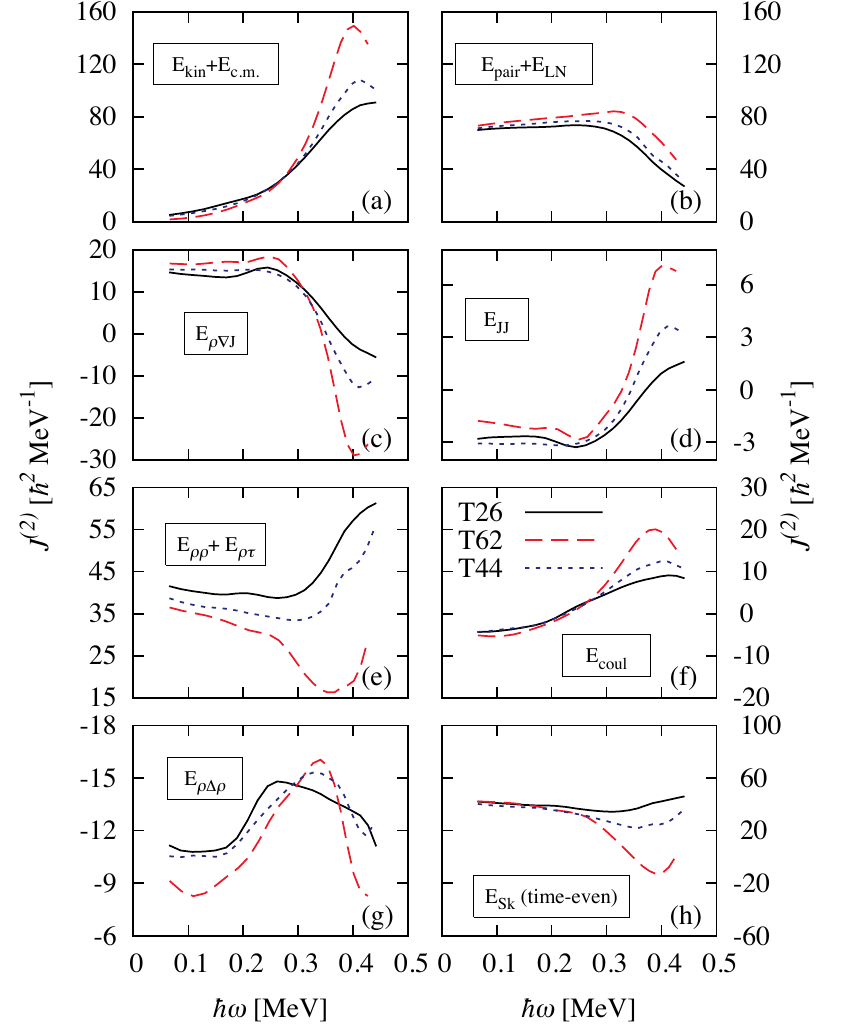}
\caption{(color online)
Same as Fig. \ref{fig:194hg:sly4t22t44:dynte} but for the T26, the T44, and the T62 parameterization. } \label{fig:194hg:t26t44t62:dynte} \end{figure} 
\begin{figure}[!htbp] 
\includegraphics
{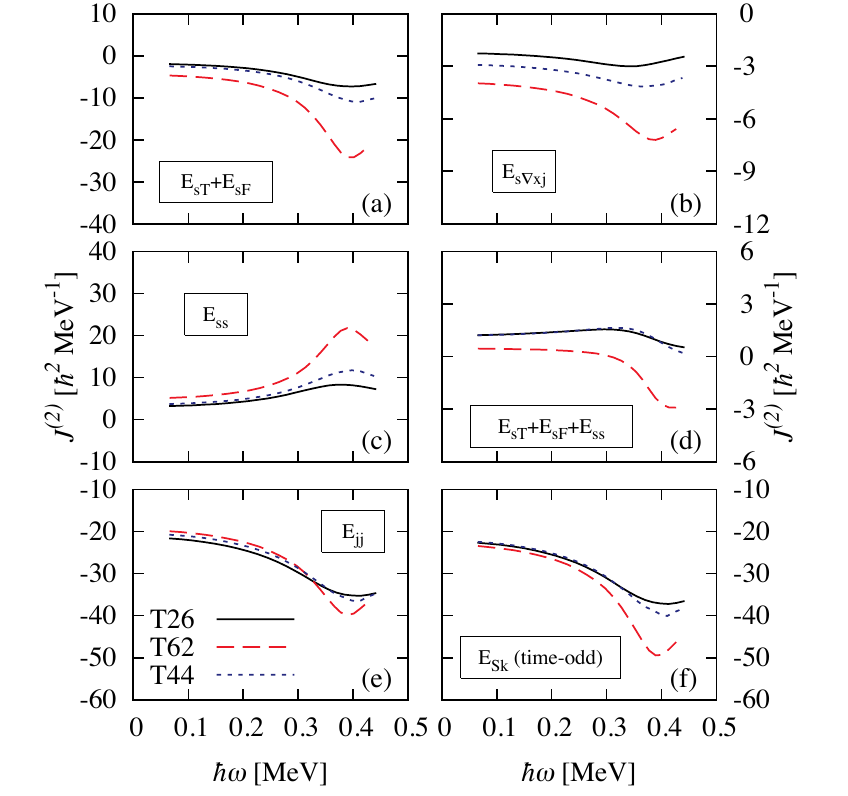} \caption{(color online)  Same as Fig. \ref{fig:194hg:sly4t22t44:dynte}, but for the time-odd terms in the Skyrme energy density functional with the T26,
the T44, and the T62 parameterization. }
\label{fig:194hg:t26t44t62:dynto}
\end{figure} %

The contributions to the dynamical moments of inertia \dyn\ of respectively the time-even and the time-odd terms in the
Skyrme EDF are presented in Figs. \ref{fig:194hg:t26t44t62:dynte} and \ref{fig:194hg:t26t44t62:dynto}. All contributions to the \dyn\ have a similar behavior for the three parameterizations,  except for the  $E_{\rho \rho}+E_{\rho\tau}$ terms.

As for the total moment of inertia, the peaks obtained in most terms at large \ho\, are the most pronounced for T62 and the smoothest for T26.
 Similar to the decomposition of the \dyn\ for SLy4, T22,
and T44, the time-even and time-odd Skyrme contributions are of equal importance because they are of the same order of magnitude. The more pronouncedly peaked the behavior of the other contributions to
the \dyn, the more the Skyrme contribution tends to counteract it. Indeed, whereas the difference in height of the peak between the T26 and the T62 parameterization is approximately 60 $\hbar^{2}$
MeV$^{-1}$ for the $E_{kin}+E_{c.m.}$ contribution, a difference that is even more intensified by the contribution of $E_{pair}+E_{LN}$ and $E_{coul}$, the difference between the total \dyn\ of the
respective parameterization in Fig. \ref{fig:194hg:t26t44t62:comparison J2 & Q} is only about 30 $\hbar$ MeV$^{-1}$. The Skyrme contributions are responsible for the decreased difference in peak-height
of the \dyn.

%
%
\subsection{Time-odd terms}

The parameters of the EDF are usually adjusted to binding energies and r.m.s. radii of doubly-magic nuclei
and on properties of infinite nuclear matter \cite{chabanat97}. As discussed in Sect. \ref{subsect:choice:cpling}, for a force-generated EDF, the coupling constants
of the time-odd part are entirely fixed by those of the time-even part although they are
rarely directly constrained by observables. By contrast, the coupling constants of the time-odd terms that are not constrained
through Galilean invariance are \emph{a priori} undetermined in a more general EDF. Their adjustment has been the subject of several studies \cite{post85,dobaczewski95,dobaczewski96,bender02,zdunczuk07,afanasjev08,margueron09,afanasjev10,pototzky10}. The possibility of using band terminating states to constrain the time-odd terms is discussed in Refs. \cite{zdunczuk07,afanasjev08} and in Ref. \cite{bender02} the effect of the spin-isopin coupling constants of the Skyrme EDFs on predictions for Gamow-Teller distributions is investigated. In the latter study, a local fit of the $C_{1}^{s}[0]$, $C_{1}^{s}[\rho_{nm}]$, $C_{1}^{\Delta s}$, and $C_{1}^{T}$ coupling constants  to the existing data was performed.

To study the effect of time-odd tensor terms in rotating nuclei, we proceed in the following way. Until now, the coupling constants of the $\vec{s}_{t}\cdot\Delta\vec{s}_{t}$ and $(\nabla\cdot\vec{s}_{t})^{2}$ terms have been set to zero for all T$IJ$ parameterizations because these  terms can induce finite-size instabilities. In Sect. \ref{subsect:results:finite size}, we have determined empirical limits between which these instabilities do not appear. To maximize the effect of these time-odd terms on \dyn, we have taken values for $C^{\Delta s}_{t}$ and $C^{\nabla s}_{t}$ close to their respective limits of stability.

The other time-odd tensor terms $\vec{s}_{t}\cdot\vec{T}_{t}$ and $\vec{s}_{t}\cdot\vec{F}_{t}$ are related to time-even tensor terms through Galilean invariance. It is therefore not desirable to vary their coupling constants directly and we have proceeded in an indirect way. We have seen
in Sect. \ref{subsect:194hg:sly4t22t44} that the contributions of $E_{ss}$ and $E_{sT}+E_{sF}$ to \dyn\, act in opposite ways.
We have therefore changed the values of $C^{s}_{t}[\rho_{nm}]$ by $\pm 50$ MeV fm$^{3}$ around their Skyrme force values at the saturation density of nuclear matter, which are  $C^{s}_{0}[\rho_{nm}] \approx 150$ MeV fm$^{3}$ and $C^{s}_{1}[\rho_{nm}] \approx 100$ MeV fm$^{3}$ for all parameterizations.
In addition, all $C^{s}_{t}$ have been taken independent of density, choosing its value at the saturation density of nuclear matter for the coupling constant that is not varied. All variations stay within the stability limits of the conditions outlined in \cite{cao10}.
The discussion of Fig.~\ref{fig:194hg:Csvar:dynto} below will indeed confirm that a variation of the $C^{s}_{t}[\rho_{0}]$ coupling constant has an indirect effect on the $\vec{s}_{t}\cdot\vec{T}_{t}$ and $\vec{s}_{t}\cdot\vec{F}_{t}$ tensor terms.

%
%
\subsubsection{SLy4}
%

\begin{figure}[!tb] 
\includegraphics{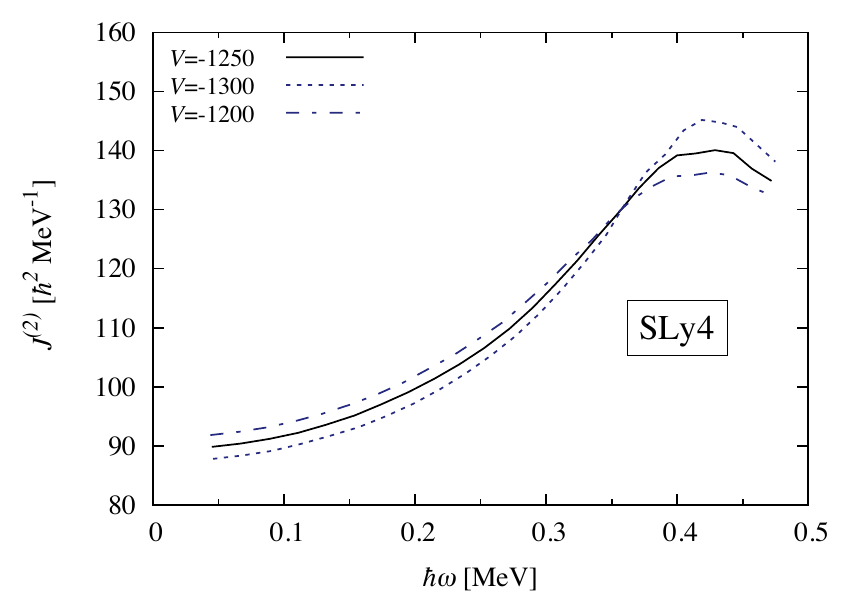}
\caption{(color online) Dynamical moments
of inertia as a function of the rotational frequency for different choices of the strength $V=V_{p}=V_{n}$ of the DDDI
pairing interaction, Eq. (\ref{eq:Epair}). The SLy4 parameterization was chosen in the particle-hole channel of the interaction and the pairing strength
is given in units MeV fm$^{-3}$.}
\label{fig:194hg:sly4:pairing variations}
\end{figure}
%
\begin{figure}[!tb]
\includegraphics {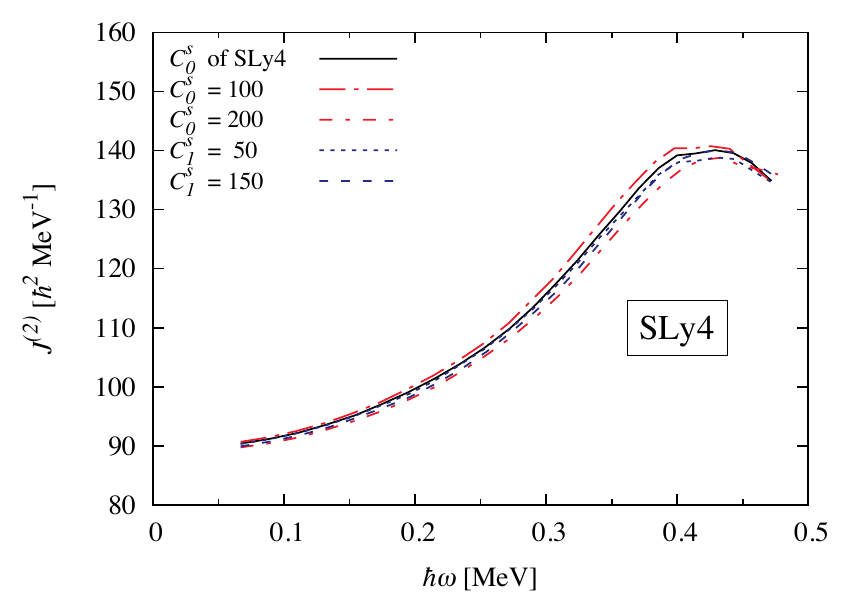}
\caption{(color online) Dynamical moments
of inertia as a function of the rotational frequency for different density-independent choices of the $C^{s}_{t}$ coupling constant. If the isoscalar $C^{s}_{0}$ is varied, then the density-independent
$C^{s}_{1}$ is chosen to be equal to $C^{s}_{1}[nm]$ and vice versa. All other coupling constants in $\mathcal{E}_{Sk} $ are determined by the SLy4 parameterization and the $C^{s}_{t}$ coupling constant is given in units MeV fm$^{3}$ .}
\label{fig:194hg:sly4:time-oddvariations}
\end{figure}

Let us first limit ourselves to the SLy4 parametrization and determine
whether the effect of the variation of the time-odd coupling constants
can be differentiated from a pairing effect. The dynamical moments of inertia
determined with reduced and increased pairing strengths are plotted in
Fig.~\ref{fig:194hg:sly4:pairing variations}.
As expected at low spins, a reduction of pairing increases \dyn\,, whereas  an increase lowers it. The height of the plateau at high
spins moves in opposite direction but it appears for similar values of \ho.

In Fig.~\ref{fig:194hg:sly4:time-oddvariations}, we show the
dependence of \dyn\ on the isoscalar $C^{s}_{0}$ and isovector
$C^{s}_{1}$ coupling constants.
 The only noticeable change in \dyn\ is a slight shift of the plateau for $C^{s}_{0}$.
The effect is small but clearly different from the effect of a variation
of the pairing strength; also, it is larger for $t=0$ than for $t=1$.

%
%
%
\subsubsection{T22, T26, T44, T62}

\begin{figure*}[!htb]
\includegraphics{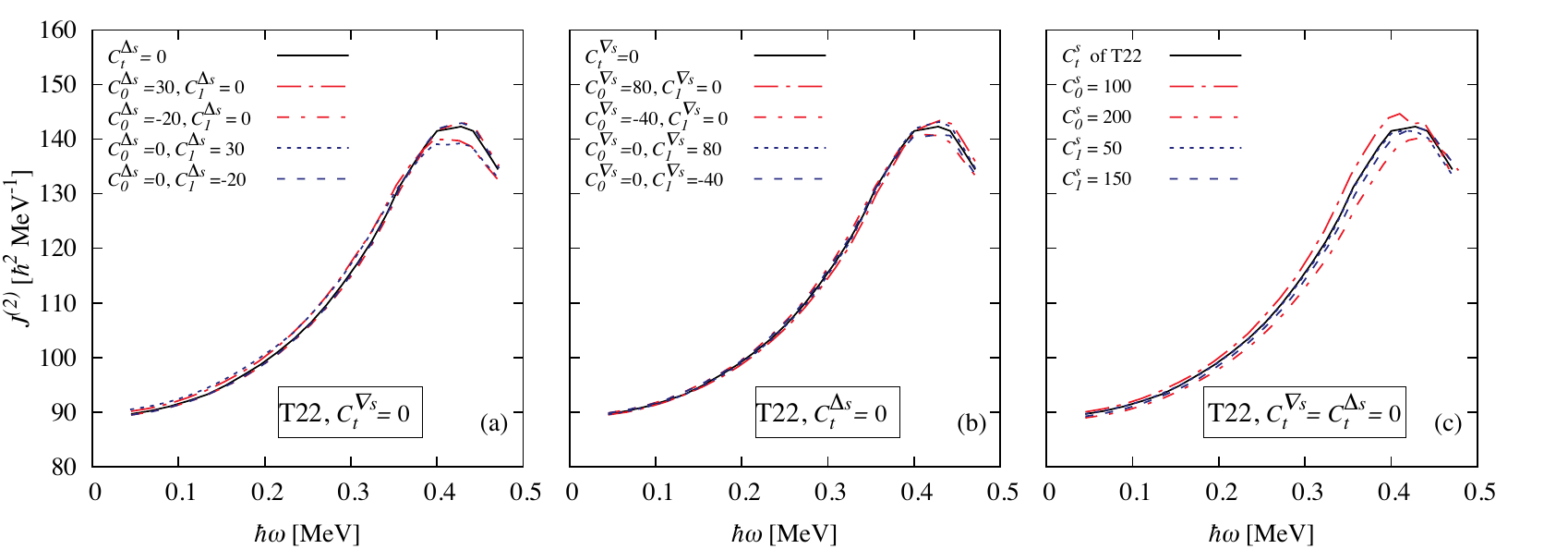}
\caption{(color online)
Dynamical moments of inertia as a function of the rotational frequency for variations of (a) $C^{\Delta s}_{t}$,  (b) the $C^{\nabla s}_{t}$, and (c) $C^{s}_{t}$ coupling constants. All coupling constants of the Skyrme EDF not explicitly mentioned in the legend are determined by the T22 parameterization. For the density-independent variations of the $C^{s}_{t}$
coupling constant (c), the same prescription as for SLy4 and explained in the caption of Fig. \ref{fig:194hg:sly4:time-oddvariations} is followed.
The $C^{\nabla s}_{t}$ and  $C^{\Delta s}_{t}$ coupling constants are expressed in units MeV fm$^{5}$ and the $C^{s}_{t}$ coupling constant is given in units MeV fm$^{3}$. }
\label{fig:194hg:t22:time-odd variations}
\end{figure*}

\begin{figure*}[!htb]
\includegraphics{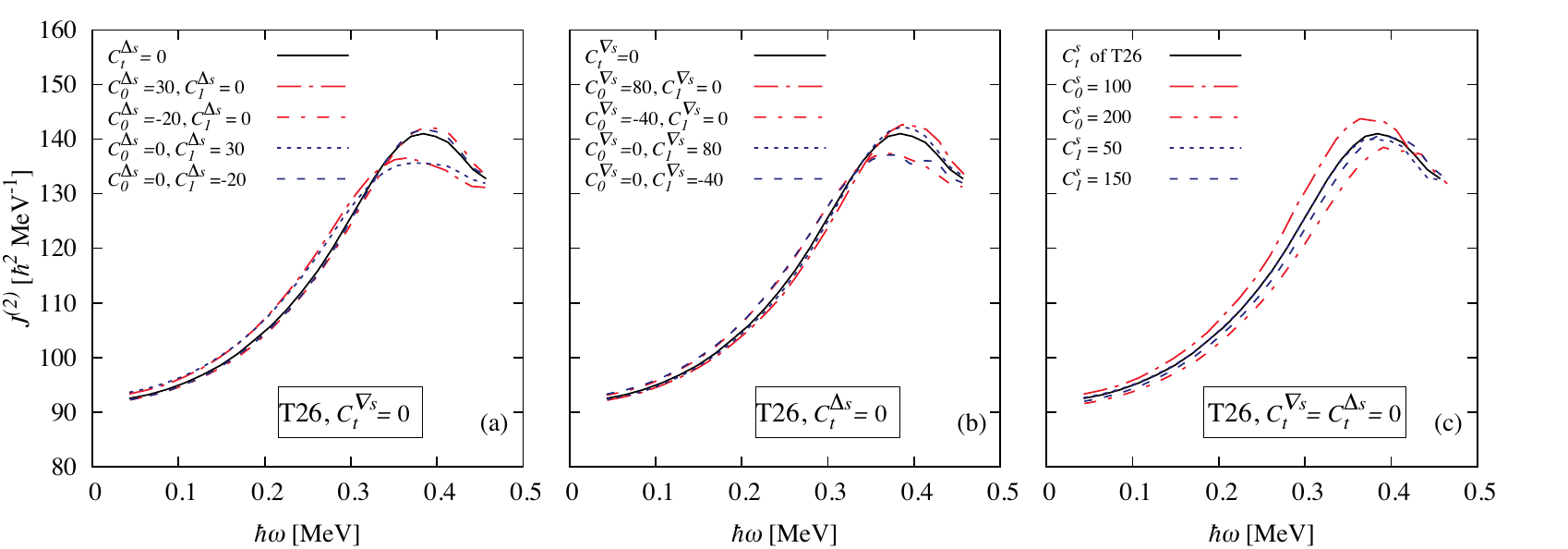}
\caption{
\label{fig:194hg:t26:time-odd variations}
(color online)
Same as Fig.~\ref{fig:194hg:t22:time-odd variations},
but for the T26 parameterization.
}
\end{figure*}

\begin{figure*}[!htb]
\includegraphics{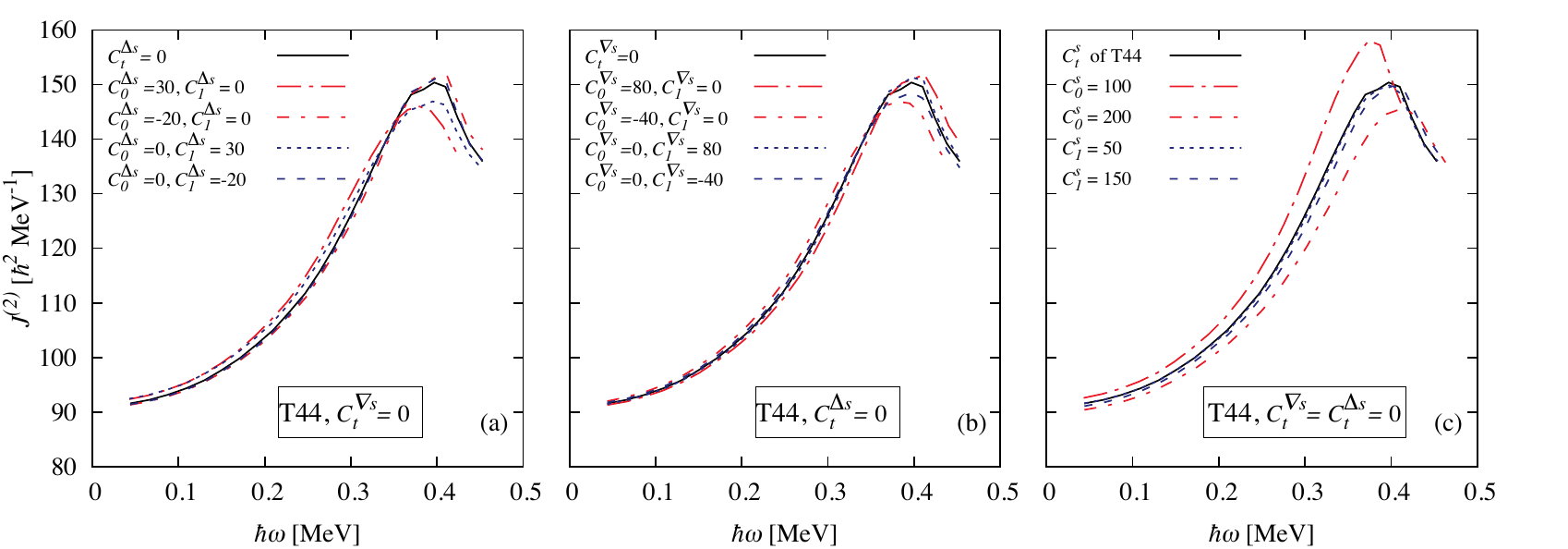}
\caption{(color online)
Same as Fig.~\ref{fig:194hg:t22:time-odd variations},
but for the T44 parameterization.}
\label{fig:194hg:t44:time-odd variations}
\end{figure*}

\begin{figure*}[!htb]
\includegraphics {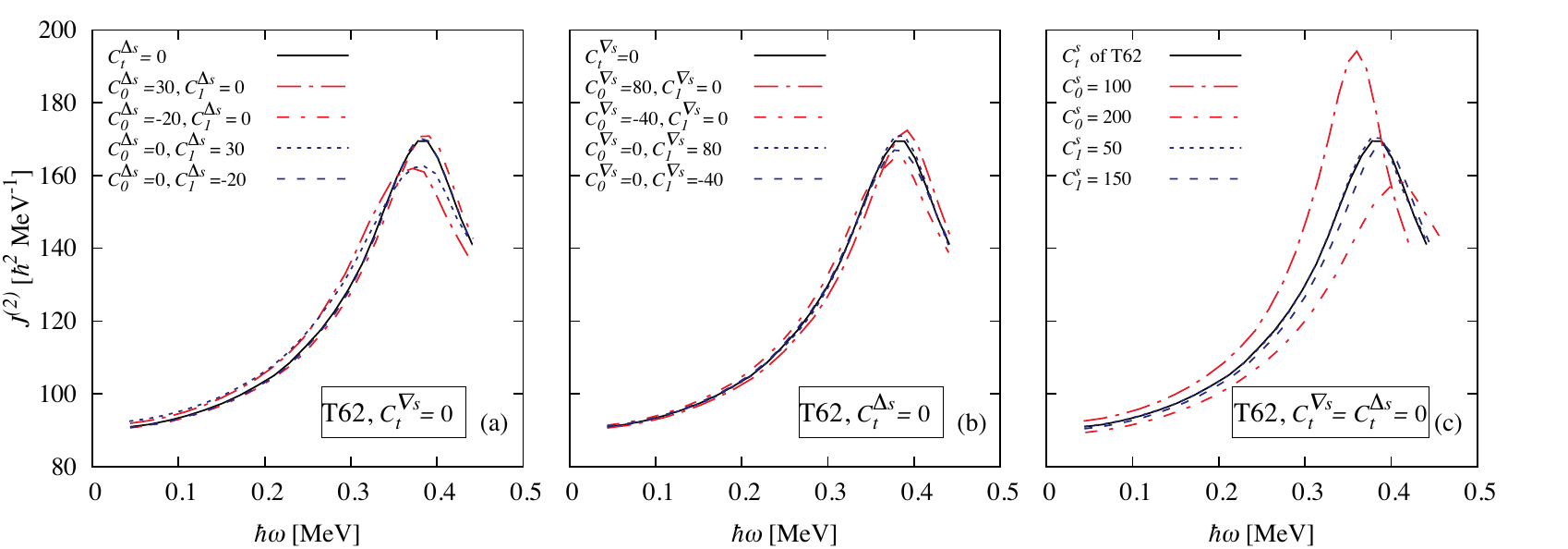}
\caption{(color online) Same as
Fig.~\ref{fig:194hg:t22:time-odd variations},
but for the T62 parameterization.
}
\label{fig:194hg:t62:time-odd variations}
\end{figure*}

\begin{table}
\begin{tabular*}{0.95\columnwidth}{@{\extracolsep{\fill}}lrrrrrr} \hline\hline
         & $g_{0}$   &$g'_{0}$  & $g_{1}$& $g'_{1}$ & $h_{0}$ & $h'_{0}$	\\
 \hline
SLy4								&	 1.387 &	0.901 &	 - &	- &	- &	-       \\
T22 								 &	0.856 & 	-0.066 & 	 0.502 & 	0.972&	 -0.100&	-0.194	\\
T26								&	0.916&	-0.074&	0.463&	 0.976 &	 0.295&	0.192	\\
T44 								&	0.400 & 	0.060 & 	 0.959 & 	0.846 &	0.198&	-0.169	\\
T62								&	-0.097&	 0.194&	1.430&	 0.715 &	0.108&	-0.536	\\
\hline
\hline
\end{tabular*}
\caption{Spin and spin-isospin
Landau parameters for SLy4, T22, T26, T44, and T62 parameterizations.}
\label{tab:landau_orig}
\end{table}

\begin{table}
\begin{tabular*}{0.85\columnwidth}{@{\extracolsep{\fill}}lrrrr} \hline\hline
 & \multicolumn{2}{c}{$g_{0}$}& \multicolumn{2}{c}{$g'_{0}$}\\
  \cline{2-3}\cline{4-5}
  &      $C^{s}_{0}$ = 100 & $ C^{s}_{0}$ = 200 & $C^{s}_{1}$ = 50 & $C^{s}_{1}$ = 150	\\\hline
  SLy4	& 0.904	&	1.808 &	0.452&	1.356\\
T22  &	0.413& 	1.328& 	 -0.514& 	 0.401\\
T26  &	0.446&	1.354&	-0.522&	0.387\\
T44 & 	-0.044& 	0.870& 	 -0.388& 	0.526\\
T62 &	-0.507&	0.415&	-0.253&	0.669\\
\hline
\hline
\end{tabular*}
\caption{Spin and spin-isospin
Landau parameters $g_{0}$ and $g'_{0}$ for the density independent variations of the $C^{s}_{t}[0]$ coupling constants of SLy4, T22, T26, T44, and T62 described in the text. The units of the $C^{s}_{t}[0]$ coupling constants is MeV fm$^{3}$. }
\label{tab:landau}
\end{table}

We now turn to interactions including a tensor term.
In Figs.~\ref{fig:194hg:t22:time-odd variations}-\ref{fig:194hg:t62:time-odd variations} we present the variations of \dyn\, as a function of  (a)  $C^{\Delta s}_{t}$, (b) $ C^{\nabla s}_{t}$, and
(c) $C^{s}_{t}$ for T22, T26, T44, and T62 respectively.
In the panels (a), one can see that the dynamical moment of inertia presents a significant decrease  at high  \ho\, for a positive value of  $C^{\Delta s}_{t}$ ($t=0$ and $t=1$) and is nearly unaffected by the $\vec{s}_{t}\cdot{\Delta \vec{s}}_{t}$ for a negative value. The result is inverted for the $(\vnabla\cdot\vec{s}_{t})^{2}$ term,  (panels (b)), for which the coupling constant has to be negative to have a visible effect.

This behavior can be understood  by looking to
Fig. \ref{fig:instabilities:laplacian isoscalar}, which is devoted to $C^{\Delta s}_{0}$ but is representative for all four coupling constants. The energy of the $\vec{s}_0\cdot \Delta \vec{s}_0$ term varies rapidly as a function of $C^{\Delta s}_{0}$  when it is positive and close to the value leading to instabilities. Owing to self-consistency effects, several other time-odd terms vary rapidly for values of $C^{\Delta s}_{0}$ close to its maximal value before instabilities appear.
For negative values of $C^{\Delta s}_{0}$, the corresponding energy behaves  much more smoothly and therefore does not affect significantly the moment of inertia.
The picture is the same for  $C^{\Delta s}_{1}$ and and for $C^{\nabla s}_{t}$, except that then, rapid changes of some energy terms are obtained for negative values of the coupling constant.

The panels (c) show how  \dyn\ is affected by variations of $C^{s}_{t}$. The changes are very similar for T22 and T26 and much larger for T44 and T62 when the $t=0$ coupling constant is varied.
Naively, one would expect that the changes of the \dyn\, with respect to the ``original'' \dyn\, (obtained with the force-generated value of $C_{t}^{s}$) should be similar for all parameterizations. Our result clearly indicates that the dynamical moment of inertia is not sensitive to the individual values of the various coupling constants but rather to some specific combinations of them.

The values of the Landau parameters (see Sect. \ref{subsect:Landau}) in the spin and the spin-isospin channels  are given in  Table~\ref{tab:landau_orig} for all parameterizations considered in this work. Table~\ref{tab:landau} presents the $g_{0}$ and $g'_{0}$ Landau parameters corresponding to the variations of $C^{s}_{t}$ that we consider in Figs.~\ref{fig:194hg:t22:time-odd variations}-\ref{fig:194hg:t62:time-odd variations}.
A close inspection of these Tables and of Figs.~\ref{fig:194hg:t22:time-odd variations}-\ref{fig:194hg:t62:time-odd variations} (c)  puts in evidence some clear trends:
\begin{itemize}
  \item For a given interaction, an increase (decrease) of $g_0$ increases (decreases) the value of \ho\, where the moment of inertia is maximum. It also broadens (sharpens) the peak of the \dyn\, curve.
  \item Similar values of $g_0$ lead to a comparable dependence of \dyn\, on \ho, as is e.g. the case for T22 and T26 for the three values of $g_0$ given in the Tables but also for T22, T26, and T44 ($C^{s}_{0}=200$ MeV fm$^{3}$), which have $g_0\approx 0.9$, or for T44 ($C^{s}_{0}=100$ MeV fm$^{3}$) and T62 where $g_0\approx -0.1$.
  \item The behavior of \dyn\, as a function of \ho\, does not depend much on $g_0$ values larger than 0.4. Its variation as of a function of $g_0$ is much larger for $g_0$ lower than 0.4.
\end{itemize}

This confirms and extends a conclusion of  Bender \emph{et al.}~\cite{bender02}. These authors have shown in a similar manner and for another Skyrme parametrization SkO' that the dynamical moment of inertia \dyn\, of the SD band of $^{152}$Dy depends mainly on the spin-isospin Landau parameter $g'_{0}$ and not so much on the actual values of the individual coupling constants $C^{s}_{1}$ and $C^{T}_{1}$. Different combinations of EDF coupling constants leading to the same value of $g_0'$ were found to change little to the \dyn. Note that these authors have also varied $C^{\Delta s}_{1}$, extending their study well outside the limits for which we find instabilities. This discrepancy might be due to the use a different technique for solving the mean-field equations, \textit{i.e.} by means of an expansion on an oscillator basis instead of using a cartesian mesh.

Note also that empirical values for $g_{0}\approx 0.4$ \cite{osterfeld92} and $1.4 \le g'_{0} \le 1.6$ \cite{bertsch81,gaarde81,suzuki82}  have been derived from M1 and Gamow-Teller response.  The values quoted in Table~\ref{tab:landau_orig} and~\ref{tab:landau} are in all cases except one very different from the empirical values.

\begin{figure}[!htbp]
\includegraphics {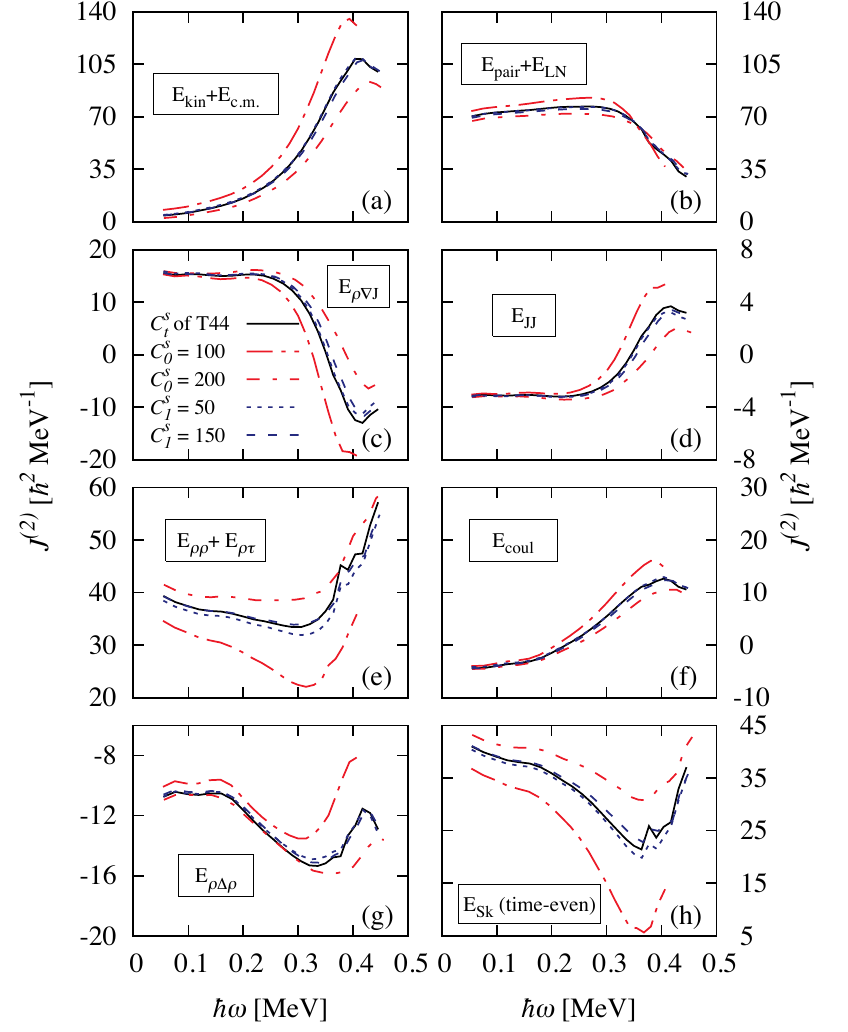}
\caption{(color online)
Contributions to the dynamical moment of inertia coming from the time-even terms of the EDF as a function of the rotational frequency for variations of the $C^{s}_{t}$ coupling constant. All coupling constants not explicitly mentioned are determined by the T44 parameterization. The $C^{s}_{t}$ coupling constants are expressed in units MeV fm$^{3}$}.
 \label{fig:194hg:Csvar:dynte}
 \end{figure}
 \begin{figure}[!htbp]
\includegraphics{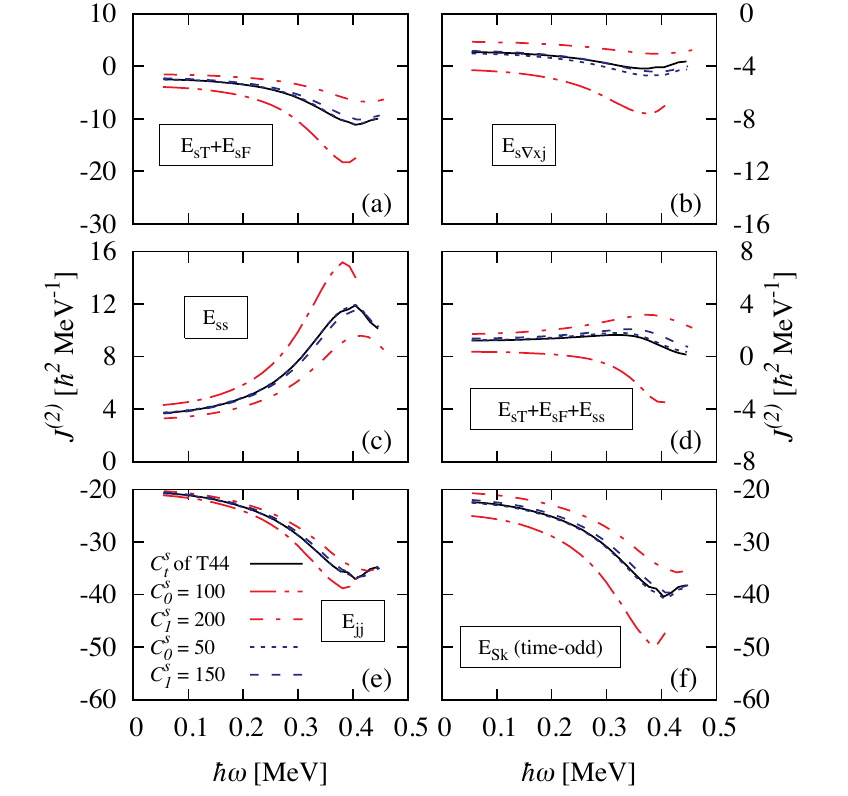}
\caption{(color online)  Same as Fig. \ref{fig:194hg:Csvar:dynte}, but for the time-odd terms in the Skyrme EDF. }
\label{fig:194hg:Csvar:dynto}
\end{figure} %

The previous analysis shows that the $C^{s}_{t}$, $C^{F}_{t}$ and $C^{T}_{t}$, coupling constants are linked by their contribution to the values of the Landau parameters $g_{0}$ and $g'_{0}$. One might wonder whether variations of $C^{s}_{t}$ affect mainly the terms in the EDF depending on the spin-density $\vec{s}_{t}$. To analyze how all terms in the EDF are affected by a variation of $C^{s}_{t}$, we have decomposed the \dyn\,  obtained for the modified T44 parameterizations in the same way as in Figs.~\ref{fig:194hg:Csvar:dynte} and ~\ref{fig:194hg:Csvar:dynto}. Differences appear only for large values of \ho. At \ho=0.4, taking $C^{s}_{0}=100$ MeV fm$^{3}$ modifies the total energy obtained with the T44 parameterization by about 600 keV. This change results from the partial cancelation of larger changes with different signs of all terms of the EDF, including
the pairing and Coulomb energy. In particular, the Skyrme time-even and time-odd contributions decrease the energy by about 900 keV and 500 keV respectively. The energy differences obtained for $C^{s}_{0}=200$ MeV fm$^{3}$ are of the same order of magnitude, but are much smaller for variations of $C^{s}_{1}$. These energy changes at high \ho\, affect the slope of the different energy contributions and therefore their contributions to the \dyn\,. These are plotted in Figs. \ref{fig:194hg:Csvar:dynte}-\ref{fig:194hg:Csvar:dynto}. One clearly sees that the modification of the coupling constant of a relatively small term, \textit{in casu} the $E_{ss}$, affects all other terms through self-consistency and how little changes in the energy can make a large difference in the \dyn.

Our analysis demonstrates that the variations of pairing and of the time-odd terms have clearly distinguishable effects on the shape of the \dyn\,. The time-odd terms influence the slope of the \dyn\, and the \ho\,  at which the plateau occurs. While the effect of variations of the $C^{\nabla s}_{t}$ and $C^{\Delta s}_{t}$ coupling constants on the \dyn\, is rather small and depends on the sign of the coupling constant, variation of the $C^{s}_{0}$ coupling constant may lead to significant changes in the \dyn.

%
%
\section{Results for the superdeformed band in \nuc{152}{Dy}}
\label{sect:dy152}

\subsection{General comments}
The superdeformed rotational bands known in $^{152}$Dy exist in a very different regime than the one of $^{194}$Hg.
 SD band in the $A\approx 150$ region have only been detected for spins higher than 20$\hbar$. For such angular momenta, pairing correlations are strongly weakened by the Coriolis anti-pairing effect. Hence, pairing is expected to play only a minor role in that region. Early studies by  Bengtsson \textit{et al.} \cite{bengtsson88} using the Nilsson-Strutinsky approach have demonstrated that the behavior of the SD bands is strongly influenced by the number of nucleons that occupy the intruder orbitals. SD bands in the Dy-region have been studied extensively within self-consistent mean-field approaches \cite{bonche96,rigollet99,bender02,afanasjev10}. Moreover, they have been used as a laboratory to study the time-odd terms in the EDF \cite{dobaczewski95,dobaczewski96,bender02}.

\subsection{General features}

%
\begin{figure}[!tb]
\includegraphics {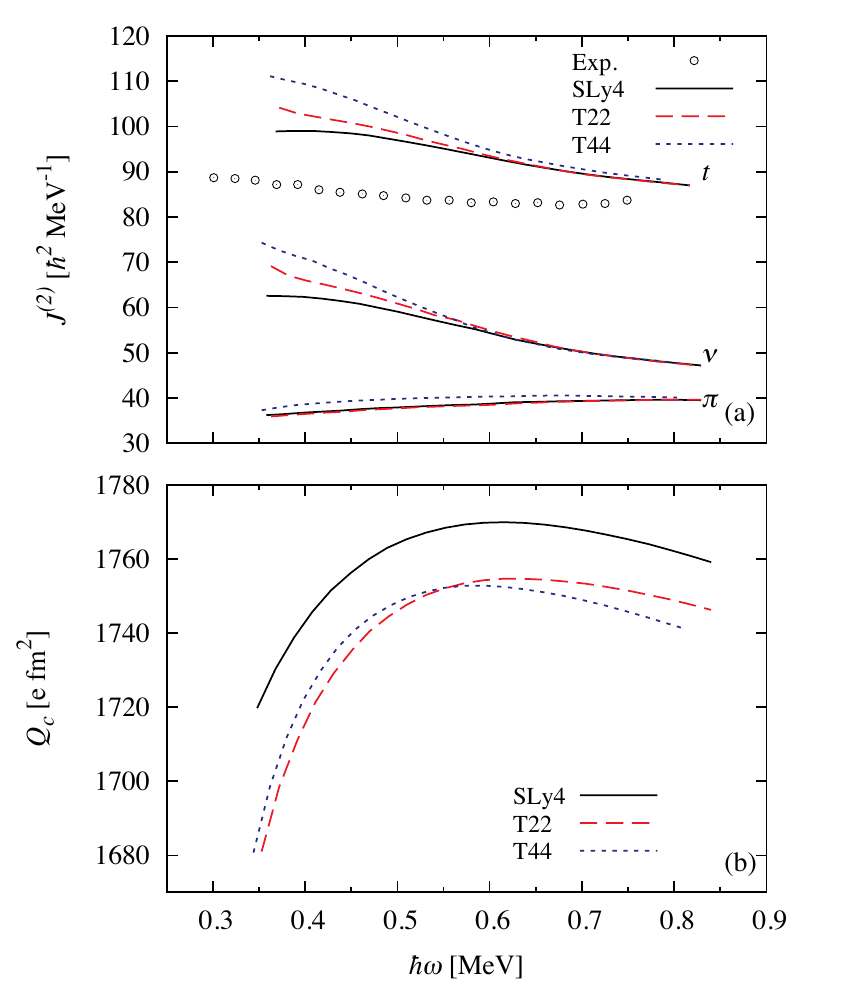}
\caption{(color online) (a) : Proton ($\pi$), neutron ($\nu$), and  total ($t$) dynamical moments of inertia  as a function of the rotational frequency for the SD ground band in $^{152}$Dy with the SLy4, the T22, and the T44 parameterization . (b) : The charge quadrupole moment in $^{152}$Dy as a function of the rotational frequency for the SLy4, the T22, and the T44 parameterization.  }
\label{fig:152dy:sly4t22t44:comparison J2 & Q}
\end{figure}
%

%
\begin{figure}[!htbp]
\includegraphics {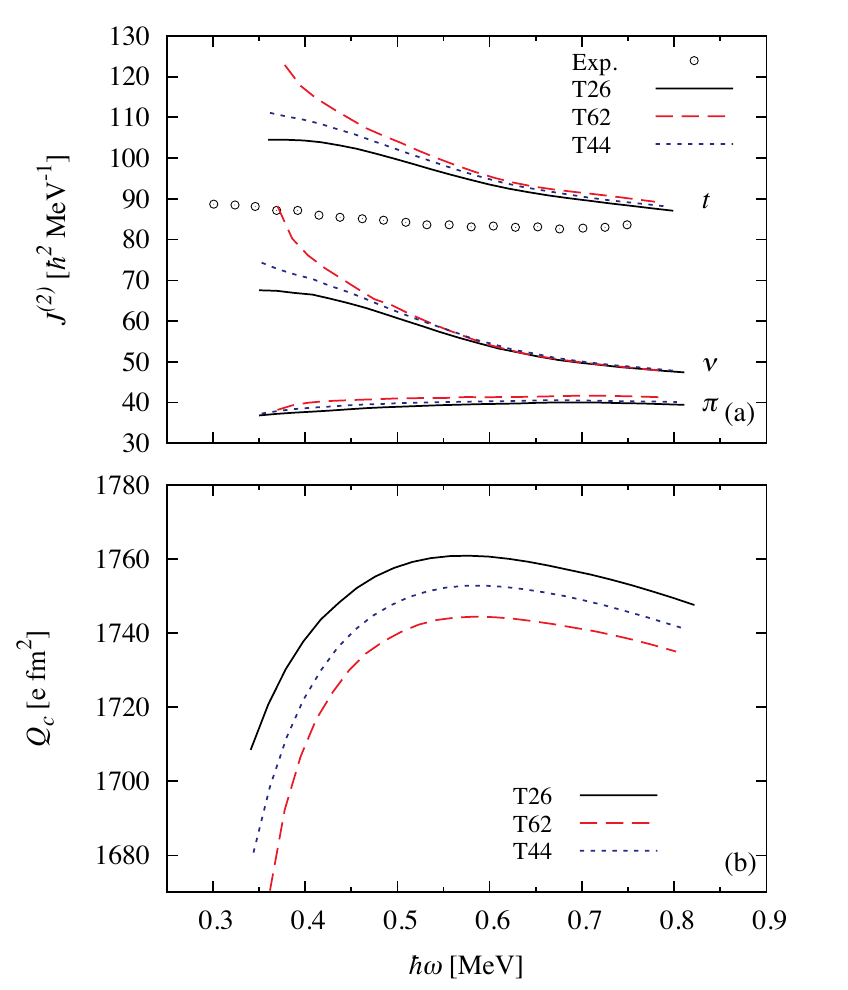}
\caption{(color online) (a) : Proton ($\pi$), neutron ($\nu$), and  total ($t$) dynamical moments of inertia as a function of the rotational frequency for the SD ground band in $^{152}$Dy with the T26, the T44, and the T62 parameterization. (b) : The charge quadrupole moment in $^{152}$Dy as a function of the rotational frequency for the T26, the T44, and the T62 parameterization.  }
\label{fig:152dy:t26t44t62:comparison J2 & Q}
\end{figure}

The dynamical moments of inertia (panel (a)) and the charge quadrupole moments (panel (b)) are shown in
Fig. \ref{fig:152dy:sly4t22t44:comparison J2 & Q} for SLy4, T22, and T44 and in Fig. \ref{fig:152dy:t26t44t62:comparison J2 & Q}  for the T26, T44, and T62. As for $^{194}$Hg, the \dyn\, calculated with SLy4 and T22 are very close, with small differences at low spin.
The main difference between the parameterizations is the presence of a strong peak at low spin for T62, which is less pronounced for T44 and absent for the other parameterizations.
   In all cases, the difference between the \dyn\, is caused by the neutrons. The $Q_{c}$ values presented in the panels (b) differ by less than 2\% and all display the same behavior, increasing until \ho=0.6 MeV after which they slowly start decreasing again.

%
\begin{figure}[!tbp]
\includegraphics {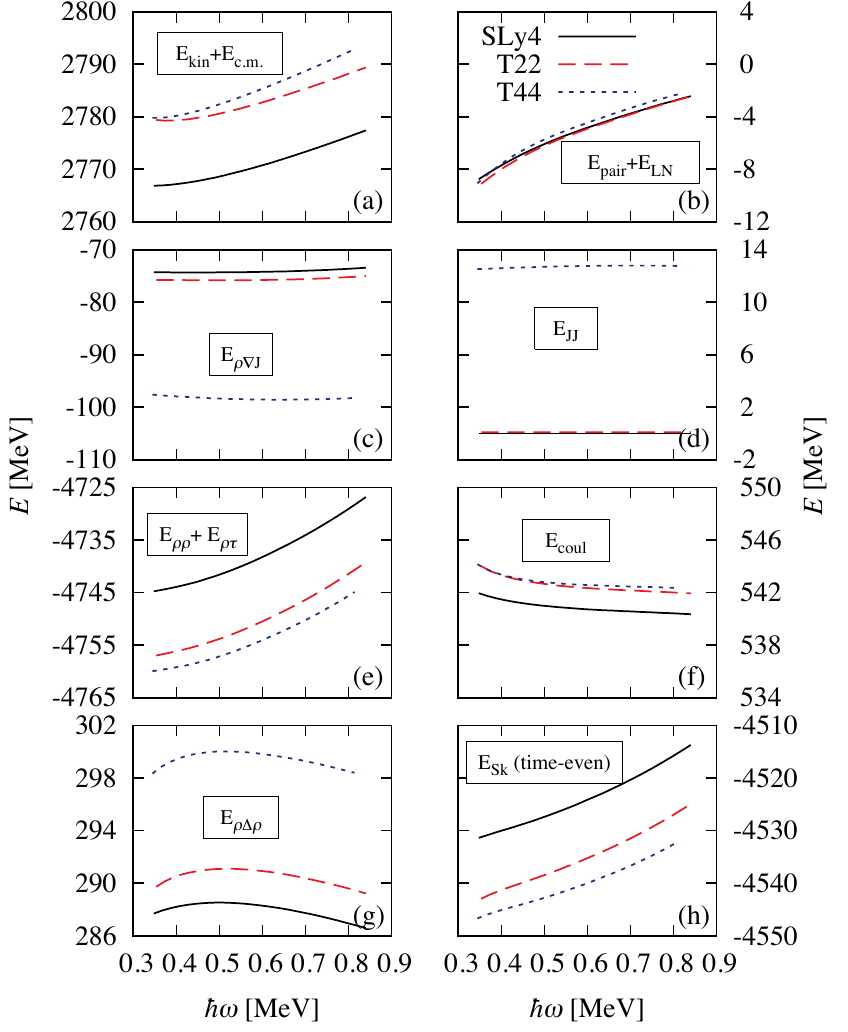}
\caption{(color online) Evolution of different terms in the energy density functional as a function of the rotational frequency for the SLy4, T22, and T44 parameterizations in the calculation of the ground state superdeformed band of $^{152}$Dy.}
\label{fig:152dy:sly4t22t44:energyte}
\includegraphics {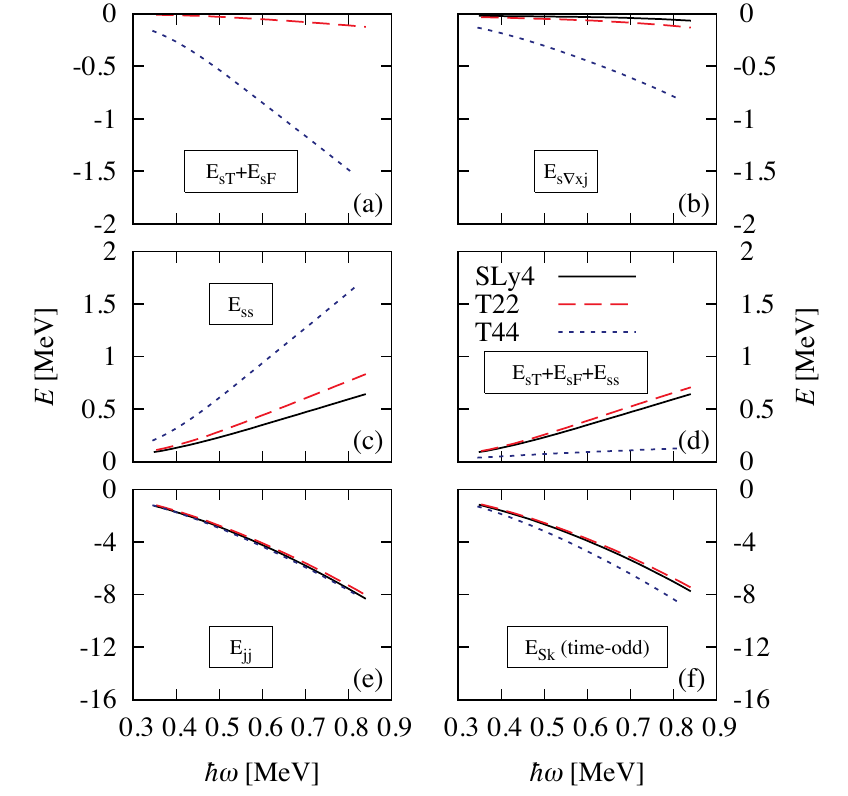}
\caption{(color online) Same as Fig. \ref{fig:152dy:sly4t22t44:energyte}, but for the time-odd terms in the Skyrme energy density functional. }
\label{fig:152dy:sly4t22t44:energyto}

\end{figure}
\begin{figure}[!tbp]
\includegraphics {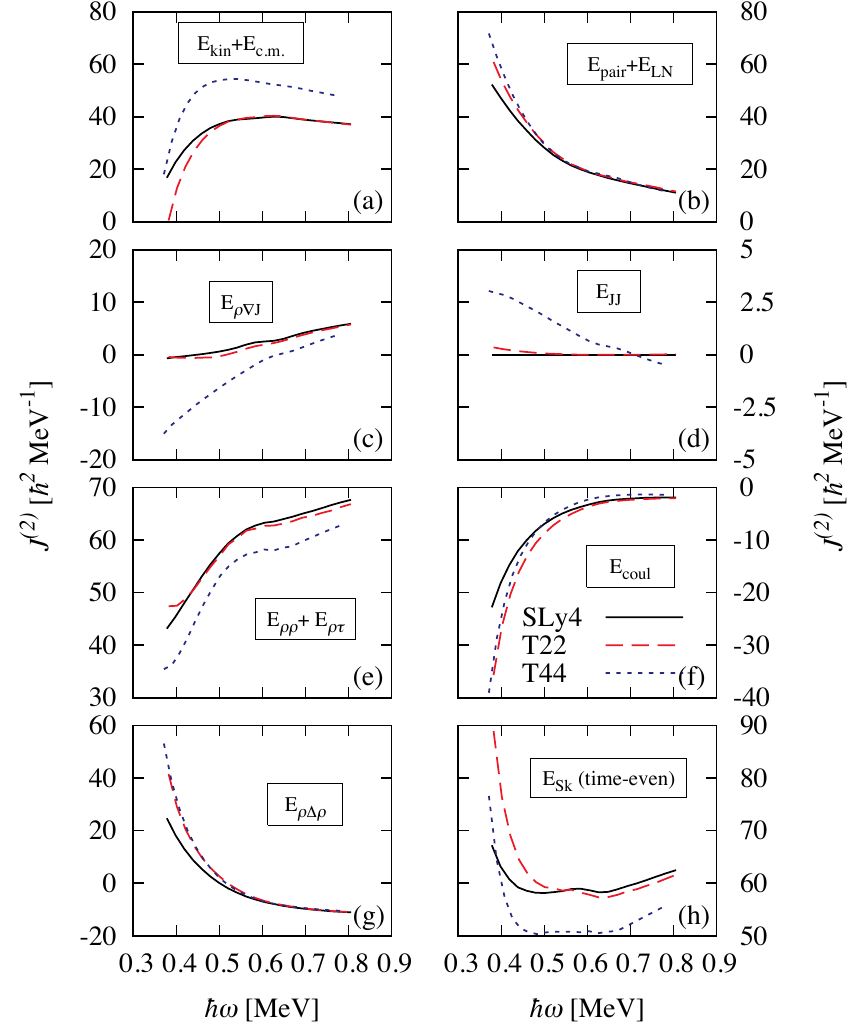}
\caption{(color online) \label{fig:152dy:sly4t22t44:dynte}Dynamical moments of inertia of different terms in the energy density functional as a function of the rotational frequency for the SLy4, T22, and T44 parameterizations  in the calculation of the ground state superdeformed band of $^{152}$Dy.}

\includegraphics {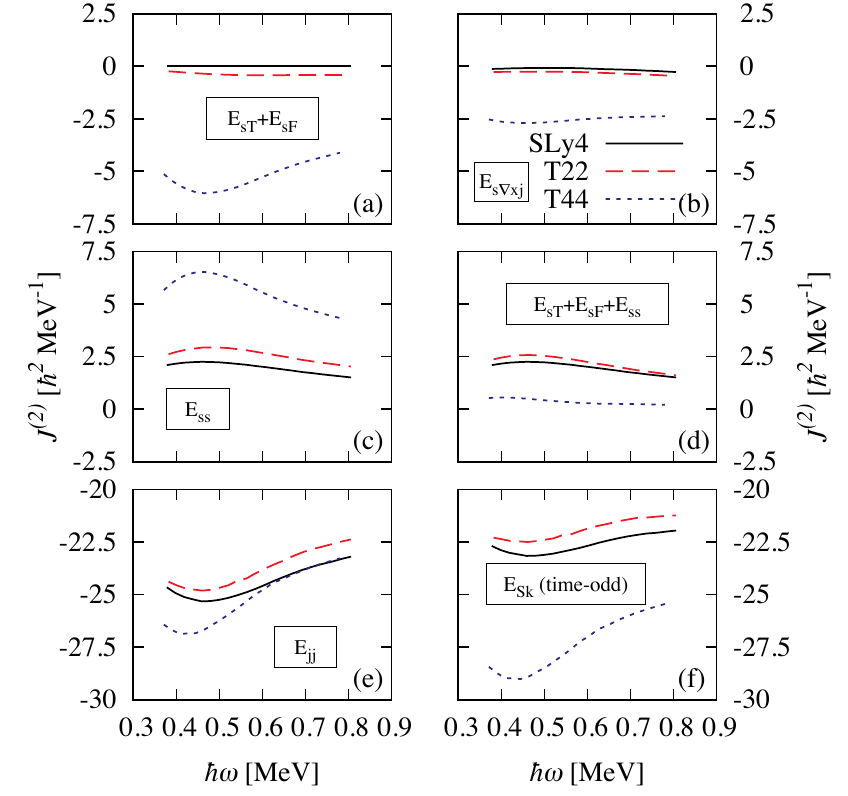}
\caption{(color online)  Same as Fig. \ref{fig:152dy:sly4t22t44:dynte}, but for the time-odd terms in the Skyrme energy density functional. }
\label{fig:152dy:sly4t22t44:dynto}

\end{figure}
%
\begin{figure}[!htbp]
\centerline{\includegraphics{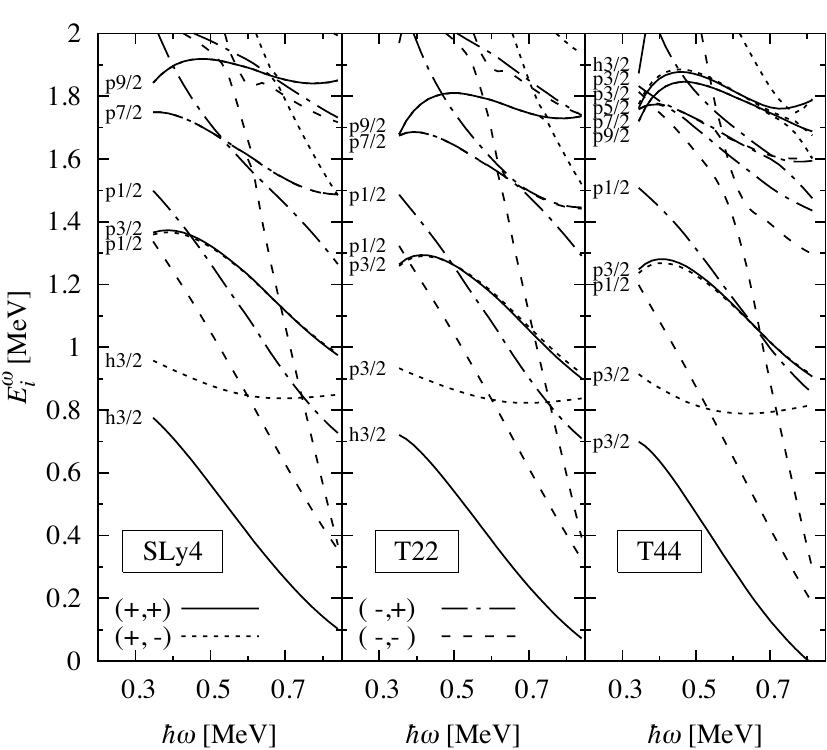}}
\caption{
\label{fig:152dy:t26t44t62:protonrouthians}
Proton quasi-particle Routhians for the ground-state superdeformed
band of $^{152}$Dy for the SLy4, the T22, and the T44 parameterization. The
(parity, signature) combinations are indicated in the figure. All Routhians are additionally
characterized by the  $j$-component of the dominant single-particle state in the quasi-particle
wave function along the axis of largest elongation at \ho=0 and by their particle (p) or hole (h) character.
}
\centerline{\includegraphics{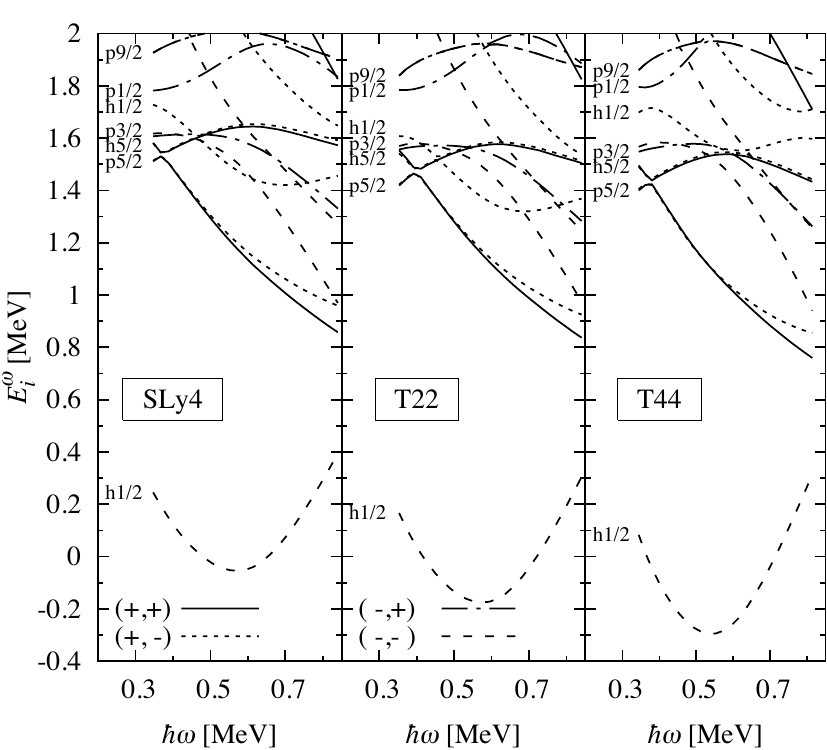}}
\caption{
\label{fig:152dy:t26t44t62:neutronrouthians}
Same as Fig. \ref{fig:152dy:t26t44t62:protonrouthians}, but for neutrons.
}
\end{figure}

The total energy and the \dyn\, are decomposed into their time-even and time-odd components in Figs. \ref{fig:152dy:sly4t22t44:energyte}-\ref{fig:152dy:sly4t22t44:energyto} and \ref{fig:152dy:sly4t22t44:dynte}-\ref{fig:152dy:sly4t22t44:dynto}, respectively, as it was done for \nuc{194}{Hg}. Decompositions are only shown for the SLy4, T22 and T44. The results for T26 and T62 are very similar.  The difference of energy between the lowest ($\langle J_{z}\rangle=28\hbar$) and highest ($\langle J_{z}\rangle=74\hbar$) states that we have calculated amounts to 26.3 MeV for SLy4 and T22 and 25.5 MeV for T44. All parts of the EDF give contributions of the same order of magnitude to this change in energy, about 12 MeV for the Skyrme EDF, 10 MeV for the kinetic energy, 7 MeV for the pairing energy and $-2$ MeV for the Coulomb energy. The Skyrme contribution can be decomposed into 18 MeV from the time-even and $-6$ MeV from the time-odd terms.

Focussing on the time-even contributions (Fig. \ref{fig:152dy:sly4t22t44:energyte}), their  \ho\, dependence is very similar for all parameterizations, which is also reflected in the corresponding contributions to \dyn\, presented in Fig. \ref{fig:152dy:sly4t22t44:dynte}. As expected, the pairing energy is smaller than in the $A\approx 190$ region, because the superdeformed bands in the $A\approx 150$ region only occur at higher spin.
 The ratio $E_{\rho\nabla J}(T22)/E_{\rho\nabla J}(T44)$ is again  approximately equal to the ratio of the corresponding coupling constants. For the time-odd terms (Fig. \ref{fig:152dy:sly4t22t44:energyto}),
the contributions of $E_{sT}$ and $E_{sF}$ terms that appear for T44 cancel out the $E_{ss}$ term, which is much larger without the inclusion of a tensor term.

The time-even and time-odd contributions to the \dyn\, (Figs. \ref{fig:152dy:sly4t22t44:dynte}-\ref{fig:152dy:sly4t22t44:dynto}), indicate that the $E_{Sk}$ (time-even) is the largest contribution  at low \ho, amounting up to 96\% of the total value for T22 and about 80\% for T44. The pairing energy $E_{pair}$ is the second largest contribution, averaging about 70\%, and is mostly canceled out by $E_{coul}$ (about -40\%) and the time-odd $E_{Sk}$ terms (about -25\%). The $E_{kin}+E_{c.m.}$ contribution is negligible at low \ho\,. With increasing \ho, the $E_{pair}+E_{LN}$ and $E_{coul}$ contributions drop to 13\% and 2\% respectively, while the $E_{kin}+E_{c.m.}$ contribution quickly grows to about 50\%. Whereas the time-odd $E_{Sk}$ contribution stays more-or-less constant with \ho\,, the time-even part of $E_{Sk}$ reduces to about 70\%.
Hence, in contrast to \nuc{194}{Hg},  the Skyrme EDF plays a more important role in the decompositions of the total energy and of the \dyn\,. Even though changes the time-even and time-odd components of the Skyrme EDF no longer almost cancel out, they still counteract each other.

Finally, the proton and neutron quasiparticle Routhians for the SLy4, T22 and T44 parameterization are presented in Figs. \ref{fig:194hg:t26t44t62:protonrouthians}-\ref{fig:194hg:t26t44t62:neutronrouthians}. Again, the low-lying quasiparticle Routhians are very similar for all parameterizations and subtle differences are observed in the alignment of the Routhians. The main difference observed between the parameterizations is the location and evolution of the neutron h $1/2^{+}$ state. When going from SLy4 to T22 and further on to T44, the  h $1/2^{+}$ quasiparticle state starts of at lower energy and the minimum becomes more pronounced.

%
%

\section{Discussion and conclusion}
\label{sect:conclusions}

We have studied the impact of the introduction of tensor terms in the Skyrme energy density functional on the dynamical moments of inertia of superdeformed bands. The excellent description of these bands by conventional EDFs was a major success of microscopic mean-field models in the 90s. Therefore, it is important to verify that the inclusion of a zero-range tensor force does not destroy this agreement. In our study, special attention was paid to the time-odd tensor terms in the Skyrme EDF, which are zero when self-consistent time-reversal symmetry is not broken, as is the case for the ground states of the spherical and deformed nuclei that were studied in Articles I and II.

In order to disentangle their respective role, we have tested a selection of four T$IJ$ parameterizations, introduced in Article~I, which represent a wide range of values for the isoscalar and isovector tensor coupling constants. As a reference without tensor terms, we have included the SLy4 parameterization. We have found that the inclusion of tensor terms in the Skyrme EDF does not change the overall behavior of the dynamical moments of inertia, although differences in slope and location of plateau do occur. This results from an intricate compensation mechanism due to the self-consistency that is implemented at two different levels in our method: in the fitting procedure of the interactions considered and in the solution of the mean-field equations.

Even though the energy contribution of the time-even $E_{JJ}$ tensor terms is in general an order of magnitude larger than that of the time-odd $E_{sT}+E_{sF}$ tensor terms, the time-odd tensor terms evolve more rapidly as a function of rotational frequency such that their contribution to \dyn\ is of the same order of magnitude as the one of the time-even tensor terms. Similarly, the Skyrme time-even and time-odd energy contributions $E_{Sk}$ typically differ by three orders of magnitude but have similar contributions to \dyn. In all cases encountered, the time-odd $E_{Sk}$ contributions to the \dyn\, partially cancels out the time-even $E_{Sk}$ contribution.

A detailed study of the time-odd terms in the Skyrme EDF has shown the following features:
\begin{enumerate}
\item We have seen that the values of the coupling constants of the time-odd tensor terms that contain derivatives of spin densities ($E_{s\Delta s}$ and $E_{\nabla s \nabla s}$) have to be chosen within strict limits to avoid finite-size instabilities. Such instabilities were encountered for all T$IJ$ parameterizations considered. Therefore, we adopted the functional point of view in our study and put their coupling constant to zero. By contrast, these instabilities are not encountered in spherical QRPA studies using the same T$IJ$ parameterizations \cite{Bai09a,Cao09a,Bai09b,cao10,Bai10a,Bai10b,Cao11a}, presumably because of the non-variational character of QRPA. A point of special interest will be the analysis of finite-size instabilities using the technique of Ref. \cite{lesinski06,davesne09} that is currently underway.
\item The effect of modifications in the strength of the pairing interactions can be clearly distinguished from modifications of the time-odd terms in the Skyrme EDF that are not restricted by symmetry considerations.
\item The energy changes that are introduced through the presence of the time-odd tensor terms $E_{sT}+E_{sF}$ are partially cancelled out by the presence of the other time-odd terms that contain the spin density $\vec{s}_{t}$. Consequently the time-odd $E_{Sk}$ is almost entirely determined by the $E_{jj}$ contribution. Variations of the $C^{s}_{t}$ coupling constant indicate that only combinations of the $E_{sT}+E_{sF}$ and $E_{ss}$ terms with similar values of $g_{0}$ and/or $g'_{0}$ act independently.
\end{enumerate}

When comparing the dynamical moments of inertia obtained for an EDF including tensor terms to the experimental ones, the agreement for T22 and T26 is as satisfactory as for SLy4. The T62 parametrization that acts only between neutrons and protons in spherical symmetry exhibits a pronouncedly peaked behavior that does not compare well with the experimental \dyn. Finally, the results obtained with the T44 parameterization, which was one of the two parameterizations leading to the best results for low-lying collective states in a RPA calculation of \nuc{208}{Pb} and \nuc{40}{Ca}~\cite{Cao11a}, are found to be in good comparison with the experimental results.

\section*{Acknowledgments}

This research was supported in parts by the PAI-P6-23 of the
Belgian Office for Scientific Policy, by the French Agence Nationale de la
Recherche under Grant No. ANR 2010 BLANC 0407 "NESQ", and by the
CNRS/IN2P3 through the PICS No. 5994. V.H. gratefully acknowledges a
postdoctoral fellowship from the F.R.S.-FNRS (Belgium) and the partial
financial support by the US DOE under grant DE-FG02-95ER-40934.

\begin{appendix}
\section{Coupling constants of the Skyrme energy density functional
in the isospin and the proton-neutron formulation}
\label{app:cpling}
In Table \ref{tab:cpling:isovspn}, we provide the relation between the coupling constants of the Skyrme EDF in the isospin formulation (\ref{eq:EF:time-even}-\ref{eq:EF:time-odd}) and those appearing in the Skyrme EDF in the proton-neutron formulation \eqref{eq:func:Skyrme:pn}.
\begin{table}[!htb]
\caption{Coupling constants in the isospin formulation of the EDF as a function of the coupling constants in the proton-neutron formulation of the EDF in the format $C=\sum_{i}a_{i}b_{i}$, where the factors $a_{i}$ are given in the Table.}\begin{center}
\begin{tabular*}{0.8\columnwidth}{@{\extracolsep{\fill}}lllllllll}
\hline\hline
		       		& $b_1$ & $b_2$  & $b_3$ & $b_4$ & $b_5$ & $b_6$ & $b_7$ & $b_8$ \\
\hline				
$C^{\rho}_{0} $			& 1   & 1/2 & 0    &  0  &  0  &  0   &  $\rho_{0}^{\alpha}$ & $\rho_{0}^{\alpha}/2$ \\
$C^{\rho}_{1} $	  	         &   0   & 1/2 & 0   & 0   &  0  &  0   &  0   & $\rho_{0}^{\alpha}/2$ \\
$C^{\tau}_{0} $	  	         &   0   &  0  &  1   & 1/2  &  0  &  0   &  0   & 0 \\
$C^{\tau}_{1} $	 	         &   0   &  0  &  0   & 1/2  &  0  &  0   &  0   & 0 \\
$C^{\Delta\rho}_{0} $        &   0   &  0  &  0   &  0    &  1  &  1/2   &  0   & 0 \\
$C^{\Delta\rho}_{1} $        &   0   &  0  &  0   & 0     &  0  &  1/2   &  0   & 0 \\
\hline
		       		& $b_9$ & $b_{9q}$  & $b_{10}$ & $b_{11}$ & $b_{12}$ & $b_{13}$ & $b_{14}$ & $b_{15}$ \\
\hline
$C^{\nabla J}_{0}$   &  1    &   1/2     &      0     &     0     &     0        &    0      &     0	    &    0     	 \\
$C^{\nabla J}_{1}$   &  0    &   1/2     &      0     &     0     &     0        &    0      &     0	    &    0     	 \\	
$C^{s}_{0}$		& 0     &   0        &     1	     &     1/2  &  $\rho_{0}^{\alpha}$ & $\rho_{0}^{\alpha}/2$  & 0 & 0\\
$C^{s}_{1}$		& 0     &   0        &     0	     &     1/2  &  0  & $\rho_{0}^{\alpha}/2$ & 0 & 0\\
$C^{T}_{0}$               &  0    &   0        &      0     &     0     &     0        &    0      &     -1	    &   -1/2     	 \\
$C^{T}_{1}$               &  0    &   0        &      0     &     0     &     0        &    0      &     0	    &    -1/2     	 \\	
\hline
		       		& $b_{16}$ & $b_{17}$  & $b_{18}$ & $b_{19}$ & $b_{20}$ & $b_{21}$&  ~  & ~   \\
\hline				
$C^{F}_{0}$	         &  -2    &   -1        &      0     &     0     &     0        &    0      &     ~	    &   ~     	 \\	 
$C^{F}_{1}$	         &    0    &   -1        &      0     &     0     &     0        &    0      &     ~	    &   ~     	 \\	 
$C^{\Delta s}_{0} $	  	         &   0   &  0  &  1   & 1/2  &  0  &  0   &  ~   & ~ \\
$C^{\Delta s}_{1} $	 	         &   0   &  0  &  0   & 1/2  &  0  &  0   &  ~  & ~ \\
$C^{\nabla s}_{0} $        &   0   &  0  &  0   &  0    &  1  &  1/2   &  ~   & ~ \\
$C^{\nabla s}_{1} $        &   0   &  0  &  0   & 0     &  0  &  1/2   &  ~  & ~ \\
\hline\hline
\end{tabular*}
\end{center}
\label{tab:cpling:isovspn}
\end{table}%

%
%

\section{Densities and currents in cr8}
\label{app:cr8}

In this appendix, we provide the expressions of the densities and currents as
they are implemented in our cranked Hartree-Fock-Bogoliubov solver cr8. This extends
the discussion of Ref.~\cite{bonche87} by the densities and currents entering the tensor
terms.

The \texttt{cr8} code uses a coordinate-space representation of the wave functions and fields.
The HFB equations are solved with the so-called two-basis method, where in an
iterative scheme the HFB Hamiltonian \eqref{eq:HFB} is diagonalized in
a single-particle basis that converges towards the eigenstates of the
mean-field Hamiltonian $h$, Eq.~\eqref{eq:h}. The densities needed to construct
the local fields are calculated in the canonical single-particle basis,
which is obtained by diagonalization of the density matrix $\rho$.
For a detailed discussion of our method of solving the cranked HFB equations
we refer to Refs.~\cite{gall94,terasaki95}.

The \texttt{cr8} code assumes triaxial symmetry of the nucleus, where all
single-particle wave functions have a plane reflection symmetry about
the $x=0$, $y=0$ and $z=0$ planes. There are several possible choices
to achieve this \cite{dobaczewski00b}. The cr8 code chooses the single-particle
wave functions $\Phi_k(\vec{r},\sigma)$ to be eigenstates of
\begin{enumerate}
\item[(i)]
parity
\begin{equation}
\label{eq:symm:p}
\hat{P} \Phi_k(\vec{r},\sigma)
= \Phi_k(-\vec{r},\sigma)
= p_k \Phi_k(\vec{r},\sigma), \qquad p_k=\pm 1 \, ,
\end{equation}
\item[(ii)] $z$ signature
\begin{align}
\label{eq:symm:rz}
\hat{R}_{z}\Phi_k(\vec{r},\sigma)
& = e^{i\pi \hat{J}_z}\Phi_k(\vec{r},\sigma)~,
	\nn\\
& = i\eta_k \Phi(\vec{r},\sigma), \qquad \eta_k=\pm 1 \, ,
\end{align}
\item[(iii)]
$y$ $T$-simplex
\begin{align}
\label{eq:symm:sty}
\hat{S}^{T}_{y}\Phi_k(\vec{r},\sigma)
& = \hat{T}\hat{P}\hat{R}_{y}\Phi_k(\vec{r},\sigma)~,
	\nn\\
& = \Phi_k(\vec{r},\sigma)~,
\end{align}	
where $\hat{T}$ is the time-reversal operator.
\end{enumerate}
A wave function is completely determined by four real functions $\Psi_{k,\alpha}$
($\alpha = 1, \ldots, 4$) that correspond to the real (Re) and imaginary (Im) parts
of the spin-up and spin-down components of $\Phi_k$. A different numbering of these
four components was adopted for wave functions of positive and negative signature
in \cite{bonche87}

\begin{align}
&\fourrow{\Psi_{k,1}(\vec{r})}{\Psi_{k,2}(\vec{r})}{\Psi_{k,3}(\vec{r})}{\Psi_{k,4}(\vec{r})}=
         \fourrow{\text{Re}~\Phi_k(\vec{r,+})}{\text{Im}~\Phi_k(\vec{r,+})}
                 {\text{Re}~\Phi_k(\vec{r,-})}{\text{Im}~\Phi_k(\vec{r,-})}
~~~\text{for $\eta_k = 1$,}\\
&\fourrow{\Psi_{k,1}(\vec{r})}{\Psi_{k,2}(\vec{r})}{\Psi_{k,3}(\vec{r})}{\Psi_{k,4}(\vec{r})}=
         \fourrow{\text{Re}~\Phi_k(\vec{r,-})}{\text{Im}~\Phi_k(\vec{r,-})}
                 {\text{Re}~\Phi_k(\vec{r,+})}{\text{Im}~\Phi_k(\vec{r,+})}
~~~\text{for $\eta_k = -1$.}
\end{align}
This choice ensures that each of the four real functions $\Psi_{k,\alpha}$
has the same definite reflection symmetry about the $x$, $y$ and $z$ planes, listed
in Table~\ref{tab:planesymmetries:wf} independently of its signature.

\begin{table}[!tb]
\caption{
Parities of the components $\Psi_{k,\alpha}$ of a wave function
$\Phi_k$ of parity $p_k$ with respect to the $x=0$, $y=0$, and $z=0$ planes.
}
\begin{center}
\begin{tabular*}{0.5\columnwidth}{@{\extracolsep{\fill}}lccc}
\hline
$\alpha$	& x	& y	& z \\
\hline
1		& $+$	& $+$	& $ p_k$ \\
2		& $-$	& $-$	& $ p_k$ \\
3		& $-$	& $+$	& $-p_k$ \\
4		& $+$   & $-$	& $-p_k$\\
\hline
\end{tabular*}
\end{center}
\label{tab:planesymmetries:wf}
\end{table}%

In our code, the local densities and currents \eqref{eq:locdensities:rho}-\eqref{eq:locdensities:T}
entering the Skyrme EDF are constructed in the canonical basis. There,
they can be expressed as
\begin{subequations}
\begin{align}
\label{eq:colocdensities:rho}
\rho_q (\vec{r})
& =  \sum_{k,\sigma}  v_k^2 \;
      \Phi^\dagger_k (\vec{r},\sigma) \Phi_k (\vec{r},\sigma)~,
      \\
\label{eq:colocdensities:tau}
\tau_q (\vec{r})
& =  \sum_{k,\sigma} v_k^2 \;
      [ \vnabla \Phi_k (\vec{r},\sigma) ]^\dagger \cdot \vnabla \Phi_k (\vec{r},\sigma)~,
      \\
      \label{eq:colocdensities:J}
J_{q,\mu \nu}(\vec{r})
& =  -\tfrac{i}{2} \sum_{k,\sigma,\sigma'} v^2_k \,
      \{   \Phi^\dagger_k (\vec{r},\sigma) \, \sigma_{\nu;\sigma,\sigma'} \,
           [ \nabla_\mu \Phi_k (\vec{r},\sigma') ]
      \nn\\
     &\quad
     - [ \nabla_\mu \Phi_k (\vec{r},\sigma) ]^\dagger \, \sigma_{\nu;\sigma,\sigma'} \,
           \Phi_k (\vec{r},\sigma')
      \}~,
      \\
\label{eq:colocdensities:j}
\vec{j}_q (\vec{r})
& =  - \tfrac{i}{2}
      \sum_{k,\sigma} v_k^2 \;
      \{   \Phi^\dagger_k (\vec{r},\sigma) \, [\vnabla \Phi_k (\vec{r},\sigma) ]
      \nn\\
      &\quad
         - [ \vnabla \Phi_k (\vec{r},\sigma) ]^\dagger \, \Phi_k (\vec{r},\sigma)
      \}~,
      \\
\label{eq:colocdensities:s}
\vec{s}_q (\vec{r})
& =  \sum_{k,\sigma,\sigma'} v_k^2  \,
      \Phi^\dagger_k (\vec{r},\sigma) \, \Phi_k (\vec{r},\sigma')\, \hat\vsigma_{\sigma,\sigma'}~,
      \\
\label{eq:colocdensities:T}
\vec{T}_{q}(\vec{r})
& =  \sum_{k,\sigma,\sigma'} v^2_k \,
      [ \vnabla \Phi_k (\vec{r},\sigma)]^\dagger
      \cdot [ \vnabla \Phi_k (\vec{r},\sigma') ] \; \hat{\vsigma}_{\sigma,\sigma'} ~,
      \\
 \label{eq:colocdensities:F}
 \vec{F}_{q} (\vec{r})
& =  \tfrac{1}{2} \sum_{k,\sigma,\sigma'} v^2_k  \,
      \Big\{
         [ \vnabla \cdot \hat{\sigmavec}_{\sigma,\sigma'} \Phi_k (\vec{r},\sigma) ]^\dagger \,
         [ \vnabla \Phi_k (\vec{r},\sigma') ]
	 \nn\\
& \quad
        +[ \vnabla \Phi_k (\vec{r},\sigma) ]^\dagger \,
         [ \vnabla \cdot \hat{\sigmavec}_{\sigma,\sigma'} \Phi_k (\vec{r},\sigma') ]
      \Big\}  ~,
\end{align}
\end{subequations}
where $v_k^2$ are the occupation probabilities and $\mu,\nu=x,y,z$. Expressed in terms of the
functions $\Psi_{k,\alpha}$, the scalar local densities take the form
\begin{subequations}
\begin{align}
\rho(\vec{r})
& = \sum_{k} v_k^2 \sum_{\alpha=1}^4 \Psi_{k,\alpha}^2 \, ,
\\
\tau(\vec{r})
& = \sum_k v_k^2 \sum_{\alpha=1}^4 (\vnabla \Psi_{k,\alpha})^2 \, ,
\end{align}
\end{subequations}
whereas the vector densities are given by
\begin{subequations}
\begin{align}
\vec{j}(\vec{r})
&=\sum_k v_k^2 \big(   \Psi_{k,1}\vnabla \Psi_{k,2}
			     - \Psi_{k,2}\vnabla \Psi_{k,1}
	  \nn\\
	&\qquad	    + \Psi_{k,3}\vnabla \Psi_{k,4}
			    -  \Psi_{k,4}\vnabla \Psi_{k,3}\big) \, ,
\\			
s_x(\vec{r})
&=\sum_k 2 v_k^2 \big(\Psi_{k,1}\Psi_{k,3} + \Psi_{k,2}\Psi_{k,4}\big) \, ,
\\
s_y(\vec{r})
&=\sum_k 2 v_k^2 \eta \big(\Psi_{k,1}\Psi_{k,4} - \Psi_{k,2}\Psi_{k,3}\big) \, ,
\\
s_z(\vec{r})
&=\sum_k  v_k^2 \eta \big(\Psi_{k,1}^2+\Psi_{k,2}^2 - \Psi_{k,3}^2-\Psi_{k,4}^2\big) \, ,
\\
T_{q x}(\vec{r})
&=  \sum_{k} 2 v^2_k \,
      \big\{  [\vnabla \Psi_{k,1}] \cdot [\vnabla \Psi_{k,3}]
      \nn\\
 &\qquad
       + [\vnabla \Psi_{k,2}] \cdot [\vnabla \Psi_{k,4}] \big\}  \, ,
\\
T_{q y}(\vec{r})
&=  \sum_{k} 2 v^2_k \eta_k \,
      \big\{  [\vnabla \Psi_{k,1}] \cdot [\vnabla \Psi_{k,4}]
            \nn\\
 &\qquad
       - [\vnabla \Psi_{k,2}] \cdot [\vnabla \Psi_{k,3}] \big\} \, ,
\\
T_{q z}(\vec{r})
&=  \sum_{k}  v^2_k \eta_k \,
      \{  [\vnabla \Psi_{k,1}] \cdot [\vnabla \Psi_{k,1}]
            \nn\\
 &\qquad
       + [\vnabla \Psi_{k,2}] \cdot [\vnabla \Psi_{k,2}]
       - [\nabla \Psi_{k,3}] \cdot [\nabla \Psi_{k,3}]
             \nn\\
 &\qquad
       - [\nabla \Psi_{k,4}] \cdot [\nabla \Psi_{k,4}] \} \, ,
\\
F_{q,\mu} (\vec{r})
& =  \sum_{k } v^2_k \,
      \big\{   \phi_{k,1} [\nabla_\mu \Psi_{k,1}]
        + \phi_{k,2} [\nabla_\mu \Psi_{k,2}]
\nn\\
& \qquad
        + \phi_{k,3} [\nabla_\mu \Psi_{k,3}]
        + \phi_{k,4} [\nabla_\mu \Psi_{k,4}]
      \big\} \, ,
\end{align}
\end{subequations}
where we defined the spinor
\begin{equation}
\vnabla \cdot \hat\sigmavec \Psi_k
=\fourrow {\phi_{k,1} }{ \phi_{k,2} }{\phi_{k,3} } { \phi_{k,4}}
 =  \fourrow {\nabla_x \Psi_{k,3}+\eta(\nabla_y \Psi_{k,4}+\nabla_z \Psi_{k,1} )}
                     {\nabla_x \Psi_{k,4}-\eta( \nabla_y \Psi_{k,3}-\nabla_z \Psi_{k,2})}
                	  {\nabla_x \Psi_{k,1} -\eta(\nabla_y \Psi_{k,2}  +\nabla_z \Psi_{k,3})}
                     {\nabla_x \Psi_{k,2}+\eta(\nabla_y \Psi_{k,1}  -\nabla_z \Psi_{k,4})}~.
\end{equation}
Finally, the spin-current tensor densities are
\begin{subequations}
\begin{align}
J_{\mu x}
 = & \sum_{k} v_k^2
      \Big\{  \Psi_{k,1} [ \nabla_\mu \Psi_{k,4} ]
            - \Psi_{k,2} [ \nabla_\mu \Psi_{k,3} ]
	\nn\\
	& \qquad
            + \Psi_{k,3} [ \nabla_\mu \Psi_{k,2} ]
            - \Psi_{k,4} [ \nabla_\mu \Psi_{k,1} ]
      \Big\} \, ,
 \\
J_{\mu y}
 = & \sum_{k} v_k^2 ~ \eta_k
      \Big\{ - \Psi_{k,1} [ \nabla_\mu \Psi_{k,3} ]
             - \Psi_{k,2} [ \nabla_\mu \Psi_{k,4} ]
	\nn\\
	& \qquad
             + \Psi_{k,3} [ \nabla_\mu \Psi_{k,1} ]
             + \Psi_{k,4} [ \nabla_\mu \Psi_{k,2} ]
      \Big\} \, ,
 \\
J_{\mu z}
 = & \sum_{k } v_k^2 ~ \eta_k
      \Big\{  \Psi_{k,1} [ \nabla_\mu \Psi_{k,2} ]
            - \Psi_{k,2} [ \nabla_\mu \Psi_{k,1} ]
	\nn\\
	& \qquad
            - \Psi_{k,3} [ \nabla_\mu \Psi_{k,4} ]
            + \Psi_{k,4} [ \nabla_\mu \Psi_{k,3} ]
      \Big\} \, .
\end{align}
\end{subequations}
for $\mu = x$, $y$, $z$. The symmetries \eqref{eq:symm:p},
\eqref{eq:symm:rz} and \eqref{eq:symm:sty} of the single-particle
wave functions impose reflection symmetries on the components of
the local densities and currents, which are listed in
Table~\ref{tab:planesymmetries:den}.

\begin{table}[t!]
\caption{Parities of the nucleon densities with respect
to the $x=0$, $y=0$, and $z=0$ planes}
\begin{center}
\begin{tabular*}{0.7\columnwidth}{@{\extracolsep{\fill}}lccc}
\hline
~									& $x$	&	$y$	&	$z$	\\
\hline
$\rho$, $\tau$							&	$+$	&	$+$	&	$+$	 \\
$s_{x}$, $T_{x}$, $F_{x}$					&	$-$	&	$+$	&       $-$	\\
$s_{y}$, $T_{y}$, $F_{y}$					&	$+$	&	$-$	&	 $-$	\\
$s_{z}$, $T_{z}$, $F_{z}$					&	$+$	&	$+$	&	 $+$	\\
$j_{x}$, $J_{xz}$, $J_{zx}$				&	$+$	&	$-$	&	$+$	 \\
$j_{y}$								&	$-$	&	$+$	&	$+$	\\
$j_{z}$, $J_{xx}$, $J_{yy}$, $J_{zz}$		&	$-$	&	$-$	&       $-$	\\
$J_{xy}$, $J_{yx}$						&	$+$	&	$+$	&	$-$	 \\
$J_{yz}$, $J_{zy}$						&	$-$	&	$+$	&	$+$	 \\
\hline
\end{tabular*}
\end{center}
\label{tab:planesymmetries:den}
\end{table}

%
%
\section{The Landau-Migdal interaction}
\label{app:LM interaction}

%
%
\subsection{General considerations}
Landau theory for normal Fermi liquids provides a framework for
the study of the long-wavelength response of a many-body system~\cite{migdal67, olsson02}. In this framework, the residual interaction is
provided by the so-called Landau interaction. It determines the
response of the system but cannot be used to calculate its ground state.
Based on this, Migdal developed the Landau-Migdal theory of finite
Fermi systems and applied it successfully to study collective
modes in atomic nuclei \cite{migdal67}.
The residual Landau-Migdal interaction acts between two
particles with momenta $\vec{q}_{1}$ and $\vec{q}_{2}$ at the Fermi surface,
$|\vec{q}_{1}|=|\vec{q}_{2}| = k_{F}$, and is given by
\begin{align}
\label{eq:app:LM interaction}
v_{res}&(\vec{q}_{1},\vec{q}_{2})
 =
 \nn\\
 N_{0}
    \Big\{&  F (\vec{q}_{1},\vec{q}_{2})
          + F'(\vec{q}_{1},\vec{q}_{2}) \, (\tauvec_{1}\cdot\tauvec_{2}) \nn \\
&         + G (\vec{q}_{1},\vec{q}_{2}) \, (\sigmavec_{1}\cdot\sigmavec_{2})
          + G'(\vec{q}_{1},\vec{q}_{2}) \, (\sigmavec_{1}\cdot\sigmavec_{2}) (\tauvec_{1}\cdot\tauvec_{2})
    \nn\\
&         + \frac{q^{2}}{k_{F}^{2}}
            \big[ H(\vec{q}_{1},\vec{q}_{2}) + H'(\vec{q}_{1},\vec{q}_{2}) \,  (\tauvec_{1}\cdot\tauvec_{2}) \big] \,
            S_{12}(\hat{\vec{q}})
    \Big\}
\, .
\end{align}
The normalization factor is defined as the average level density
$N_{0} \equiv 2 k_{F} m^{*}/ \hbar^{2}\pi^{2}$ at at the Fermi momentum
$k_{F} = (\tfrac{3}{2}\pi^{2}\rho_{0})^{1/3}$,
with $m^{*}_0$ being the isoscalar effective mass. The tensor operator
\begin{equation}
S_{12} (\hat{\vec{q}})
= 3 \, (\sigmavec_{1} \cdot \hat{\vec{q}}) \, (\sigmavec_{2} \cdot \hat{\vec{q}})
  - \sigmavec_{1} \cdot \sigmavec_{2}
\end{equation}
depends also on the angle between the direction $\hat{\vec{q}}=\vec{q}/|\vec{q}|$
 of the momentum transfer $\vec{q} = \vec{q}_{1}-\vec{q}_{2}$ and the direction of
the particles' spins. Because the single-particle momenta are restricted to the
Fermi surface, the parameters $F$, $F'$, $G$, $G'$, $H$, $H'$ depend only on
the angle between $\vec{q}_{1}$ and $\vec{q}_{2}$ and can be expanded into Legendre
polynomials, i.e.\
\begin{equation}
F = \sum_{\ell} f_{\ell} \, P_{\ell} (\cos \theta)
\, ,
\end{equation}
and similar for the other parameters.

%
%

\subsection{The Landau parameters derived from the Skyrme energy density functional}
\label{app:landau parameters skyrme}

To establish the relationship between the Landau parameters in \eqref{eq:app:LM interaction} and the coupling constants of the Skyrme EDF in symmetric nuclear matter, we follow the procedure outlined in \cite{liu91,lesinski06}. Starting from the Skyrme energy density functional, the residual particle-hole interaction in coordinate space is obtained as
\begin{align}
\langle  \vec{r}'_{1}\sigma_{1}'\tau_{1}',& \vec{r}'_{2}\sigma_{2}'\tau_{2}'|V_{ph}|\vec{r}_{1}\sigma_{1}\tau_{1}, \vec{r}_{2}\sigma_{2}\tau_{2}\rangle =\nn\\
&\frac{\delta^{2}\mathcal{E}_{Sk}}
{\delta\rho(\vec{r}_{1}\sigma_{1}\tau_{1},\vec{r}'_{1}\sigma'_{1}\tau'_{1})
\delta\rho(\vec{r}_{2}\sigma_{2}\tau_{2},\vec{r}'_{2}\sigma'_{2}\tau'_{2})}\, .
\end{align}
From this, the momentum-space matrix elements of the residual Landau interaction \eqref{eq:app:LM interaction} are obtained through
the substitutions $\vnabla_{1}=i\vec{q}_{1}$, $\vnabla_{2}=i\vec{q}_{2}$, $\vnabla'_{1}=-i\vec{q}_{1}$, and $\vnabla'_{2}=-i\vec{q}_{2}$
in the Landau limit where initial and final momenta $\vec{q}_{1}$ and $\vec{q}_{2}$ are both on the Fermi surface. This results in momentum-space matrix elements of the form
\begin{align}\label{eq:app:res interaction}
\langle \vec{q}_{1},\vec{q}_{2}|V_{ph}&|\vec{q}_{1},\vec{q}_{2}\rangle
 = \nn\\
    \Big\{  &W^{ss} (\vec{q}_{1},\vec{q}_{2})
          + W^{sv}(\vec{q}_{1},\vec{q}_{2}) \, (\tauvec_{1}\cdot\tauvec_{2}) \nn \\
&       + \sum_{\mu,\nu}W_{\mu\nu}^{vs} (\vec{q}_{1},\vec{q}_{2}) \, \sigma^{\mu}_{1}\sigma^{\nu}_{2}\nn\\
 &      + \sum_{\mu,\nu} W_{\mu\nu}^{vv}(\vec{q}_{1},\vec{q}_{2}) \,(\tauvec_{1}\cdot\tauvec_{2})\, \sigma^{\mu}_{1}\sigma^{\nu}_{2}
    \Big\}
\, .
\end{align}
with
\begin{align}
W^{ss}(\vec{q}_{1},\vec{q}_{2})
		=&\, 2C_{0}^{\rho}[\rho_{0}] + 4 \frac{\partial C^{\rho}_{0}}{\partial\rho_{0}}\rho_{0}
		+  \frac{\partial^{2} C^{\rho}_{0}}{\partial\rho_{0}^{2}}\rho^{2}_{0}\nn\\
		&+ C_{0}^{\tau} (\vec{q}_{1}-\vec{q}_{2})^{2}\\
W^{vs}(\vec{q}_{1},\vec{q}_{2})
		=&\, 2C_{1}^{\rho}[\rho_{0}]
		+ C_{1}^{\tau} (\vec{q}_{1}-\vec{q}_{2})^{2}\nn\\	
W_{\mu\nu}^{vs} (\vec{q}_{1},\vec{q}_{2})	
		=&\, \Big(2C^{s}_{0}[\rho_{0}] + C^{T}_{0} (\vec{q}_{1}-\vec{q}_{2})^{2}\Big)\delta_{\mu\nu}\nn\\
		  & + C^{F}_{0} 	(\vec{q}_{1}-\vec{q}_{2})_{\mu} (\vec{q}_{1}-\vec{q}_{2})_{\nu}	\\
W_{\mu\nu}^{vv} (\vec{q}_{1},\vec{q}_{2})	
		=&\, \Big(2C^{s}_{1}[\rho_{0}] + C^{T}_{1} (\vec{q}_{1}-\vec{q}_{2})^{2}\Big)\delta_{\mu\nu}\nn\\
		  & + C^{F}_{1} 	(\vec{q}_{1}-\vec{q}_{2})_{\mu} (\vec{q}_{1}-\vec{q}_{2})_{\nu}		\, .	
\end{align}
The tensor operator $S_{12}(\hat{q})$ is easily recognized in \eqref{eq:app:res interaction} when we rewrite $W_{\mu\nu}^{vs}$ and $W_{\mu\nu}^{vv}$
\begin{align}
W_{\mu\nu}^{vs} (\vec{q}_{1},\vec{q}_{2})	
		=&\, \Big(2C^{s}_{0}[\rho_{0}] + (C^{T}_{0}+\frac{1}{3} C^{F}_{0}) (\vec{q}_{1}-\vec{q}_{2})^{2}\Big)\delta_{\mu\nu}\nn\\
		  & + \frac{1}{3}C^{F}_{0} q^{2} \big(3\hat{q}_{\mu} \hat{q}_{\nu}-\delta_{\mu\nu}\big)	\\
W_{\mu\nu}^{vv} (\vec{q}_{1},\vec{q}_{2})	
		=&\, \Big(2C^{s}_{1}[\rho_{0}] + (C^{T}_{1}+\frac{1}{3} C^{F}_{1}) (\vec{q}_{1}-\vec{q}_{2})^{2}\Big)\delta_{\mu\nu}\nn\\
		  & + \frac{1}{3}C^{F}_{1} q^{2} \big(3\hat{q}_{\mu} \hat{q}_{\nu}-\delta_{\mu\nu}\big)	\, .
\end{align}
Because $\vec{q}_{1}$ and $\vec{q}_{2}$ are both on the Fermi surface, $(\vec{q}_{1}-\vec{q}_{2})^{2}$ can be rewritten as $2k_{F}(1-\cos\theta)$, with $\theta$ being the angle between $\vec{q}_{1}$ and $\vec{q}_{2}$. A straightforward comparison between
 \eqref{eq:app:LM interaction} and  \eqref{eq:app:res interaction} then finally gives us the relation between the Landau parameters and the coupling constants in the Skyrme EDF
 \begin{align}
f_{0}
&= N_{0} \Big[ 2C_{0}^{\rho}[\rho_{0}] + 4 \frac{\partial C^{\rho}_{0}}{\partial\rho_{0}}\rho_{0}
		+  \frac{\partial^{2} C^{\rho}_{0}}{\partial\rho_{0}^{2}}\rho^{2}_{0}
		+ 2 C_{0}^{\tau} k_{F}^{2} \Big]\, ,
		\nonumber\\
f_{1}
&= -2 N_{0} C_{0}^{\tau} k_{F}^{2} 	\, ,
	\nonumber\\
f'_{0}
&= N_{0} \Big[ 2C_{1}^{\rho}[\rho_{0}] +  2 C_{1}^{\tau} k_{F}^{2} \Big] \, ,
	\nonumber\\
f'_{1}
&= -2 N_{0} C_{1}^{\tau} k_{F}^{2} \, ,
	\nonumber\\
g_0
& = 2 N_0 \big[ C^s_0 + \big(C^T_0 + \tfrac{1}{3} \, C^F_0 \big) \, k_F^2
         \big] \, ,
       \nonumber  \\
g^\prime_0
& = 2 N_0 \big[ C^s_1 + \big(C^T_1 + \tfrac{1}{3} \, C^F_1 \big) \, k_F^2
         \big] \, ,
       \nonumber  \\
g_1
& =  - 2 N_0 \; \big( C^T_0 + \tfrac{1}{3} \, C^F_0 \big) \, k_F^2 \, ,
       \nonumber  \\
g^\prime_1
& =  - 2 N_0 \; \big( C^T_1 + \tfrac{1}{3} \, C^F_1 \big) \, k_F^2 \, ,
       \nonumber  \\
h_0
& =   \tfrac{1}{3} \, N_{0}  \, C^F_0 \, k_F^2 \, ,
        \nonumber \\
h^\prime_0
& =   \tfrac{1}{3} \, N_{0}  \, C^F_1 \, k_F^2 \, .
\end{align}

\end{appendix}
%
%

\end{document}